\documentclass[12pt,a4paper]{article}

\usepackage[T1]{fontenc}
\usepackage{amsmath,amssymb,url}
\usepackage{graphicx} 
\usepackage{booktabs}
\usepackage{array}
\usepackage{xspace}
\usepackage{units}
\usepackage{xcolor}
\usepackage{acronym}
\usepackage{multirow}

\usepackage{hyperref}
\hypersetup{%
    colorlinks,
    citecolor=blue,
    filecolor=black,
    linkcolor=black,
    urlcolor=black
}

\usepackage{cleveref}

\makeatletter
\newif\if@preliminary
\@preliminaryfalse
\def\preliminary{\@preliminarytrue}
\def\preprintno#1{\def\@preprintno{#1}}
\def\address#1{\def\@address{#1}}
\def\email#1#2{\thanks{\tt #1@{}#2}}
\def\abstract#1{\def\@abstract{#1}}
\renewcommand\abstractname{ABSTRACT}
\newlength\preprintnoskip
\newlength\abstractwidth
\@titlepagetrue
\renewcommand\maketitle{\begin{titlepage}%
  \let\footnotesize\small
  \hfill\parbox{\preprintnoskip}{%
  \begin{flushright}\@preprintno\end{flushright}}\hspace*{1cm}
  \vskip 60\p@
  \begin{center}%
    {\Large\bf\boldmath \@title \par}\vskip 1cm%
    {\sc\@author \par}\vskip 3mm%
    {\@address \par}%
    \if@preliminary
      \vskip 2cm {\large\sf PRELIMINARY DRAFT \par \@date}%
    \fi
  \end{center}\par
  \@thanks
  \vfill
  \begin{center}%
    \parbox{\abstractwidth}{\centerline{\abstractname}%
    \vskip 3mm%
    \@abstract}
  \end{center}
  \end{titlepage}%
  \setcounter{footnote}{0}%
  \let\thanks\relax\let\maketitle\relax
  \gdef\@thanks{}\gdef\@author{}\gdef\@address{}%
  \gdef\@title{}\gdef\@abstract{}\gdef\@preprintno{}
}%
\def\@citex[#1]#2{\if@filesw\immediate\write\@auxout{\string\citation{#2}}\fi
  \def\@citea{}\@cite{\@for\@citeb:=#2\do
    {\@citea\def\@citea{,\penalty\@m}\@ifundefined
       {b@\@citeb}{{\bf ?}\@warning
       {Citation `\@citeb' on page \thepage \space undefined}}%
\hbox{\csname b@\@citeb\endcsname}}}{#1}}
\def\citerange{\@ifnextchar [{\@tempswatrue\@citexr}{\@tempswafalse\@citexr[]}}
\def\@citexr[#1]#2{\if@filesw\immediate\write\@auxout{\string\citation{#2}}\fi
  \def\@citea{}\@cite{\@for\@citeb:=#2\do
    {\@citea\def\@citea{--\penalty\@m}\@ifundefined
       {b@\@citeb}{{\bf ?}\@warning
       {Citation `\@citeb' on page \thepage \space undefined}}%
\hbox{\csname b@\@citeb\endcsname}}}{#1}}
\long\def\@makecaption#1#2{%
  \sbox\@tempboxa{#1: \emph{#2}}%
  \ifdim \wd\@tempboxa >\hsize
    #1: \emph{#2}\par
  \else
    \hbox to\hsize{\hfil\box\@tempboxa\hfil}%
  \fi
  \vskip\belowcaptionskip}
\def\fmslash{\@ifnextchar[{\fmsl@sh}{\fmsl@sh[0mu]}}
\def\fmsl@sh[#1]#2{%
  \mathchoice
    {\@fmsl@sh\displaystyle{#1}{#2}}%
    {\@fmsl@sh\textstyle{#1}{#2}}%
    {\@fmsl@sh\scriptstyle{#1}{#2}}%
    {\@fmsl@sh\scriptscriptstyle{#1}{#2}}}
\def\@fmsl@sh#1#2#3{\m@th\ooalign{$\hfil#1\mkern#2/\hfil$\crcr$#1#3$}}
\makeatother

\newcommand\ltap{\
  \raise.3ex\hbox{$<$\kern-.75em\lower1ex\hbox{$\sim$}}\ }
\newcommand\gtap{\
  \raise.3ex\hbox{$>$\kern-.75em\lower1ex\hbox{$\sim$}}\ }

\newcommand\simge{\mathrel{%
   \rlap{\raise 0.511ex \hbox{$>$}}{\lower 0.511ex \hbox{$\sim$}}}}
\newcommand\simle{\mathrel{
   \rlap{\raise 0.511ex \hbox{$<$}}{\lower 0.511ex \hbox{$\sim$}}}}

\newcommand\be{\begin{equation}}
\newcommand\ee{\end{equation}}
\newcommand\bea{\begin{eqnarray}}
\newcommand\eea{\end{eqnarray}}
\newcommand\ba{\begin{array}}
\newcommand\ea{\end{array}}

\newcommand\yt{\ensuremath{y_t}\xspace}

\newcommand\epem{e^+e^-}

\newcommand\jb{\ensuremath{j_{b}}\xspace}
\newcommand\jbbar{\ensuremath{j_{\bar b}}\xspace}

\newcommand\pTtoprec{\ensuremath{p_{{\rm T}, W^+\jb}}\xspace}
\newcommand\mtoprec{\ensuremath{m_{W^+\jb}}\xspace}

\newcommand\pTblp{\ensuremath{p_{{\rm T},  \ell^+ \jb}}\xspace}
\newcommand\pTbb{\ensuremath{p_{{\rm T}, b\bar b}}\xspace}

\newcommand\mblp{\ensuremath{m_{\ell^+\jb}}\xspace}
\newcommand\mbl{\ensuremath{m_{\ell\jb}}\xspace}
\newcommand\thetablp{\ensuremath{\cos\theta_{\ell^+\jb}}\xspace}
\newcommand\thetablm{\ensuremath{\cos\theta_{\ell^-\jbbar}}\xspace}
\newcommand\thetabl{\ensuremath{\cos\theta_{\ell\jb}}\xspace}
\newcommand\mtt{\ensuremath{m_{t\bar t}}\xspace}
\newcommand\mttrec{\ensuremath{m_{ W^+W^-\jb\jbbar }}\xspace}

\def\bq{\begin{equation}}
\def\eq{\end{equation}}
\def\ba{\begin{eqnarray}}
\def\ea{\end{eqnarray}}

\newcommand{\Op}{\mathcal{O}}

\newcommand{\cw}{c_w}
\newcommand{\sw}{s_w}

\newcommand{\GeV}{{\ensuremath\rm GeV}}

\newcommand{\fb}{{\ensuremath\rm fb}}

\newcommand{\ValMeV}[1]{\unit[#1]{MeV}}
\newcommand{\ValGeV}[1]{\unit[#1]{GeV}}
\newcommand{\Valfb}[1]{\unit[#1]{fb}}

\newcommand{\wz}{\textsc{Whizard}\xspace}
\newcommand{\va}{\textsc{Vamp}\xspace}
\newcommand{\po}{\textsc{Powheg}\xspace}
\newcommand{\om}{\textsc{O'Mega}\xspace}
\newcommand{\ol}{\textsc{OpenLoops}\xspace}
\newcommand{\re}{\textsc{Recola}\xspace}
\newcommand{\go}{\textsc{GoSam}\xspace}
\newcommand{\hepmc}{\textsc{HepMC}\xspace}
\newcommand{\rv}{\textsc{Rivet}\xspace}
\newcommand{\fastjet}{\textsc{FastJet}\xspace}
\newcommand{\munich}{\textsc{Munich}\xspace}

\newcommand{\sqrts}{\ensuremath{\sqrt{s}}\xspace}

\newcommand{\mis}{\ensuremath \text{mism}}

\newcommand{\ttbar}{\ensuremath{t \bar t}\xspace}
\newcommand{\ttbarh}{\ensuremath{\ttbar H}\xspace}
\newcommand{\bbww}{\ensuremath{W^+ W^- b \bar{b}}\xspace}
\newcommand{\llllbb}{\ensuremath{\mu^+ \nu_\mu e^- \bar\nu_e b \bar{b}}\xspace}
\newcommand{\bbwwH}{\ensuremath{\bbww H}\xspace}
\newcommand{\llllbbH}{\ensuremath{\llllbb H}\xspace}
\newcommand{\ttbarH}{\ttbarh\xspace}
\newcommand{\tth}{{\ttbarh}\xspace}
\newcommand{\ttH}{{\ttbarh}\xspace}
\newcommand{\wwbb}{\bbww\xspace}
\newcommand{\wbwb}{\bbww\xspace}
\newcommand{\eett}{\ensuremath{\epem \to \ttbar}\xspace}
\newcommand{\eewwbb}{\ensuremath{\epem \to \bbww}\xspace}
\newcommand{\eewwbbH}{\ensuremath{\epem \to \bbwwH}\xspace}
\newcommand{\eellllbb}{\ensuremath{\epem \to \llllbb}\xspace}
\newcommand{\eellllbbH}{\ensuremath{\epem \to \llllbbH}\xspace}
\newcommand{\eetth}{\ensuremath{\epem \to \ttbarh}\xspace}
\newcommand{\wwbbH}{\bbwwH\xspace}

\newcommand{\ri}{\mathrm{i}}
\newcommand{\GF}{{G_\mu}}

\newcommand{\kT}{k_{\mathrm{T}}}

\newcommand{\rR}{\mathrm{R}}
\newcommand{\Pt}{\ensuremath{\mathrm{t}}\xspace}

\newcommand{\PW}{\ensuremath{\mathrm{W}}\xspace}

\newcommand{\PZ}{\ensuremath{\mathrm{Z}}\xspace}
\newcommand{\PH}{\ensuremath{\mathrm{H}}\xspace}

\newcommand{\beqar}{\begin{eqnarray}}
\newcommand{\eeqar}{\end{eqnarray}}
\newcommand{\beq}{\begin{equation}}
\newcommand{\eeq}{\end{equation}}
\newcommand{\bit}{\begin{itemize}}
\newcommand{\eit}{\end{itemize}}

\newcommand\order[1]{\ensuremath{\mathcal{O}\left(#1\right)}}
\newcommand\Rcite[1]{Ref.~\cite{#1}}
\newcommand\RRcite[1]{Refs.~\cite{#1}}

\def\refeq#1{\mbox{(\ref{#1})}}

\usepackage[babel]{csquotes}
\usepackage[style=numeric,
	maxnames=3,
	minnames=3,
	backend=bibtex,
	citestyle=numeric-comp,
	url=false,
	maxbibnames=5
]{biblatex}
\makeatletter
\def\blx@maxline{77}
\makeatother

\DeclareFieldFormat{journaltitle}{{#1},}
\DeclareFieldFormat[inbook,thesis]{title}{{#1}\addperiod}                                             
\DeclareFieldFormat[article]{title}{\mkbibemph{#1}}                                                   
\DefineBibliographyStrings{english}{%
    backrefpage  = {see p.}, 
    backrefpages = {see pp.} 
}
\DeclareFieldFormat{eprint}{%
  \iffieldundef{eprintclass}                                                                          
    {}
    {\thefield{eprintclass}/}%
    #1%
}
\DeclareFieldFormat{eprint:arxiv}{[\printfield{eprint}]}
\DeclareFieldFormat[article]{volume}{\textbf{#1}\space}                                             
\AtEveryBibitem{\clearfield{month}}

\renewbibmacro{in:}{}                                                                                 

\def\reldiagwidth{0.27}
\def\relplotwidth{0.48}

\newacro{NWA}{narrow-width-approximation}
\newacro{FtW}{finite-top-width}

\newacro{NLL}{next-to-leading-logarithmic}

\begin{document}

\date{\today}

\preprintno{DESY 16-140 \\ SI-HEP-2016-24 \\ ZU-TH 30/16   }

\title{NLO QCD Predictions for off-shell $\ttbar$ and $\ttbarh$ Production and Decay at a Linear Collider}

\author{Bijan Chokouf\'{e} Nejad\email{bijan.chokoufe}{desy.de}$^a$,
  Wolfgang Kilian \email{kilian}{physik.uni-siegen.de}$^b$,
  Jonas M. Lindert \email{lindert}{physik.uzh.ch}$^c$,\\
  Stefano Pozzorini \email{pozzorin}{physik.uzh.ch}$^c$,
  J\"urgen Reuter\email{juergen.reuter}{desy.de}$^a$,
  Christian Weiss\email{christian.weiss}{desy.de}$^{a,b}$}
  
\address{\it%
$^a$DESY Theory Group, \\
  Notkestr. 85, D-22607 Hamburg, Germany
\\[.5\baselineskip]
$^c$ Emmy-Noether-Campus, \\
Walter-Flex-Str. 3, 57068 Siegen, Germany
\\[.5\baselineskip]
$^b$ Physik-Institut, Universit\"at Z\"urich, \\
Winterthurerstrasse 190, CH-8057 Z\"urich, Switzerland
}

\abstract{%
  We present predictions for $\ttbar$ and $\ttbarh$ production and
  decay at future lepton colliders including non-resonant and
  interference contributions up to next-to-leading order (NLO) in
  perturbative QCD. The obtained precision predictions 
   are necessary for a future precise determination of the top-quark
   Yukawa coupling, and allow for top-quark phenomenology in the
   continuum at an unprecedented level of accuracy. Simulations are
   performed with the automated NLO Monte-Carlo framework \wz
   interfaced to the \ol matrix element generator. 
}

\maketitle

\tableofcontents

\section{Introduction}
\label{sec:introduction}
The top quark is the heaviest particle of the Standard Model
(SM), and its detailed study offers great potential to probe the
electroweak, flavor and Higgs sector. The close connection between the
Higgs boson and the top quark is most apparent for the (meta-)
stability of the electroweak vacuum, which crucially depends on $m_t$
and $m_H$~\cite{Ellis:2009tp, Degrassi:2012ry, Bednyakov:2015sca}.
A precise determination of top-quark properties is thus a powerful
opportunity to find possible hints of new physics and has far reaching
consequences for our understanding of the universe. However, at hadron
colliders like the LHC many quantities in the top-quark sector, like
the top-quark mass, forward-backward asymmetry or the top Yukawa
coupling can only be measured with a limited precision. A future
linear lepton collider, such as the proposed International Linear
Collider~(ILC)~\cite{Baer:2013cma, Behnke:2013lya} 
or Compact Linear Collider~(CLIC)~\cite{Linssen:2012hp},
on the other hand will reach unprecedented precision in the
electroweak and top sector.

With respect to top physics, the two most interesting processes 
to be studied in lepton collisions are top-pair production with and
without an associated Higgs boson. Top-pair production
allows to measure the top-quark mass at threshold
in a theoretically well defined short distance scheme, like the
1S~\cite{hep-ph/9904468} or PS 
scheme~\cite{hep-ph/9804241}, with uncertainties at or below 100
MeV~\cite{Seidel:2013sqa,Horiguchi:2013wra,1603.04764}.
Associated $\ttbar H$ production is our best handle to measure the top
Yukawa coupling with per cent level precision, see
e.g.~\Rcite{Agashe:2013hma,1409.7157}. Obviously, these physical
parameters can only be extracted with this level of precision when the
theoretical uncertainties at least match their experimental 
counterparts.

Due to the relatively large top width, which comes exclusively from the 
decay into a bottom quark and a W boson, top quarks decay before they
can form bound states. 
The produced W boson decays further via hadronic or leptonic channels,
whereas the bottom quark hadronizes and can be identified as a tagged
jet. Especially in the clean lepton collider environment, the charge
of the b-jet can be reconstructed with reasonably high
efficiency~\cite{1604.08122}. A consistent treatment of the associated
finite width effects is both a conceptionally as well as
computationally nontrivial problem. Within the so-called narrow-width
approximation (NWA), top quarks are produced on-shell and decay
subsequently according to their (spin correlated) branching ratios. 
Higher-order QCD predictions for on-shell top-pair production are
well-known, the current best predictions being N$^3$LO~\cite{Kiyo:2009gb} at the inclusive and NNLO at the fully differential level~\cite{1410.3165}.
First NLO electroweak corrections have been obtained in
\Rcite{Fleischer:2003kk}.
For top-pair production in association with a Higgs boson, there are comprehensive studies of NLO QCD corrections available in \Rcite{Dittmaier:1998dz}.
First inclusive combined electroweak and QCD corrections have been
computed in \Rcite{Belanger:2003nm}, followed by an in-depth study in
\Rcite{hep-ph/0309274}. 

While computationally simple, the NWA has the obvious drawback that 
various non-resonant background processes are not included.
For off-shell \ttbar or \tth production, however, especially
single-top resonances can contribute significantly and can hardly
be distinguished experimentally from
double-resonant contributions~\cite{1411.2355}. Furthermore, off-shell
effects can only be treated approximatively via a Breit-Wigner
parameterization, as in \Rcite{hep-ph/9504434}. 
Non-resonant contributions and finite width effects 
can be consistently taken into account employing the complex-mass
scheme~\cite{Denner:2005fg}, which guarantees gauge invariance at NLO
-- at the price of increased computational complexity.
Such a calculation for the process \eewwbb{} at NLO QCD has first been
presented in \Rcite{0802.4124}. It has recently been reevaluated
in~\Rcite{1511.02350}, with the aim of extracting the top-quark width via
ratios of single- to double-resonant signal regions. 

In this paper, we study top-pair and Higgs associated top-pair
production and decay including non-resonant contributions, off-shell
effects and interferences at NLO. The simulation is done with the
multi-purpose event generator \wz~\cite{0708.4233, hep-ph/0102195},
which has been extended to perform automated NLO calculations.
In this framework, we compare the on-shell processes \eett\ and \eetth\
with the off-shell processes \eewwbb\ and \eewwbbH. At the
differential level, the full processes including leptonic decays are
considered, i.e. \eellllbb\ and \eellllbbH. To our knowledge, NLO
studies of \eewwbbH\ or the complete off-shell processes $\epem \to
b\bar{b} 4f$ or $\epem \to b \bar{b} 4f H$ have not been performed
previously in the literature.  In contrast, at hadron colliders
off-shell top-pair production has been studied
in~\RRcite{Bevilacqua:2010qb, Denner:2010jp,
1207.5018,Heinrich:2013qaa, Frederix:2013gra,1312.0546}. Furthermore,
employing the resonance-aware method of \Rcite{1509.09071},
the process $pp \to b\bar{b} 4f$ has been matched consistently to
parton showers, as presented recently in \Rcite{Jezo:2016ujg}.
For hadron colliders, corresponding NLO QCD corrections to top-quark
pair production in association with a Higgs boson~\cite{1506.07448} or
a jet~\cite{Bevilacqua:2015qha} including leptonic decays have also
been studied.

While at hadron colliders top-pairs originate from QCD production,
at lepton colliders they are produced via electroweak interactions.
This implies that a fixed-order computation of the off-shell processes
at a lepton collider comprises a considerably larger set of
irreducible electroweak background processes. Such processes involve
(very) narrow resonances, like e.g. $H\,\to\,b \bar{b}$. 
In NLO computations, resonances with very small widths can severely hamper
the quality of the infrared (IR) subtraction and consequently influence the
convergence and quality of the integration. In order to have these
resonance effects under control, in \wz{} we have implemented an
automatized version of the resonance-aware scheme of
\Rcite{1509.09071}. This is also a prerequisite for a future
consistent matching of off-shell processes with parton showers.

Besides the phenomenological relevance of the presented results
-- in particular for top quark mass measurements in the continuum and
measurements of the top Yukawa coupling -- this paper demonstrates the
progress on \wz{} as a fully automated NLO event generator.  
\wz{} has for a long time been a (high-multiplicity) tree-level event
generator, where besides its usage in all areas of lepton collider
physics, its focus on hadron colliders had been mostly on beyond the
Standard Model (BSM) physics. NLO QCD corrections have only been
considered for the explicit study of $pp \to
b\bar{b}b\bar{b}$~\cite{0910.4379, 1105.3624}.  
Furthermore, NLO QED effects have been studied on fixed order as well as by 
resumming soft photons for chargino production at the
ILC~\cite{hep-ph/0607127, 0803.4161}. Apart from \wz, various
collaborations are including generic NLO simulations into their event
generators. 
This has been made possible by tremendous advances in the automation of the
computation of one-loop amplitudes during the last decade. Publicly
available one-loop providers (OLPs) such as
\textsc{Helac-1Loop}~\cite{vanHameren:2009dr},
\ol~\cite{Cascioli:2011va}, \go~\cite{1404.7096}, 
\re~\cite{1211.6316,1605.01090}
or {\sc MadLoop}~\cite{Hirschi:2011pa}
can compute arbitrary virtual matrix elements in the SM, though in
practice limited by computing power. 
Complete NLO QCD support has so far been achieved within the frameworks of
\textsc{Helac-NLO}~\cite{Bevilacqua:2011xh},
\textsc{Madgraph5\_aMC@NLO}~\cite{1405.0301},
\textsc{Sherpa}~\cite{0811.4622} and \textsc{Herwig7}~\cite{1512.01178}.
Finally, we want to remark on a topic that is both closely related to fixed
order NLO predictions and important for the description of \ttbar{} and
$t\bar{t}H$.
At threshold, these processes actually require the inclusion of bound state
effects that can be treated in non-relativistic QCD\@.
Here, the exchange of soft gluons leads to Coulomb singularities that have to
be resummed. In order to take this into account, \wz{} ships with 
the \textsc{Toppik} program~\cite{hep-ph/9904468}, which can be used
to compute resummed form factors up to \ac{NLL} accuracy, as presented in
\cite{1411.7318}. Inclusive NNLO calculations at threshold have been
compared in~\Rcite{Hoang:2000yr}. The state-of-the-art of fixed-order
corrections has been recently improved to
$\text{N}^3\text{LO}$~\cite{Beneke:2015kwa} and at the
resummed level to NNLL~\cite{1309.6323}. However, these results are
only accurate in the threshold region, while \wz\ can describe both
the threshold and continuum domain by using a smooth matching 
approach. \wz{}'s NLO capabilities are used hereby to compute the
radiative corrections to the top decay in a factorized approach as
well as to obtain the full $\wwbb$ process at NLO\@.
A preliminary status thereof is presented in
\Rcite{1601.02459,Reuter:2016ohp}, while an in-depth study of the
threshold matching in \wz{} is in
preparation~\cite{chokoufe:2016future}.  In this paper, we focus on a 
fixed-order description of the continuum, while pointing out 
regions where threshold effects become important. 

The paper is organized as follows.
In section~\ref{sec:setup}, we describe the setup of the calculation.
In section~\ref{sec:resonances}, we address the issue of
resonance-aware subtraction and its implementation in \wz. 
In section~\ref{sec:pheno}, the phenomenology of \ttbar{} and \tth{}
is briefly reviewed. The employed input parameters, scale choices and
phase-space cuts as well as an overview of the performed validations
can be found in section~\ref{sec:paramter_setup_and_validation}.
The main phenomenological results of this paper can be found in
section~\ref{sec:inclusive_predictions}
and~\ref{sec:differential_predictions}, where in section
\ref{sec:inclusive_predictions} we focus on results at the inclusive
level, while in section \ref{sec:differential_predictions}
corresponding differential predictions are investigated. We discuss
scale variations for the NLO QCD corrections, show results for
polarized lepton beams and discuss the influence of the NLO QCD
corrections on the extraction of the top Yukawa coupling. 
Additional differential results as well as technical details on the
resonance-aware subtraction can be found in the appendix.
Our conclusions are presented in section~\ref{sec:conclusions}.

\section{Setup of the calculation}
\label{sec:setup}
The predictions presented in this paper are obtained with the
automated Monte Carlo framework \wz{} combined with the amplitude
generator \ol{}. As in this paper we introduce this framework for the
first time, in the following, we give a short introduction to both
programs, starting with a discussion of the event generator \wz{} and
its treatment of next-to-leading order QCD corrections in
section~\ref{ssec:whizard}. This is followed by a description of \ol{}
in section~\ref{ssec:openloops}. 

\subsection{The {\wz{}} event generator at next-to-leading order}
\label{ssec:whizard}

\wz~\cite{0708.4233, hep-ph/0102195} is a multi-purpose event
generator for both lepton and hadron colliders. At leading-order, it
can deal with arbitrary SM processes, as well as a multitude of BSM
processes (e.g. generated from automated tools like in
\Rcite{Christensen:2010wz}). Moreover, it can perform simulations for  
a broad class of processes at next-to-leading order. The modern
release series (v2) has been developed to meet the demands of LHC
physics analysis, while its generic treatment of beam-spectra and
initial-state photon radiation makes it especially well suited for
lepton collider physics. The program has a modular structure and
consists of several subcomponents, the most important being
\om~\cite{hep-ph/0102195}, \va~\cite{hep-ph/9806432} and
\textsc{Circe}~\cite{hep-ph/9607454}: \om{} computes multi-leg
tree-level matrix elements as helicity amplitudes in a recursive way
that avoids Feynman diagrams. \va{} is used for Monte-Carlo
integration and grid sampling. It combines the multi-channel
approach~\cite{hep-ph/9405257} with the classic \textsc{Vegas}
algorithm~\cite{Vegas1976} to automatically integrate cross sections
with non-factorizable singularities. The \textsc{Circe} package can be
used to create and evaluate lepton beam spectra.

\wz{} can be used for event generation on parton level as well as for the
subsequent shower and hadronization.
For this purpose, it has its own analytical~\cite{Kilian:2011ka} as
well as $k_T$-ordered parton shower, and a built-in interface to
\textsc{Pythia6}. Color information is treated in \wz{} using the
color-flow formalism~\cite{1206.3700}.

The generic NLO framework in \wz{} builds upon the FKS subtraction
scheme~\cite{hep-ph/9512328}, which partitions the phase space into
regions where only one divergent configuration is present. 
This divergence is then regulated using plus-distributions.
FKS subtraction allows for the application of \wz{}'s optimized
multi-channel phase-space generator for the underlying Born
kinematics, from which real kinematics are generated. It is also 
very well suited to the matching procedures employed, as described
below. First preliminary results of the NLO functionality of \wz\ have
been presented in \RRcite{1510.02666,Reuter:2016ohp}. 
\wz{} supports \ol{} and \go{} (an interface to \re\ is being
developed) as one-loop matrix element providers as well as 
for the computation of color- and spin-correlated Born matrix
elements. At tree-level, they can also be used as alternatives to
\om{}.

For event generation, \wz{} can produce weighted fixed-order NLO QCD
events that are written to \hepmc{}~\cite{Dobbs:2001ck} files. This
allows for flexible phenomenological fixed order studies, especially
in combination with \rv{}'s~\cite{1003.0694} generic event analysis
capabilities. Matching to parton showers is achieved with an
independent implementation~\cite{1510.02739} of the \po{} matching
method~\cite{hep-ph/0409146}. 

Apart from scattering processes, \wz{} is also able to compute decay
widths for \mbox{$1 \to N$} processes at NLO\@. The final-state phase space
is built in the usual fashion, whereas the initial-state phase space
is adapted for decays. \wz\ constructs the gluon momentum separately and
then applies a recursive reassignment of the virtualities of the
intermediate particles. Computing decay widths directly in \wz{}
allows for a consistent treatment of the top width, which has to be
recomputed according to the physical parameters and the process
definition as discussed in~\cref{ssec:input_parameters}.

\subsection{Virtual matrix elements from \ol}
\label{ssec:openloops}
All necessary Born and one-loop amplitudes together with the color and
helicity correlators required within the FKS subtraction are provided
by the publicly available \ol{} program~\cite{hepforge}. It is based
on a fast numerical algorithm for the generation of Born and one-loop
scattering amplitudes by means of a hybrid tree--loop recursion that
generates cut-open loops as functions of the circulating loop
momentum~\cite{Cascioli:2011va}.  
Combined with the~{\sc CutTools}~\cite{Ossola:2007ax} OPP reduction~\cite{Ossola:2006us}
library and the {\sc OneLOop} library~\cite{vanHameren:2010cp}
or with the {\sc Collier}~\cite{Denner:2016kdg} tensor integral reduction 
library based on \RRcite{Denner:2002ii,Denner:2005nn,Denner:2010tr},
the employed recursion permits to achieve very high CPU performance
and a high degree of numerical stability.  A sophisticated stability
system is in place to rescue the small number of potentially unstable
phase space points via a re-evaluation at quadruple precision.  

Within \ol, ultraviolet (UV) and infrared (IR) divergences are
dimensionally regularized and take the form of poles in $(4-D)$. However, all ingredients of 
the numerical recursion are handled in four space-time dimensions.
The missing $(4-D)$-dimensional contributions---called $R_2$ rational 
terms---are universal and can be restored from process-independent 
effective counterterms~\cite{Ossola:2008xq,Binoth:2006hk,Bredenstein:2008zb,
Draggiotis:2009yb,Garzelli:2009is,Garzelli:2010qm,Garzelli:2010fq,Shao:2011tg}.

The strong coupling constant is renormalized in the 
$\overline{\text{MS}}$ scheme, and heavy quark contributions can be decoupled 
via zero-momentum subtraction in a flexible way, depending on the number of active flavors in the evolution of $\alpha_S$. 
Unstable particles with a finite width are by default treated via an automated implementation of the complex-mass scheme~\cite{Denner:2005fg}.

The publicly available \ol amplitude library includes all relevant matrix
elements to compute NLO QCD corrections, including color- and
helicity-correlations and real radiation as well as loop-squared amplitudes,
for more than a hundred LHC processes. Many libraries for lepton collisions can 
easily be taken from this LHC library, as any crossing of external particles is automatically done 
when a library is loaded. For example, the one-loop library to be used for the process 
$e^+ e^- \to jj$ is \texttt{ppll}. For many other processes, especially for those
with massive quarks in the final state, dedicated lepton collider
libraries have been added to the public \ol amplitude repository,
which will be further extended in the near future~\footnote{Details
can be found at http://openloops.hepforge.org. One-loop amplitudes for
lepton-collider processes that might not yet be available can be 
easily added to the \ol repository upon request.}. 

\begin{figure}[tp]
\centering
  \includegraphics[width=\reldiagwidth\textwidth]
                  {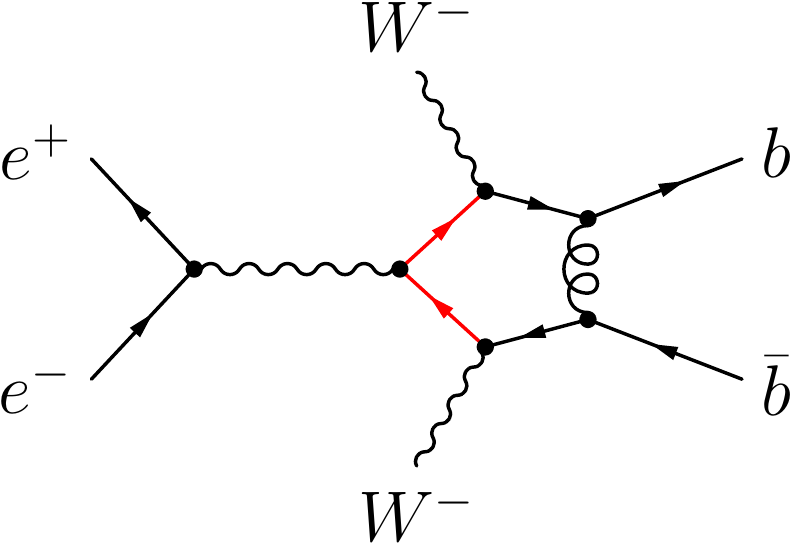}
    \quad
  \includegraphics[width=\reldiagwidth\textwidth]
                  {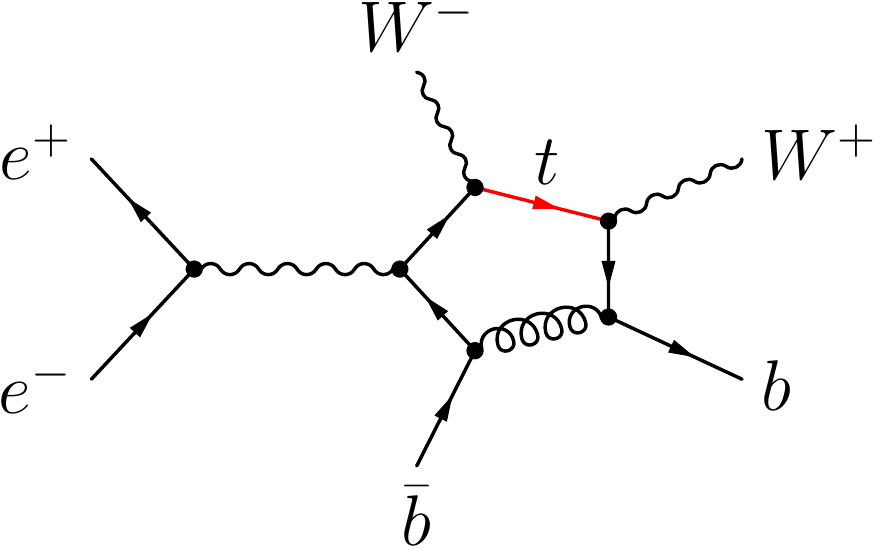}
    \includegraphics[width=\reldiagwidth\textwidth]
                    {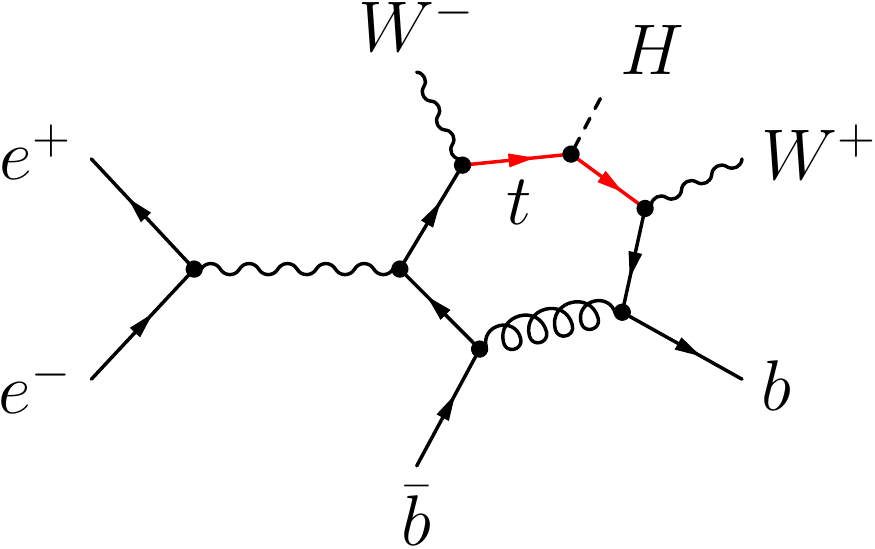}
  \caption{Example pentagon diagrams contributing to the $\bbww$
    final-state process containing one or two (leftmost diagram) top
    resonances and an hexagon diagram contributing for  $\bbwwH$
    production.} 
  \label{fig:Diags:loops}
\end{figure} 

The \wz{}+\ol interface is based on the BLHA standard~\cite{Alioli:2013nda}. Moreover, 
this new interface can use a modification of this standard allowing the computation of polarized amplitudes.
To this end, the process registry can contain dedicated entries for each polarization configuration
 of initial or final state particles. This implements an automated  NLO
setup which allows to study effects of beam polarization - an important feature
at future linear colliders like the ILC.

Table~\ref{tab:diagrams} lists information about the computational complexity with respect to the one-loop amplitudes of the processes studied in this paper. Note that the total number of diagrams is not decisive for the
computational effort in the \ol recursion formalism. Instead, the crucial point is the
maximal number of n-point functions involved. For the $bb(W \to
l\nu)(W \to l\nu)$ processes discussed in this paper, the
most complex integrals stem from pentagon diagrams, examples for which
are depicted in fig. \ref{fig:Diags:loops}.  
Also shown in fig. \ref{fig:Diags:loops} is a hexagon diagram contributing to the associated Higgs production process also discussed in this paper.
Concerning the complexity of the amplitudes, due to the reduced number
of contributing helicity structures the calculation of the off-shell
processes including leptonic decays are less involved compared to the
corresponding processes with on-shell W-bosons - despite the increased
number of diagrams. 

\begin{table}[htbp]
   \caption{Overview of loop matrix elements at NLO QCD for the studied
     processes. Shown are the number of one-loop diagrams, the maximal
     number of loop propagators and the number of helicity
     structures (assuming charged leptons to be massless). 
   \vspace{1em}} 
   \label{tab:diagrams}
   \centering
   \begin{tabular}{c c c c}
      \toprule{}%
      $\epem \to$ & $n_{\text{loop diags}}$ & Max. prop. & $n_{\text{hel}}$ \\
      \midrule{}%
      $t\bar{t}$ & 2 & 3 & 16 \\
      $\bbww$ & 157 & 5 & 144 \\
      $b\bar{b}\bar{\nu}_e e^- \nu_\mu \mu^+$ & 830 & 5 & 16 \\
      \midrule{}%
      $t\bar{t}H$ & 17 & 4 & 16 \\
      $bW^+\bar{b}W^-H$ & 1548 & 6 & 144 \\
      $b\bar{b} \bar{\nu}_e e^- \nu_\mu \mu^+ H$ & 7436 & 6 & 16 \\
      \bottomrule{}
   \end{tabular}
\end{table}

\section{Resonance-aware FKS subtraction}
\label{sec:resonances}

The standard approach to compute automated NLO corrections 
can be very inefficient if QCD radiation off partons
originating from the decay of
a resonance is present. In our case, 
this issue arises from resonant subprocesses of type
$Z/H \rightarrow b\bar{b}$ and $t \rightarrow Wb$.
As discussed for the first time in \Rcite{1509.09071}, 
the problem is due to the fact that the momentum of the
resonant particle can be different in the Born phase space and the corresponding
real phase spaces with one additional gluon momentum\footnote{%
Likewise, in Catani-Seymour subtraction~\cite{hep-ph/9605323}, one single real-emission
phase space is mapped to different $N$-particle phase spaces. Consequently, 
the resonance mismatch also appears in this approach.}.
The real-subtracted contribution to the NLO matrix element contains
$N+1$-particle matrix elements with corresponding kinematics, as well
as Born matrix elements with factorized kinematics in the subtraction
terms. In collinear and/or soft regions, it is crucial that both 
terms agree well. However, the presence of resonances significantly
affects their cancellation, and hence the convergence of the
integration. 

To understand this more in-depth, consider the $H \to b\bar{b}$
splitting with the very narrow Higgs resonance $\Gamma_H =
\mathcal{O}(\ValMeV{1})$. This occurs as a Higgsstrahlung 
background process to $\epem \to \wwbb$ and its decays. Thus, the
squared matrix element of the total process contains a term with the
contribution of the squared Higgs  
propagator,
\begin{equation}
  D_H^{\text{Born}} = \left[(\bar{p}_{bb}^2 - m_H^2)^2 +
    m_H^2\Gamma_H^2\right]^{-1}, 
  \label{eqn:higgs_prop_born}
\end{equation}
where $\bar{p}_{bb}^2$ denotes the invariant mass of the
$b\bar{b}$-pair in the Born phase space. The Higgs propagator in the
corresponding real squared matrix element takes 
the form
\begin{equation}
  D_H^{\text{Real}} = \left[(p_{bbg}^2 - m_H^2)^2 + m_H^2\Gamma_H^2\right]^{-1},
  \label{eqn:higgs_prop_real}
\end{equation}
where now the Higgs virtuality is made up by the invariant mass of the
$b\bar{b}$-system and the additional gluon, $p_{bbg}^2$. 
Let the change of the Higgs virtuality from the Born to the real phase
space be described by $\Delta_{bbg}$, such that
\begin{equation}
  p_{bbg}^2 = \bar{p}_{bb}^2 + \Delta_{bbg}^2.
  \label{eqn:delta_bbg}
\end{equation}
The explicit form of $\Delta_{bbg}^2$ does not depend on the process, but on the subtraction scheme.
In the FKS approach, the real phase space is constructed
in such a way that the invariant mass of the recoiling system and the
emitter-radiation system are conserved separately. Thus,
$\Delta_{bbg}^2$ consists of boosts and projections of the Born
momenta.

Either way, we can define $\varepsilon = \bar{p}_{bb}^2 - m_H^2$ and,
for the ratio of weights associated with the $H\to b \bar b$ 
resonances in the emission matrix element and in the related
subtraction terms, we 
find
\begin{equation}
  \mathcal{D} := \frac{D_H^{\text{Born}}}{D_H^{\text{Real}}} 
     = 1 + \frac{\Delta_{bbg}^4 + 2\Delta_{bbg}^2 \varepsilon}{\varepsilon^2 + m_H^2\Gamma_H^2}
     \stackrel{\varepsilon \to 0}{=} 1 + \frac{\Delta_{bbg}^4}{m_H^2\Gamma_H^2}.
  \label{eqn:higgs_prop_ratio}
\end{equation}
For the real and subtraction terms to match, it is required that
$\mathcal{D} \approx 1$ in the soft as well as the collinear limit. 
At the resonance, $\varepsilon \to 0$, we see that this condition is fulfilled
if $\Delta_{bbg}^4 \ll m_H^2\Gamma_H^2$.
We immediately see that this poses a problem in the collinear limit,
since $\Delta_{bbg}^4$ can become large if a hard-collinear gluon is
emitted. However, also in the soft limit a significant mismatch can
occur if the denominator $m_H^2 \Gamma_H^2$ is sufficiently 
small. This is definitely the case for $H \to b\bar{b}$, with 
$m_H^2
\Gamma_H^2 = (\ValGeV{0.720})^4$,
while for $t \to W b$ the problem is
less severe with 
$m_t^2\Gamma_t^2=(\ValGeV{15.4})^4$.
As already noted in \cref{sec:introduction}, the problem that $H\to
b\bar{b}$ is contained in the off-shell \ttbar{} process is unique to
the lepton collider, as here at LO 
the production is of $\order{\alpha^2}$ instead of
$\order{\alpha_s^2}$ at hadron colliders. 
For our study, we have 
addressed
the problem of narrow 
resonances
implementing the modified FKS subtraction procedure presented recently
in \Rcite{1509.09071} for generic processes in \wz.  This
implementation is briefly outlined in the following. 
More in-depth information and validation can be found in
appendix~\ref{sec:bbmumu_example}.

In the so-called {\it resonance-aware} FKS approach, in addition to
being partitioned into distinct singular regions, the phase space is
also separated into resonance regions, according to the resonance
structures
of the process. In each extended singular region, the real
phase space is constructed in such a way that the invariant mass of
the particles which originate from the same resonance is kept
fixed. In this way, the shift $\Delta_{bbg}$ in 
eq.~\eqref{eqn:higgs_prop_ratio} is exactly zero by construction, and
hence $\mathcal{D} = 1$.  
This approach makes use of
modified FKS mappings 
which are evaluated in the rest
frame of the corresponding resonance. This leads to the problem that
the sum over all singular regions does not reproduce the full real
matrix element any more. As shown in \Rcite{1509.09071}, this can be
solved by introducing a new component to the integration, the
so-called soft mismatch. 
In \wz{}, the integration of the soft mismatch is automatically
performed and is included as an additional contribution next to Born,
real and virtual components when the resonance-aware FKS subtraction is
activated. Related technical details
can be found in Appendix~\ref{sec:bbmumu_example}.

In the resonance-aware FKS approach, the standard FKS projectors
$\mathcal{S}_\alpha$ are extended by resonance projectors
$\mathcal{P}_{\alpha'}$, with $\mathcal{P}_{\alpha'} \to 1$ if the phase
space is close to the resonance associated with the resonance history
$\alpha'$. They thus map out this particular resonance structure.
Motivated by the narrow-width limit of a resonant process,
$\mathcal{P}_{\alpha'}$ is proportional to the Breit-Wigner factors
of a given resonance structure.

In \wz{}, resonance information is generated for every simulation, already at
leading order. This information is used 
by the multi-channel integrator \va, where all relevant 
resonance structures are sampled in order to enhance the performance.
We use exactly these resonance structures to set up resonance-aware FKS subtraction. 
Thus, in principle, each of \wz{'s} integration channels could be 
identified with the resonance histories, also using the internal
mappings used in the construction of the Born phase-space. However, 
we decided to introduce resonance histories using the projectors of~\cite{1509.09071}
completely independent of the Monte Carlo integration channels.

The implementation of the resonance-aware FKS subtraction led to a
restructuring of the \hepmc output of weighted fixed-order NLO
events. In earlier \wz{} versions \cite{1510.02666}, 
different phase space points and weights were assigned to each
singular region $\alpha_r$. However, when resonances are included,
different $\alpha_r$ can be associated with the same real phase space
(e.g.\ in the case of $Z/H \to b\bar{b}$), which leads to 
an unnecessary abundance of real-emission events in the event
output. Therefore, in the most recent \wz{} version, a real-emission
event is created for each distinct phase-space structure, which is defined
by its emitter and the decaying particles. Each of these phase-space
structures can be associated to multiple singular regions, over which
it is summed to obtain the complete real weight of the event.
Moreover, the soft mismatch is included in the subtraction weight.  

Employing the resonance-aware FKS subtraction scheme for off-shell top-pair 
production and decay in leptonic collisions 
is not trivial, since 
resonance histories where the gluon is emitted from the production process, 
i.e.~from the top before it decays, cannot be associated to any valid FKS
sector. In proton--proton collisions, the consistent resonance-aware
treatment of such resonance structures, which requires mappings that
preserve simultaneously the invariant masses of the $W^+ b$ and
$W^-\bar b$ pairs without the emitted gluon, is guaranteed through FKS
sectors associated with the initial-state quark or gluon
emitters. However such FKS sectors are not present in the case of
uncolored initial states. Thus the extension of the resonance-aware
approach to $e^+e^-$ collisions requires 
a dedicated treatment for the case of QCD radiation that is emitted 
by unstable colored particles before they decay.

While this issue deserves more detailed studies that we have deferred to the future,
for the study of the off-shell processes $\eewwbb(H)$ and $\eellllbb(H)$, 
presented in this paper, 
we use only the  two resonance histories $Z \to b\bar{b}$ and $H \to b\bar{b}$.
Employing the implementation of the resonance-aware subtraction scheme with these resonance histories
we observe a decent convergence of the numerical integration at the inclusive and differential
level. 

Finally we want to note, that the resonance-aware FKS subtraction scheme,
including a definite resonance history assignment in the event output, 
enables a consistent matching of fixed-order NLO predictions  with
parton shower generators for processes with intermediate
resonances~\cite{1509.09071,Jezo:2016ujg}. 
To this end, all relevant resonance histories should be taken into account.

\section{Phenomenology of \texorpdfstring{$\boldsymbol{t\bar{t}}$}{t tbar} and
  \texorpdfstring{$\boldsymbol{{t}\bar{t}H}$}{t tbar H} production and decay}
\label{sec:pheno}

\subsection[Phenomenology of \texorpdfstring{$t\bar{t}$}{t tbar} production and decay]
  {Phenomenology of $\boldsymbol{t\bar{t}}$ production and decay}
\label{sec:pheno:ttbar}
\begin{figure}[tb]
\centering
  \includegraphics[width=\reldiagwidth\textwidth]
                  {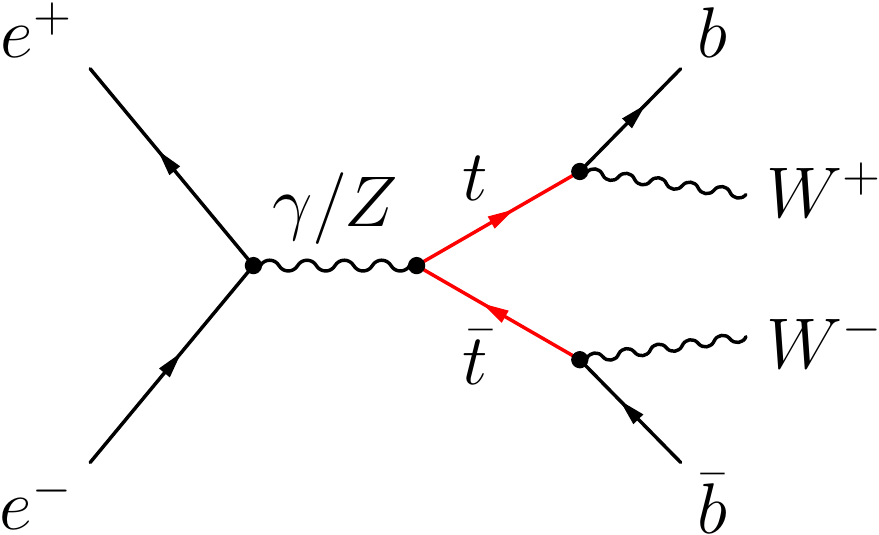}\qquad 
  \includegraphics[width=\reldiagwidth\textwidth]
                  {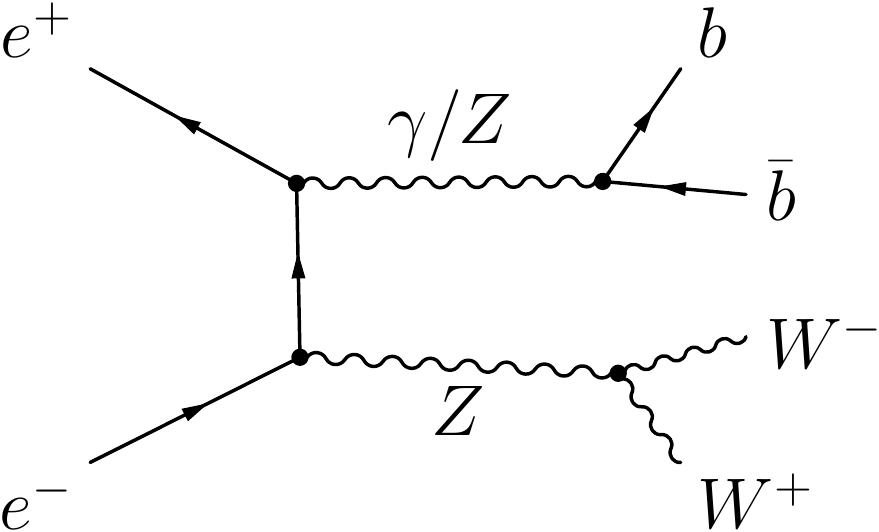}\\
  \includegraphics[width=\reldiagwidth\textwidth]
                  {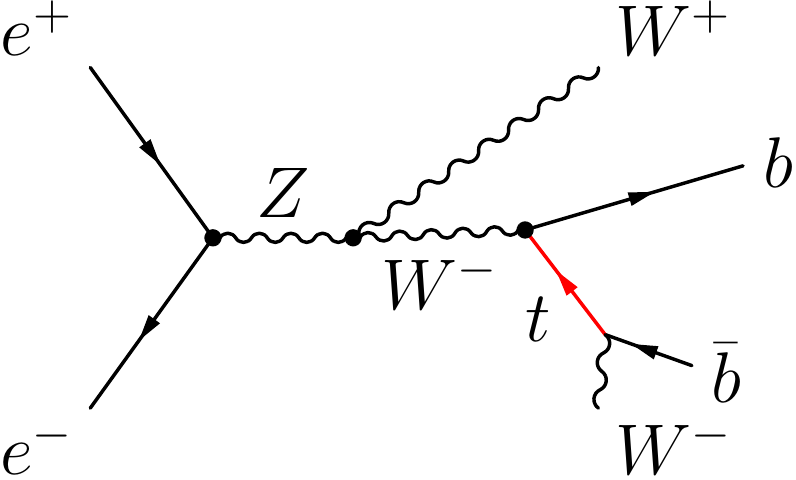}\qquad
  \includegraphics[width=\reldiagwidth\textwidth]
                  {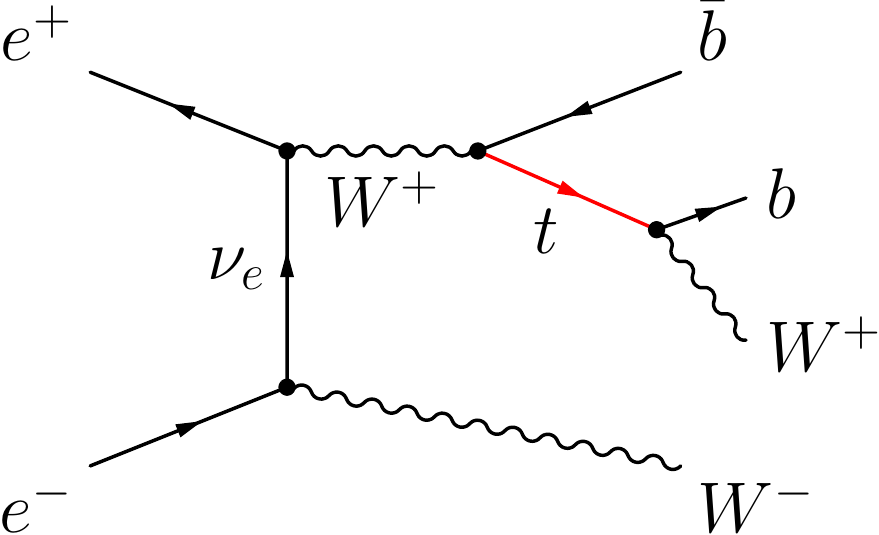}
  \caption{The double-resonant signal diagram (top left) besides
    example non-resonant (top right) and s- and t-channel single-top
    diagrams (bottom left and right, respectively) of the process
    \eewwbb.} 
  \label{fig:Diags:bbww}
\end{figure}

In this study we want to investigate NLO QCD perturbative corrections in top-quark pair production at lepton colliders modeling off-shell and interference effects at increasing levels of precision. To this end we will consider the following related $2\to2, 2\to4 $ and $2\to6$  processes,
\begin{align}
\epem & \to \ttbar \,,  \label{eq:tt}  \\
\epem & \to  \wwbb \,, \label{eq:wwbb} \\
\epem & \to  \llllbb \label{eq:4l2b}  \,,
\end{align}
where we treat the bottom quarks as massive. Top quarks almost exclusively decay
via $t\to bW^+$, such that the process in eq.~\refeq{eq:wwbb} can be understood
as the top-quark pair production process of eq.~\refeq{eq:tt} including
top-quark decays. Beyond the narrow width approximation, i.e. including
off-shell effects for the produced top quarks, the process of
eq.~\refeq{eq:wwbb}, however, receives besides doubly-resonant (signal)
top-quark contributions, also contributions from non-resonant and
single-resonant (background) diagrams together with their interference. Example
diagrams for all three production mechanisms are shown in
\cref{fig:Diags:bbww}. The sub-dominant single-top diagrams always occur via a
fermion line between the two external bottom quarks. Thanks to the finite bottom
mass even non-resonant contributions from diagrams with a $\gamma \to b \bar b$
splitting, like the one in the top right of \cref{fig:Diags:bbww}, can be
integrated over the whole phase space without the necessity for cuts.

At the NLO QCD level, the calculation of the process in eq.~\refeq{eq:wwbb}
includes corrections to top-quark pair production and also to the top decays
together with non-factorizable corrections, which are formally of the order of
$\Op\left(\alpha_S \Gamma_t /m_t\right)$. Diagrammatically such non-factorizable
contributions interconnect production and decay stage of the signal process or
the two individual decays, as for example depicted in
\cref{fig:Diags:loops}(left). At the same time NLO interference effects with
single-resonant and non-resonant contributions and also spin correlations in the
top decay are consistently taken into account.

\begin{figure}[tb]
\centering
  \includegraphics[height=.2\textwidth]{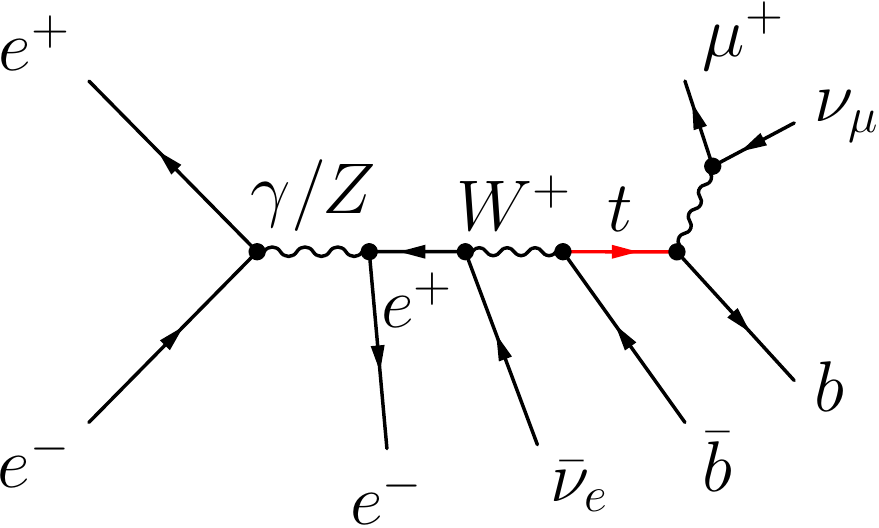}
     \qquad
  \includegraphics[height=.2\textwidth]{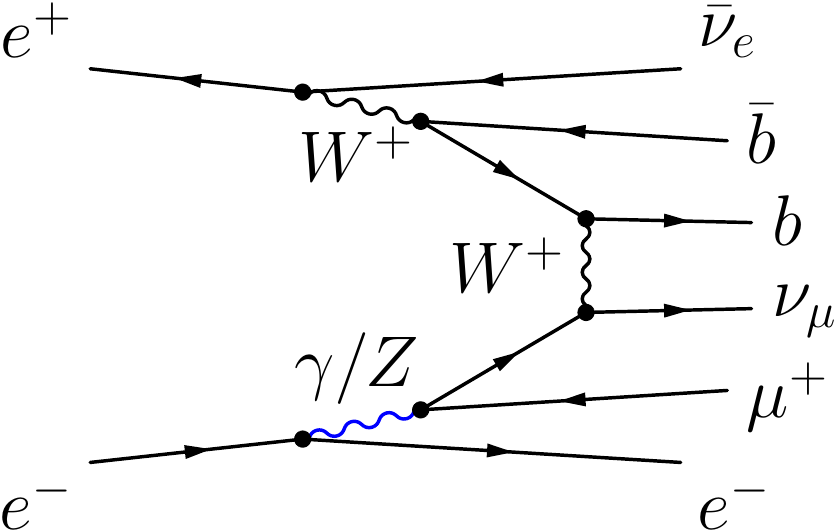}
  \caption{Possible topologies of the full process. The blue line indicates a
  potentially soft photon that gives rise to a leading-order singularity.}
  \label{fig:Diags:bblnulnu}
\end{figure}

In order to make contact with experimental signatures and to further
increase theoretical precision, the process in eq. \refeq{eq:4l2b}
introduces -- beyond the top-quark decays -- also leptonic decays of
the W-bosons including respective 
off-shell effects. Due to the purely EW nature of the leptonic W-boson decays,
from a perturbative point of view these additional decays do not increase the
computational complexity compared to the process with on-shell W-bosons, i.e.
the one of eq.~\refeq{eq:wwbb}. However, besides the more involved phase space
integration, the number of contributing diagrams increases substantially due to
additional single- and non-resonant contributions, as illustrated in
\cref{fig:Diags:bblnulnu}. Notabene, in the case of decays with
initial-state lepton flavor, diagrams like the one on the
right of \cref{fig:Diags:bblnulnu} show a singularity and can not be
integrated over the whole phase space without cuts. For brevity, here
we focus on the different lepton flavor case but an analysis for the
very similar same flavor case can easily be performed with the
publicly available \wz{}+\ol framework. Hadronic top-quark decays will
be investigated in the future.

\begin{figure}
\centering
  \includegraphics[width=\reldiagwidth\textwidth]{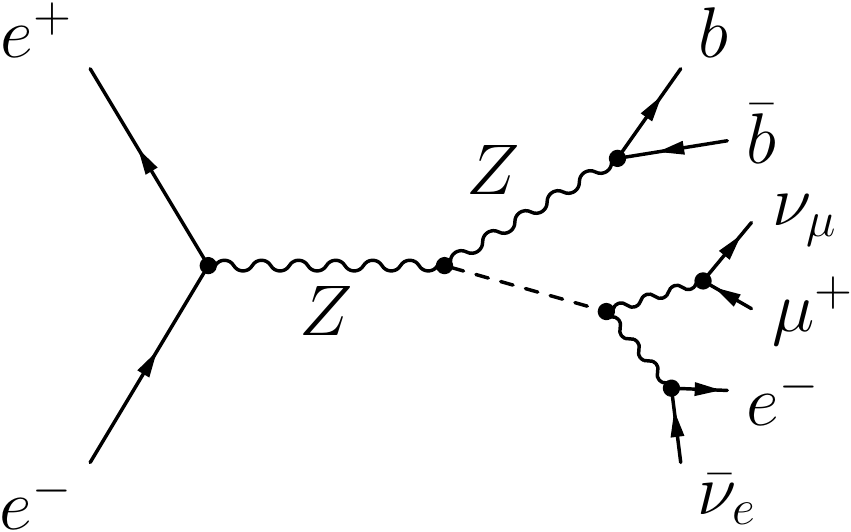}
    \qquad
  \includegraphics[width=\reldiagwidth\textwidth]{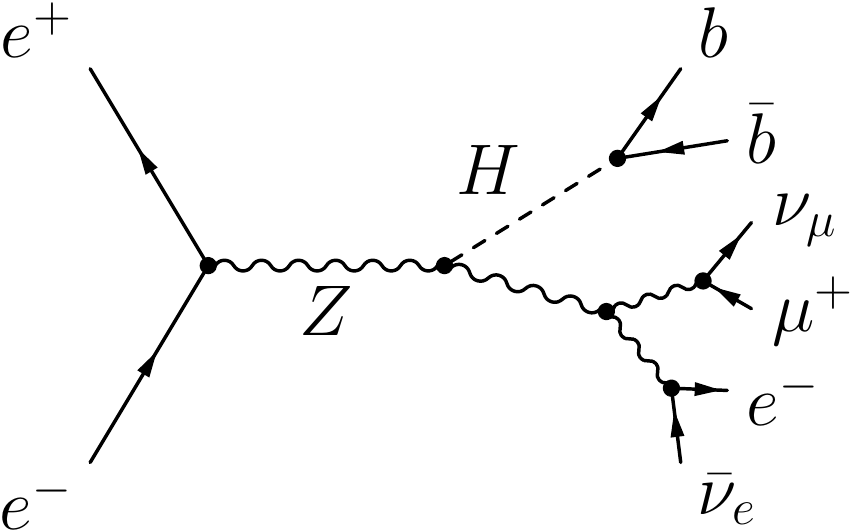}
  \caption{Contributions to the process \eellllbb{} involving a $Z$ or $H$
    resonance, treated via the resonance-aware FKS subtraction.}
\label{fig:Diags:bblnulnu-zh-resonances}
\end{figure}

The off-shell processes of eqs.~\refeq{eq:wwbb}-\refeq{eq:4l2b}
contain diagrams with $Z/H \to b\bar{b}$ splittings, as for example
depicted in \cref{fig:Diags:bblnulnu-zh-resonances}. Due to the small
intermediate widths, the integration of such contributions benefits
strongly from the extended resonance-aware FKS subtraction, described
in section~\ref{sec:resonances}. For the technical reasons discussed
there in detail, we only apply the resonance-aware FKS subtraction for
the intermediate $Z/H$ resonances, but not for the top resonances.

\subsection[Phenomenology of \texorpdfstring{$t\bar{t}H$}{t tbar H}
  production and decay] 
  {Phenomenology of $\boldsymbol{t\bar{t}H}$ production and decay}
\label{sec:pheno:tth}

Similar to top-quark pair production, we consider the following related $2\to3,
2\to5 $ and $2\to7$ processes for the associated production of a Higgs boson
together with a top-quark pair with increasing level of precision with respect
to off-shell, non-resonant and interference effects,
\begin{align}
\epem & \to \ttbarH \,,  \label{eq:ttH}  \\
\epem & \to  \wwbbH \,, \label{eq:wwbbH} \\
\epem & \to  \llllbbH \label{eq:4l2bH}  \,,
\end{align}
where again all b-quarks are treated as massive.

\begin{figure}[tb]
\centering
  \includegraphics[width=\reldiagwidth\textwidth]
                  {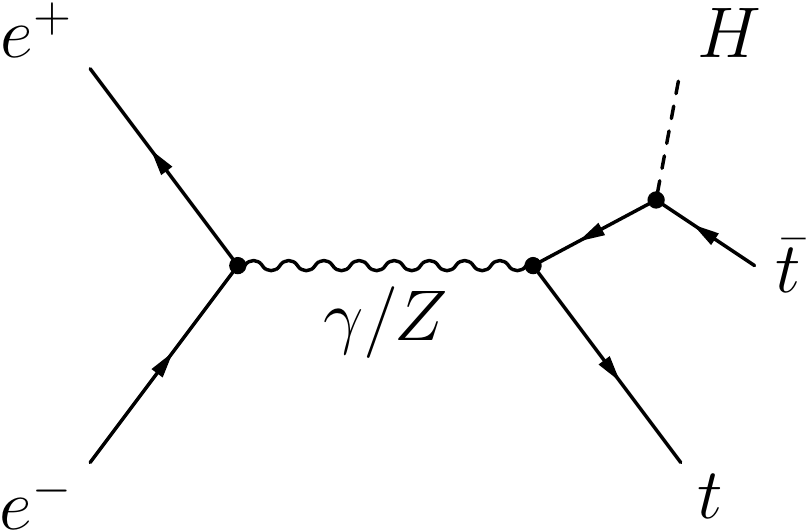}\qquad
  \includegraphics[width=\reldiagwidth\textwidth]
                  {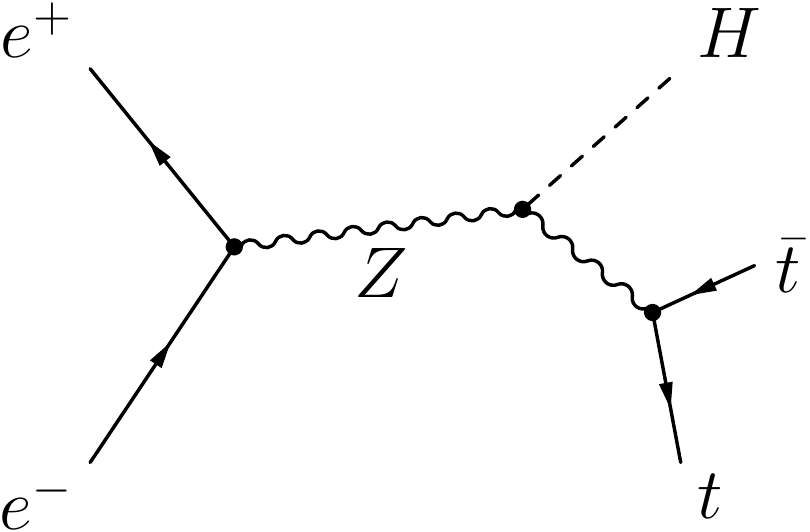}
  \caption{Contributing diagrams to \ttH production: associated
    production of a Higgs boson and a top quark pair and
    Higgsstrahlung with an off-shell $Z^\ast \to t \bar t$ decay.} 
  \label{fig:Diags:tth}
\end{figure}

The diagrams involved in these process are very similar to those of
the corresponding \ttbar production processes, apart form the
additional Higgs boson that couples now to all massive internal or
external particles ($t,b,W^\pm,Z,H$). Already on the level of the  
on-shell processes of eq.~\eqref{eq:ttH} this results into two
competing contributions, as depicted in \cref{fig:Diags:tth}. The
diagram on the left of \cref{fig:Diags:tth} is proportional to the
top Yukawa coupling $y_t$ and will be denoted as \ttH signal
contribution, while the diagram on the right can be considered as
irreducible Higgsstrahlung background in the $ZH$ channel with an
off-shell  $Z^\ast \to t \bar t$ decay. 

\begin{figure}
\centering
  \includegraphics[width=\reldiagwidth\textwidth]{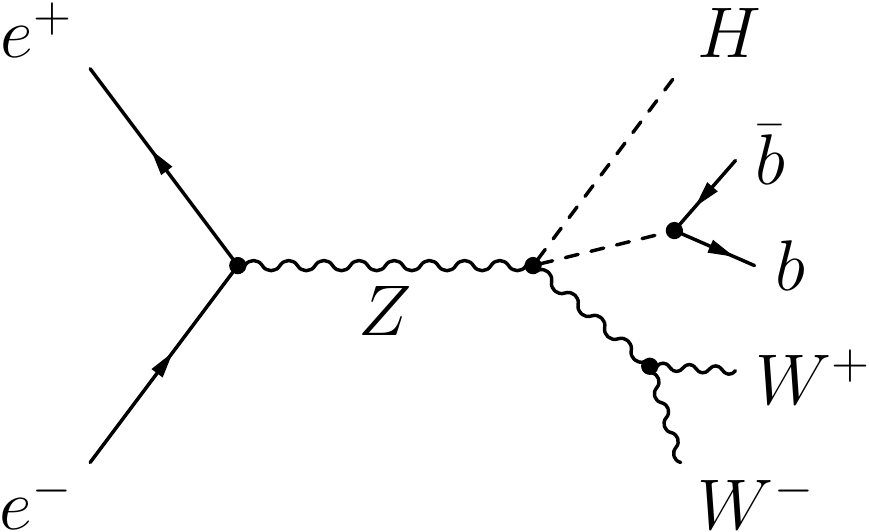}
  \caption{A representative non-resonant diagram contributing to \wwbbH production via a quartic $ZZHH$-coupling.}
  \label{fig:diag:hbbww-quartic}
\end{figure}

Furthermore, at the level of the off-shell processes of
eqs. \refeq{eq:wwbbH} and \refeq{eq:4l2bH} -- besides topologies
already present for the corresponding \ttbar processes with an
additional attached Higgs boson --, new  contributions arise from
quartic EW couplings as illustrated in
fig.~\ref{fig:diag:hbbww-quartic}. In such contributions, as before,
the tiny width of the Higgs boson requires a resonance-aware
subtraction scheme to yield a converging integration at NLO over the
whole phase space.

\section{Setup and validation}
\label{sec:paramter_setup_and_validation}
\subsection{Input parameters, scale choice and phase-space cuts} 
\label{ssec:input_parameters}
As input parameters, we use the following gauge-boson, quark and Higgs
masses~\cite{Agashe:2014kda}, 
\begin{align*}
  m_Z &= \ValGeV{91.1876}\,, & m_W &= \ValGeV{80.385}\,, \\	
  m_b &= \ValGeV{4.2}\,,     &  m_t &= \ValGeV{173.2}\,, \\
  m_H &= \ValGeV{125}\,.
\end{align*}
The electroweak couplings are derived from the gauge-boson masses and the Fermi constant,
$\GF=\unit[1.1663787\times10^{-5}]{\GeV^{-2}}$, in the $G_{\mu}$-scheme. 
The CKM matrix is assumed to be trivial, which is for the most relevant element
of our computation ($V_{tb}$) consistent with the measured value
($1.021\pm0.032$~\cite{Agashe:2014kda}).
Furthermore, using the precisely measured value of $G_\mu$ automatically absorbs
important electroweak corrections to the top decay.
For the strong coupling constant we use $\alpha_s(m_Z) = 0.1185$ and a two-loop
running including $n_f = 5$ active flavors.

With this setup, the gauge boson and top widths are computed directly with \wz
at LO and NLO.
In the NLO computation, we use the mass of the decaying particle as
renormalization scale.
The obtained LO and NLO gauge boson widths are
\begin{align}
  \label{eq:Zwidth}
  \Gamma_Z^{\text{LO}} &= \ValGeV{2.4409}, 	&
  \Gamma_Z^{\text{NLO}} &= \ValGeV{2.5060},  \\ 
  \label{eq:Wwidth}
  \Gamma_W^{\text{LO}} &= \ValGeV{2.0454}, 	&
  \Gamma_W^{\text{NLO}} &= \ValGeV{2.0978}.  
\end{align}
In our calculation we use $\Gamma_Z$ and $\Gamma_W$ 
at NLO throughout, i.e.~also for off-shell cross sections at LO. 
This ensures that the effective $W$ and $Z$ leptonic branching ratios
that result from $e^+e^-\to b\bar b4 f(H)$ matrix elements are always
NLO accurate. In contrast, in order to guarantee that $t\to Wb$
branching ratios remain consistently equal to one at LO and NLO,
off-shell matrix elements and the top-decay width need to be evaluated
at the same perturbative order.
For the top width we employ two distinct sets of values: one for the
on-shell decay $t \to W^+ b$ and one for the off-sell decay  $t \to
f\bar f b$, as also detailed in~\Rcite{1207.5018}. The value used for
the off-shell top decay includes decays into three lepton generations
and two quark generations. It also involves the $W$ width, for which
we use the previously computed NLO value. The numerical values are 
\begin{align}
  \label{eq:twidth_onshell}
  \Gamma_{t\to Wb}^{\text{LO}} &= \ValGeV{1.4986}, & \Gamma_{t\to
    Wb}^{\text{NLO}} &= \ValGeV{1.3681}, \\ 
  \label{eq:twidth_offshell}
  \Gamma_{t\to f\bar fb}^{\text{LO}} &= \ValGeV{1.4757}, 	&
  \Gamma_{t\to f\bar fb}^{\text{NLO}} &= \ValGeV{1.3475}. 
\end{align}
The Higgs width is set to $\Gamma_H = \ValMeV{4.143}$.

In the determination of the off-shell top width and in all calculations
presented in this paper, intermediate massive particles are treated in the
complex-mass scheme~\cite{Denner:2005fg}.
This leads to a gauge invariant treatment of finite width effects as well as
perturbative unitarity~\cite{1406.6280}.
On the technical side, it necessitates complex-valued renormalized masses
\beqar\label{eq:complexmasses}
\mu^2_i=M_i^2-\ri\Gamma_iM_i \qquad\mbox{for}\;i=\PW,\PZ,\Pt,\PH\;,
\eeqar 
that imply for consistency a complex-valued weak mixing angle
\beq\label{eq:defsintheta}
\sw^2=1-\cw^2=1-\frac{\mu_\PW^2}{\mu_\PZ^2}\,.
\eeq
For the electromagnetic coupling in the $G_\mu$ scheme we set
\beq\label{eq:alpha_from_cms}
\alpha_e = \frac{\sqrt{2}}{\pi}G_\mu \left| \mu_W^2 s_w^2 \right|,
\eeq
which gives $\alpha_e^{-1} = 132.16916$.

For the on- and off-shell $\ttbar$ and $\ttbarh$ processes that we consider in
this paper, the renormalization scale $\mu_R$ is set to
\beqar\label{eq:RFscales} 
\mu_{\rR}=\xi_{\rR}\mu_0,
\quad\mbox{with}\quad 
\mu_0=\left\{ 
\begin{array}{ll}
        m_t & \mbox{for ~$\ttbar$~ processes}\\
        m_t+m_H & \mbox{for ~$\ttbarh$~ processes}
\end{array} \right.
\quad\mbox{and}\quad 
\frac{1}{2}\le \xi_{\rR}\le 2\,.
\eeqar
Our default scale choice corresponds to $\xi_{\rR}=1$
and theoretical uncertainties are probed by scale variations.
This is obviously no complete assessment of the theoretical errors, for the LO
e.g.\ we do not obtain an uncertainty band, but it is our best handle on
perturbative QCD uncertainties.

Thanks to the finite $b$ quark mass all on- and off-shell $\ttbar$ and $\ttbarh$
processes considered in this paper can in principle be integrated over the whole
phase space.
However, for processes with final state electrons or positrons a
singularity emerges for small photon energy transfers, as depicted on the
right-hand side of fig.~\ref{fig:Diags:bblnulnu}.
To avoid this, we apply a mild phase-space cut for these processes
\begin{equation}
	\label{eq:CoulombCut}
	\sqrt{\left(k_{e^{\pm}}^{\text{in}} - k_{e^{\pm}}^{\text{out}}\right)^2} > \ValGeV{20}.
\end{equation}

For the definition of jets we employ the generalized $\kT$ algorithm
(\texttt{ee-genkt} in \fastjet)~\cite{Cacciari:2008gp,Cacciari:2011ma}
with $R=0.4$ and $p=-1$. We tag $b/\bar b$-jets according to their
partonic content and denote them as \jb and \jbbar. Similarly, in the
on-shell processes $\eett$ and $\eetth$, we identify the top quark
with the jet containing a top quark. In the discussion of differential
cross sections in section \ref{sec:differential_predictions} we always
require at least two b-tagged jets.\footnote{%
Since we do not impose any kinematic restriction on b-jets, requiring
two $b$-jets amounts to a lower bound for their $\Delta R$ separation.}
No further phase-space restrictions are applied. 

\subsection{Validation}
\label{ssec:validation}
To validate the new automated subtraction within \wz{}, we have performed
various cross checks.
All of the following checks have been performed at the per mil level, i.e.
differences are all at the few per mil level and within two standard
deviations of the MC integration. The NLO top-quark width computed by
\wz{} has been cross-checked both with the value used in
\Rcite{1207.5018} and the analytical formulae~\cite{Jezabek:1988ja,
  Jezabek:1988iv}. Total cross sections for simple $2 \to 2$
processes, like $e^+e^- \to q\bar q$ and $e^+e^- \to t \bar t$, have
been validated against analytical calculations. For $\eewwbb$, we have
performed an in-depth cross check with various other results and
generators. The total cross section corresponding to the study of
\Rcite{1511.02350}, therein computed with
\textsc{Madgraph5\_aMC@NLO}~\cite{1405.0301}, has been reproduced. 
Moreover, we find excellent agreement between \wz{},
\textsc{Sherpa}~\cite{0811.4622} and 
\munich{}~\footnote{\munich{} is the abbreviation of
``MUlti-chaNnel Integrator at Swiss (CH) precision''---an automated
  parton level NLO generator by S.~Kallweit. In preparation.} for the
parameter set given in~\cref{ssec:input_parameters}. Note, that both
\textsc{Sherpa} and \munich use CS
subtraction~\cite{hep-ph/9605323,Catani:2002hc}, while
\textsc{Madgraph5\_aMC@NLO} and \wz\ use FKS
subtraction~\cite{hep-ph/9512328}. The resonance-aware NLO
calculation was validated internally, comparing the result with a
computation based on the traditional FKS subtraction (see also
appendix \ref{sec:bbmumu_example}). To this end, we used large widths
in order to avoid problems with the traditional FKS approach.

\section{Numerical predictions for inclusive cross sections}
\label{sec:inclusive_predictions}

\subsection{Integrated cross sections and scale variation}
\label{ssec:predictions:integrated}
\begin{figure}
\centering
   \includegraphics[width=\relplotwidth\textwidth]{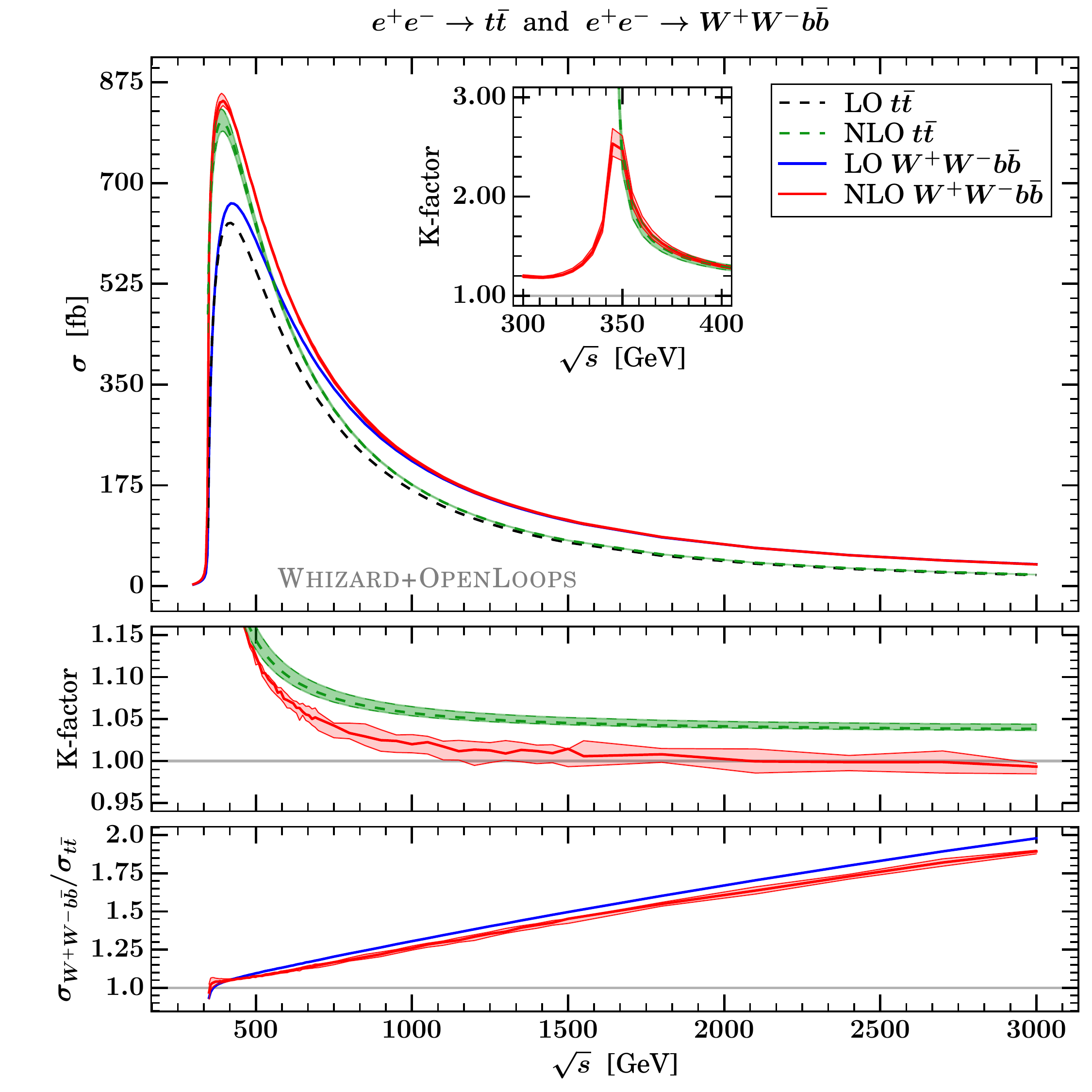}
         \quad
   \includegraphics[width=\relplotwidth\textwidth]{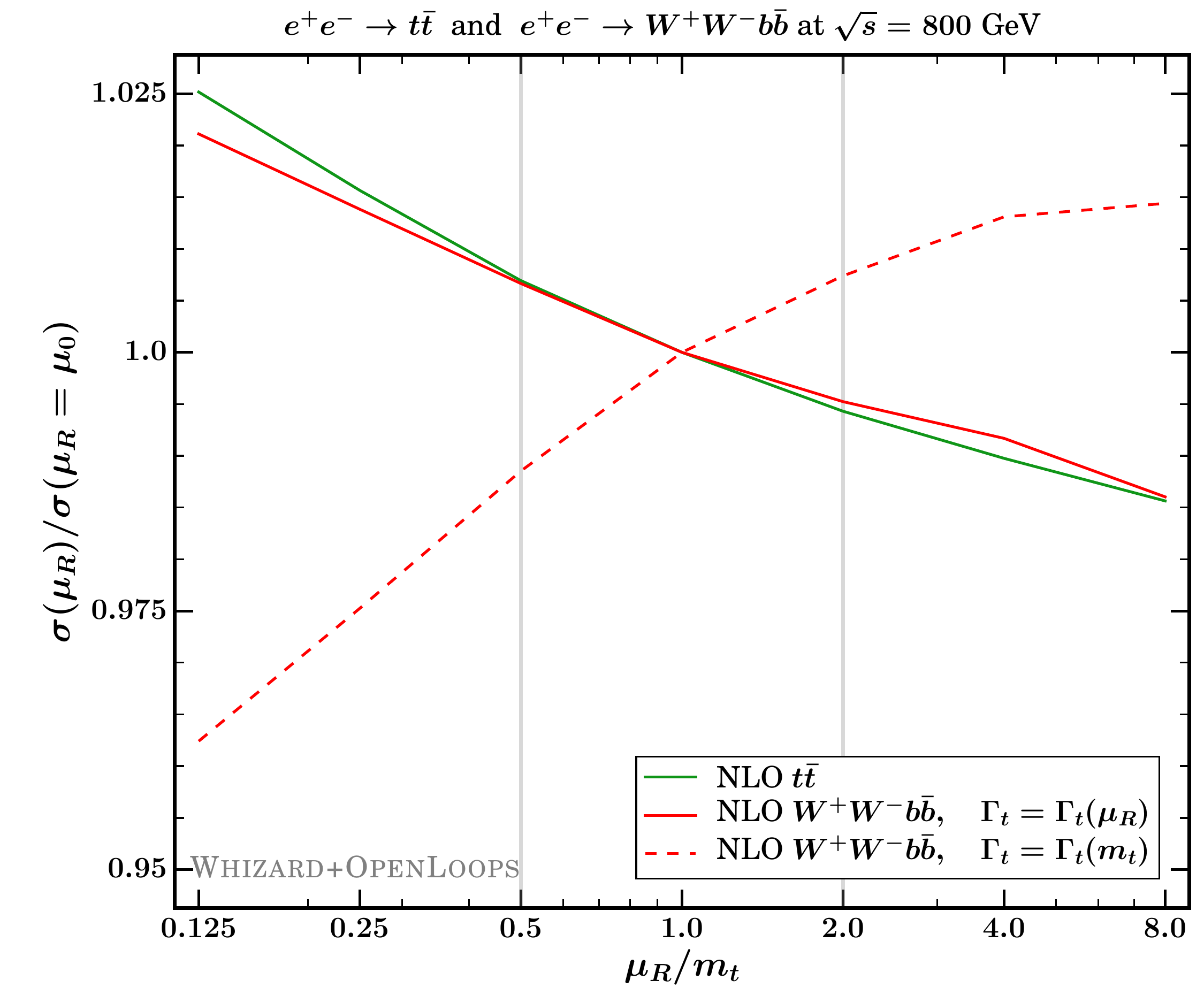}
   \caption{Total cross section for on-shell and off-shell $t\bar{t}$
     production  as a function of $\sqrts$ and $\mu_R$. In the lower
     panels of the left plot, we show the K-factor for $t\bar t$ and
     $\wbwb$ in green and red, respectively, as well as the ratio of
     off-shell to on-shell results for LO and NLO in blue and red.} 
\label{fig:tt-xsec-scan}
\end{figure}
We start our discussion of the numerical results with an investigation
of the NLO QCD corrections to inclusive top quark pair-production
cross sections depending on the center-of-mass energy \sqrts of the
leptonic collisions. In the left plot of \cref{fig:tt-xsec-scan} we
show inclusive LO and NLO cross sections for the on-shell process
\eett and the off-shell process \eewwbb together with the
corresponding K-factor ratios, defined as  
\begin{align}
K^{\text{NLO}} = \sigma^{\rm NLO}/\sigma^{\rm LO}\,.
\end{align}
Right above the production threshold $\sqrts = 2m_t$, both LO and NLO cross sections are strongly enhanced, and 
in the limit $\sqrts \to 2m_t$ the NLO corrections to the on-shell process \eett diverge due to non-relativistic threshold
corrections, which manifest themselves as large logarithmic
contributions to the virtual one-loop matrix element. Instead, in the
off-shell process \eewwbb the Coulomb singularity is regularized by
the finite top-quark width, and the NLO corrections
remain finite. However, threshold corrections introduce a distinct
peak in the NLO corrections at $\sqrts = 2m_t$ with a maximum K-factor
of about 2.5. Below threshold the cross section drops sharply, but
QCD corrections remain significant. Far above threshold the NLO
corrections are rather small for both the on-shell and the off-shell
processes. For \eett, the
corrections remain positive for all \sqrts. In
fact, for large center-of-mass energies, the effect of the top quark
mass becomes negligible and the corrections approach the universal
leptonic massless quark pair-production correction factor $\alpha_s /
\pi$. In contrast, the NLO corrections to \eewwbb decrease
significantly faster for large center-of-mass energies, are at the
per cent level for $\sqrts=\ValGeV{1500}$, and come close to zero at 
 $\sqrts=\ValGeV{3000}$.
This corresponds to the fact that the non-resonant irreducible
background and interference contributions grow with energy relative to
the $t\bar t$ signal contribution, which receives purely positive
corrections. Our results suggest that at $\sqrts=\ValGeV{800}$,
positive corrections to the signal process and negative corrections to
the background are of the same order of magnitude and partially cancel
each other. This leads to very small NLO QCD corrections. However, at
this level the currently unknown and possibly large NLO EW corrections
to \eewwbb have to be included as well for reliable
predictions. Comparing off-shell to on-shell cross sections, we see
that they are about equal at threshold, but at $\sqrts=\ValGeV{800}$
the off-shell prediction is about $20\%$ larger. 

In the right panel of \cref{fig:tt-xsec-scan} we show the variation
for $\sqrts=\ValGeV{800}$ of the \eett and \eewwbb NLO predictions
with respect to the renormalization scale $\mu_{\rm R}$ in the
interval $\mu_{\rm R}=[1/8, 8] \cdot m_t$. Within the error band $[m_t
/ 2, 2 m_t]$ predictions for $t\bar t$  and \wbwb{} with fixed
top-quark width, 
$\Gamma_t=\Gamma_t(\mu_{\rm R}=m_t)$, vary at the level of a few
per cent, however with an opposite slope. To understand this behavior,
we show the scale variation of the off-shell process additionally with
a scale-dependent width, $\Gamma_t(\mu_{\rm R})$. With such a 
consistent setting of the width according to the input parameters,
including $\mu_{\rm R}$, scale variations in the off-shell process are
very similar to the on-shell one. We note that the scale dependence in
the top width is in principle a higher-order 
effect, such that both approaches are in principle valid to
estimate missing higher order effects by means of scale
variations. However, in order to properly recover the narrow width limit the
parameter settings for the width in the propagator and the decay part
of the matrix element have to match, including the scale setting.

\begin{figure}
\centering
   \includegraphics[width=\relplotwidth\textwidth]{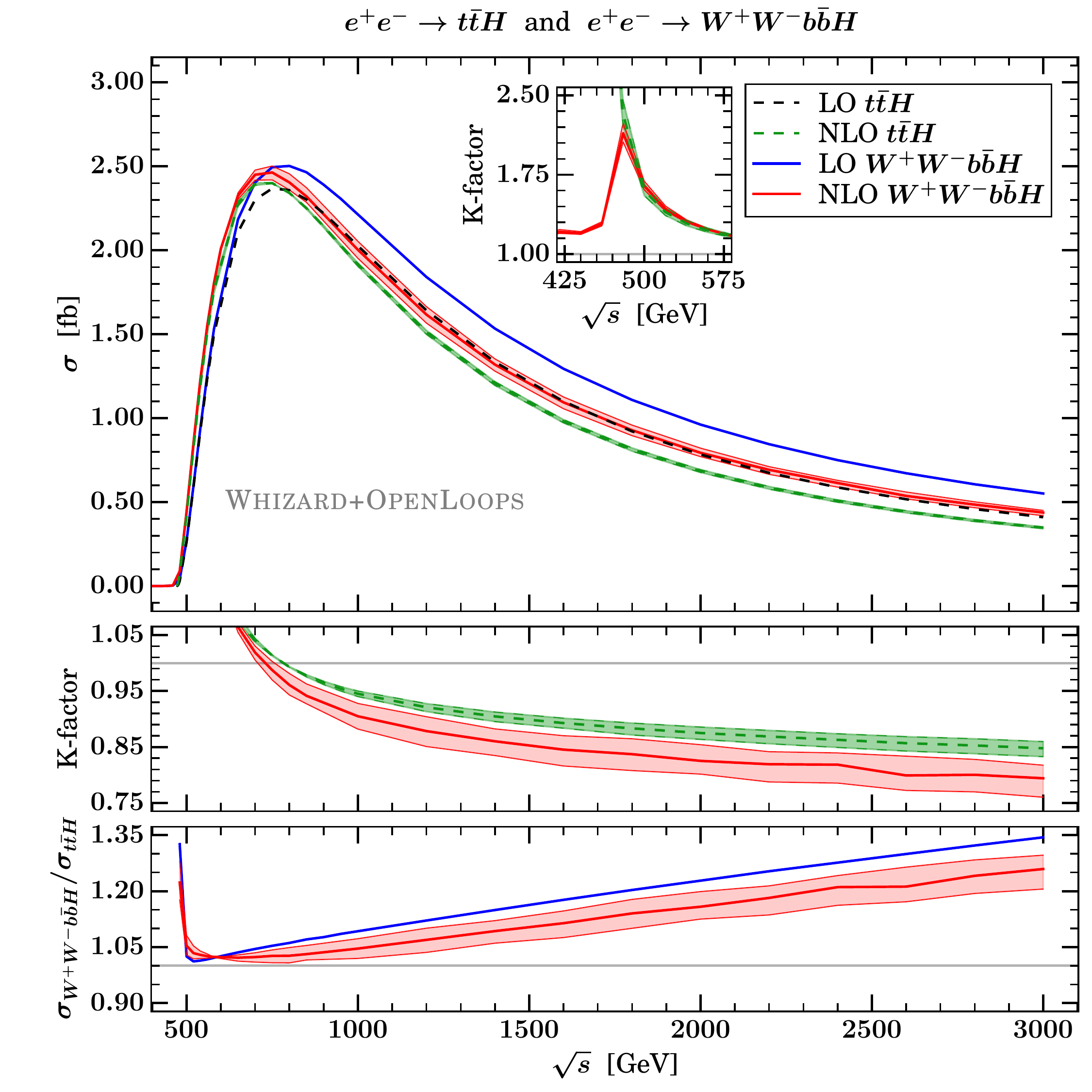}
         \quad
   \includegraphics[width=\relplotwidth\textwidth]{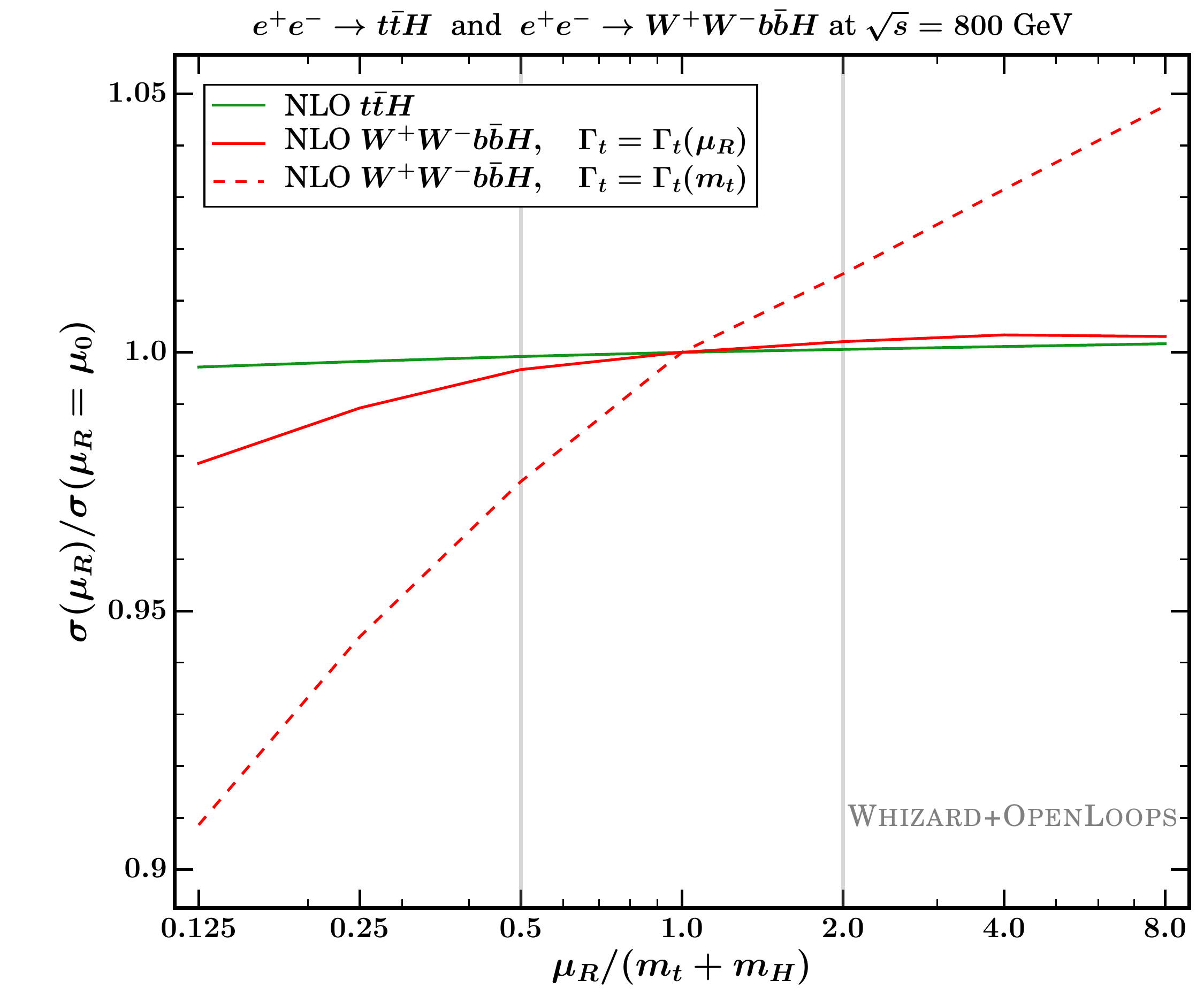}
   \caption{Total cross section of on-shell and off-shell $t\bar{t}H$ production
   subject to $\sqrts$ and $\mu_R$. Extra panels as in \cref{fig:tt-xsec-scan}.}
\label{fig:tth-xsec-scan}
\end{figure}

Inclusive cross sections for Higgs associated top-pair production are
shown in the left panel of~\cref{fig:tth-xsec-scan}. Also here we
observe an enhancement of the cross sections with a maximum located at
around $\sqrts=\ValGeV{800}$, i.e. far above the production threshold
at $2m_t + m_H \approx \ValGeV{471}$, where
$\sigma_{\text{incl.}}(\sqrts=\ValGeV{800}) \approx
\Valfb{2.4}$. Again, NLO QCD corrections are sizeable due to
non-relativistic Coulomb enhancements close to the production
threshold.  For the off-shell process \eewwbbH the corrections reach
$+100\%$ and remain large but finite below threshold, while for the
on-shell process they diverge close to threshold. Around the maximum
of the cross sections, NLO corrections vanish for both, the on-shell
and the off-shell process. Above this maximum, the NLO corrections
turn negative, yielding corrections at $\sqrts=\ValGeV{3000}$ of up to
$-15\%$ for the on-shell process \eetth and up to $-20\%$ for the
off-shell process \eewwbbH. Again one should also consider how the
off-shell cross sections behave relative to their on-shell
counterparts. While at LO the \eewwbbH cross section decreases
considerably slower with energy compared to the on-shell process
\eetth, at NLO the corrections to the off-shell process are 
more sizeable and negative with respect to the on-shell case,
yielding comparable inclusive cross sections for the on-shell and
off-shell process. Still, at $\ValGeV{3000}$ the off-shell inclusive
cross section is about $20\%$ smaller then the on-shell one.

In the right panel of~\cref{fig:tth-xsec-scan}, we display
renormalization scale variations at $\sqrts=\ValGeV{800}$ for Higgs
associated top-pair production. For this center-of-mass energy scale
variation uncertainties in \eetth are negligible (induced by vanishing
NLO QCD corrections), while in \eewwbbH{} 
with the standard choice $\Gamma_t=\Gamma_t(\mu_{\rm R}=m_t)$ they
amount to several per cent in the considered variation band. Similar
to the $t\bar t$ case, we also show scale variations taking
consistently into account the scale dependence in the top-quark width.
Here, the behavior of the off-shell process is very similar to 
 the on-shell one.

\begin{table}[tbp]
 \def\arraystretch{1.2}
  \caption{LO and NLO inclusive cross sections and K-factors for \eett
    and \eewwbb for various center-of-mass energies. Uncertainties at
    NLO are due to scale variation.} 
  \label{tab:xsec-incl-tt}
  \begin{center}
    \begin{tabular}{c c c c c c c c c}
      \toprule{}%
       & & \multicolumn{3}{c}{\eett} & & \multicolumn{3}{c}{\eewwbb} \\
      \sqrts [\GeV] & & $\sigma^{\text{LO}}[\fb]$ & $\sigma^{\text{NLO}}[\fb]$ & K-factor & & $\sigma^{\text{LO}}[\fb]$ & $\sigma^{\text{NLO}}[\fb]$ & K-factor \\
      \midrule{}%
      $500$  & & $548.4$ & $627.4^{+1.4\%}_{-0.9\%}$ & $1.14$ & & $600.7$ & $675.1^{+0.4\%}_{-0.8\%}$ & $1.12$ \\
      $800$  & & $253.1$ & $270.9^{+0.8\%}_{-0.4\%}$ & $1.07$ & & $310.2$ & $320.7^{+1.1\%}_{-0.7\%}$ & $1.03$ \\
      $1000$ & & $166.4$ & $175.9^{+0.7\%}_{-0.3\%}$ & $1.06$ & & $217.2$ & $221.6^{+1.1\%}_{-1.0\%}$ & $1.02$ \\
      $1400$ & & $86.62$ & $90.66^{+0.6\%}_{-0.2\%}$ & $1.05$   & & $126.4$ & $127.9^{+0.7\%}_{-1.5\%}$ & $1.01$ \\
      $3000$ & & $19.14$ & $19.87^{+0.5\%}_{-0.2\%}$ & $1.04$ & & $37.89$ & $37.63^{+0.4\%}_{-0.9\%}$ & $0.993$  \\ 
      \bottomrule{}
    \end{tabular}
  \end{center}
\end{table}

\begin{table}[tbp]
 \def\arraystretch{1.2}
  \caption{LO and NLO inclusive cross sections and K-factors for
    \eetth and \eewwbbH for various center-of-mass
    energies. Uncertainties at NLO are due to scale variation.} 
  \label{tab:xsec-incl-tth}
  \begin{center}
    \begin{tabular}{c c c c c c c c c}
      \toprule{}%
       & & \multicolumn{3}{c}{\eetth} & & \multicolumn{3}{c}{\eewwbbH} \\
      \sqrts [\GeV] & & $\sigma^{\text{LO}}[\fb]$ & $\sigma^{\text{NLO}}[\fb]$ & K-factor & & $\sigma^{\text{LO}}[\fb]$ & $\sigma^{\text{NLO}}[\fb]$ & K-factor \\
      \midrule{}%
      $500$ & & $0.26$ & $0.42^{+3.6\%}_{-3.1\%}$ & $1.60$ & & $0.27$ & $0.44^{+2.6\%}_{-2.4\%}$ & $1.63$ \\
      $800$ & & $2.36$ & $2.34^{+0.1\%}_{-0.1\%}$ & $0.99$ & & $2.50$ & $2.40^{+2.1\%}_{-1.9\%}$ & $0.96$ \\
      $1000$ & & $2.02$ & $1.91^{+0.5\%}_{-0.5\%}$ & $0.95$ & & $2.21$ & $2.00^{+2.5\%}_{-2.5\%}$ & $0.90$ \\
      $1400$ & & $1.33$ & $1.21^{+0.9\%}_{-1.0\%}$ & $0.90$ & & $1.53$ & $1.32^{+2.6\%}_{-3.0\%}$ & $0.86$ \\
      $3000$ & & $0.41$ & $0.35^{+1.4\%}_{-1.8\%}$ & $0.84$ & & $0.55$ & $0.44^{+2.9\%}_{-4.3\%}$ & $0.79$ \\
      \bottomrule{}
    \end{tabular}
  \end{center}
\end{table}

Finally, in tables~\ref{tab:xsec-incl-tt} and~\ref{tab:xsec-incl-tth}
we list inclusive cross sections for \ttbar and \ttbarH (both on- and
off-shell) processes, respectively, for several representative
center-of-mass energies. Listed uncertainties are due to scale
variations, where we employ the fixed top-width,
$\Gamma_t=\Gamma_t(\mu_{\rm R}=m_t)$. In section
\ref{sec:differential_predictions} we will continue our discussion of
NLO corrections for top-pair and Higgs associated top-pair production
at the differential level. There we will focus on
$\sqrts=\ValGeV{800}$, as here cross sections are largest for \tth
production, which should offer the best condition for a precise
determination of the top Yukawa coupling, as discussed in the
following section. While consider this as a viable running scenario
for a precision measurement, one should keep in mind that for other
energies the NLO QCD corrections will be larger in general, at least
at the inclusive level. 

\subsection{Determination of the top Yukawa coupling}
\label{ssec:yukawa}

\def\ytsm{y_t^{\mathrm{SM}}}
\def\xit{\xi_t}
A precise measurement of Higgs associated top-pair production allows for 
the direct determination of the top-quark Yukawa coupling \yt at the
per cent level~\cite{Agashe:2013hma,1409.7157}. This allows -- next to
the measurement of the $ttZ$ coupling -- for decisive probes of many
new physics models, as significant deviations from the Standard Model
value $\ytsm = \sqrt{2} m_t / v$ are predicted in many such models,
e.g.\ in generic two Higgs-doublet models, the MSSM or composite Higgs
or Little Higgs models. A per cent level measurement of \yt is 
feasible at future high-energy lepton colliders, as the $ttH$ and
$\bbwwH$ cross sections are quite sensitive to $y_t$.
\begin{figure}[tbp]
	\centering
	\includegraphics[width=\relplotwidth\textwidth]
                        {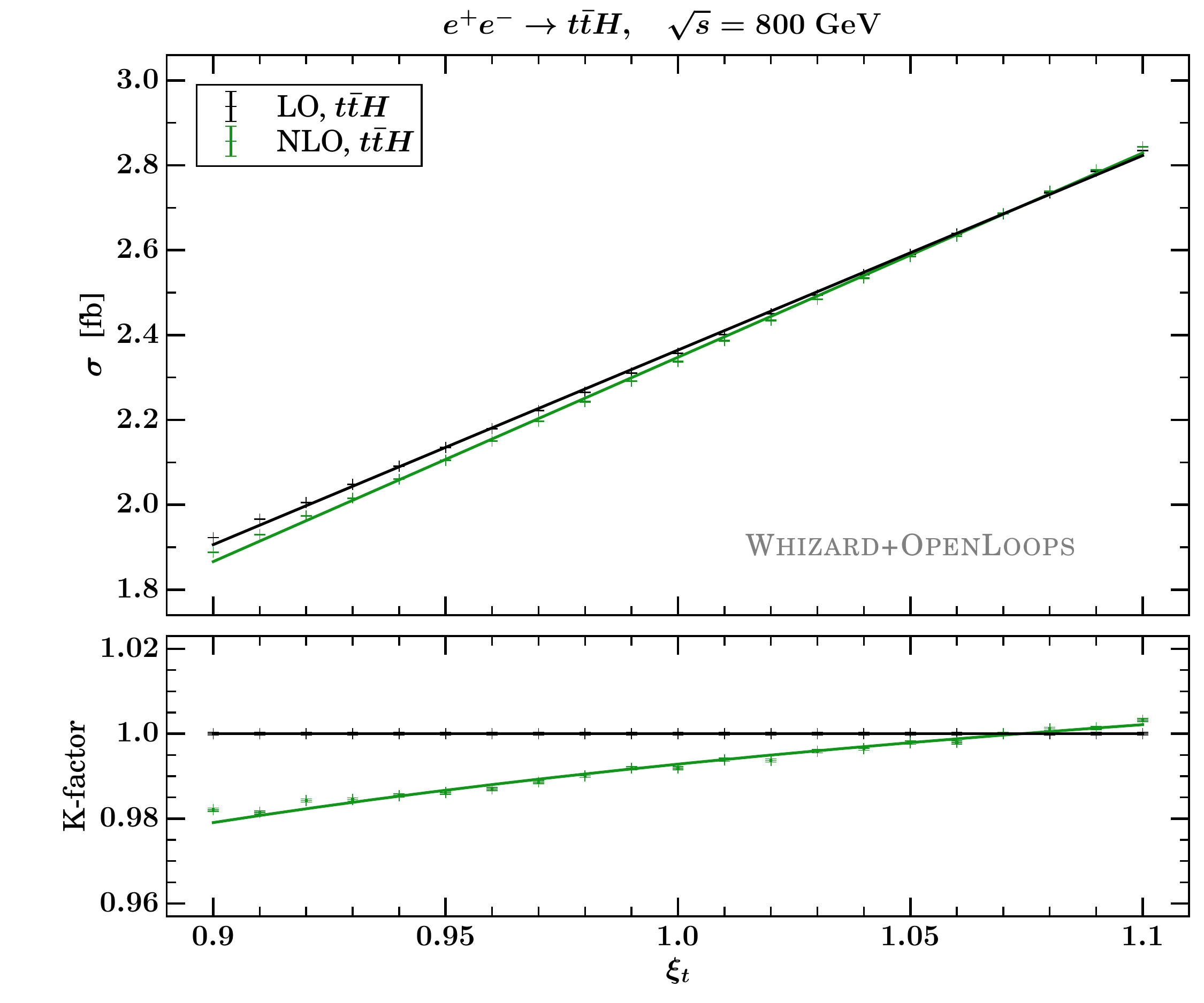} 
	                \quad
	\includegraphics[width=\relplotwidth\textwidth]
                        {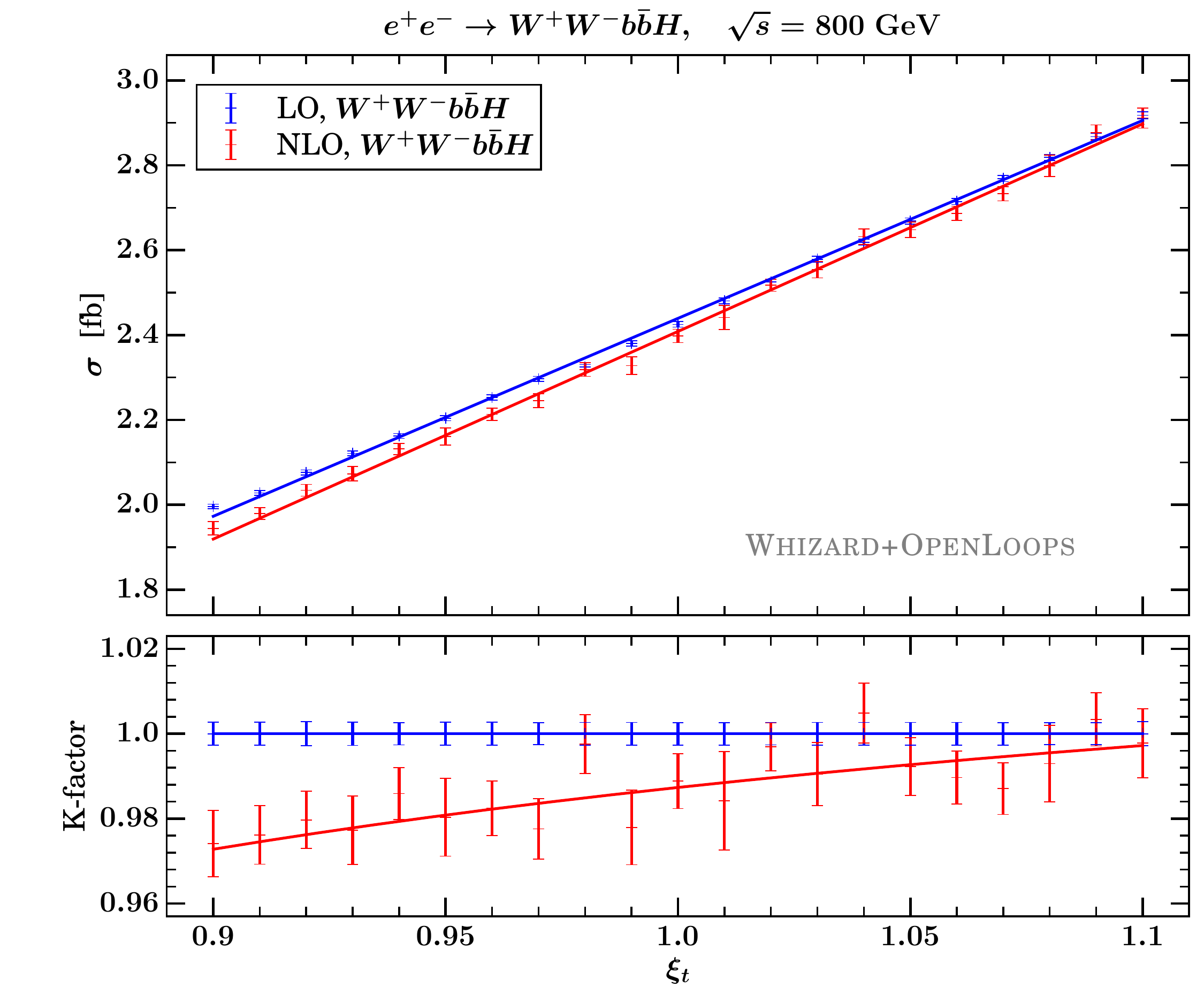}
  \caption{The \eetth{} and \eewwbbH{} LO and NLO cross sections as a
    function of the top Yukawa coupling modifier $\xi_t=y_t/y_t^{\rm
      SM}$, as well as a linear fit used to determine the coefficient
    $\kappa$ as described in the text, \eqref{eq:yukawa-sensitivity}.}   
	\label{fig:scan-yukawa-ttH}
\end{figure}
The sensitivity of the \tth processes (on- and off-shell) on \yt is
commonly expressed in terms of~\cite{1409.7157,1307.7644}  
\begin{equation}
  \label{eq:yukawa-sensitivity}
  \frac{\Delta y_t}{y_t} = \kappa \frac{\Delta \sigma}{\sigma}.
\end{equation}
In this way, the relative accuracy on the measured cross section
can directly be related to a relative accuracy on the top Yukawa
coupling. Since the $y_t$-dependence of the cross section is
approximately quadratic, $\kappa$ is close to 0.5.
More precisely, parameterizing deviations of the top-Yukawa coupling from
its SM value as $y_t=\xit \cdot \ytsm$ we can write the total cross section as
$\sigma(\xit)=\xit^2\cdot S + \xit\cdot I+ B$, where $S$ and $B$ 
denote $\ttbar H$ signal\footnote{More precisely, the $S$ term corresponds 
to squared $e^+e^-\to\ttbar H$ 
amplitudes excluding Higgsstrahlung contributions.} and 
background contributions, respectively, while $I$ stands for
interference terms. The $y_t$-sensitivity of $\ttbar H$ cross sections
can be determined via a linear fit of $\sigma(y_t)$, which corresponds
to
\begin{equation}
  \label{eq:yukawa-fromfit}
\kappa = \lim_{\xit\to 1} 
\sigma(\xit)\,\left[\frac{\mathrm{d}\sigma(\xit)}{\mathrm{d}\xit}\right]^{-1}\;=
\frac{S+I+B}{2S+I} = \frac{1}{2} +  \frac{I/2 + B}{2S+I}.
\end{equation}
Note that whereas $B$ is strictly positive, we can make no statement
about the sign of $I$. Eq. \eqref{eq:yukawa-fromfit} -- making the
quite general assumption that the signal dominates over the
interference, $-I < 2S$ -- shows that $\kappa<0.5$ can only be
realized via sufficiently large and negative interference
contributions, $I<-2 B$. From the above reasoning, we see that
$\kappa$ quantifies the contamination from the Higgsstrahlung
subprocess into \eetth, and, for off-shell processes, of any
additional background subprocess including contributions proportional
to the $HWW$ coupling.

In table~\ref{tab:yukawa-slope} we list the values of $\kappa$
corresponding to the LO and NLO fits shown in
fig. \ref{fig:scan-yukawa-ttH}. As expected, all listed
$\kappa$-values are close to $0.5$. 
\begin{table}[htbp]
 \def\arraystretch{1.2}
  \caption{The parameter $\kappa$ as defined in eq.~\eqref{eq:yukawa-fromfit} for \eetth and \eewwbbH at LO and NLO for $\sqrts = \ValGeV{800}$.}
   \label{tab:yukawa-slope}
   \begin{center}
      \begin{tabular}{c c c c c c c}
        \toprule{}%
        $\epem \to$ & & $\kappa^{\rm LO}$ & & $\kappa^{\rm NLO}$ & & $\kappa^{\rm NLO}/\kappa^{\rm LO}$  \\
        \midrule{}%
        \tth   & & 0.514 & & 0.485 & & 0.943\\
        \wwbbH & & 0.520 & & 0.497 & & 0.956\\
        \bottomrule{}
      \end{tabular}
      \end{center}
\end{table}
For \eetth at LO the Higgsstrahlung contribution induces a value
$\kappa > 0.5$. For the off-shell process \eewwbbH we observe a
slightly larger value compared to the on-shell process, originating
from additional irreducible backgrounds. The NLO QCD corrections to
$\kappa$ turn out to be significant. They decrease $\kappa$ by $6.0\%$
and $4.6\%$ compared to LO for the on- and off-shell case,
respectively. This can be understood from a different behavior of the
signal and background contributions with respect to QCD corrections. 
From table~\ref{tab:yukawa-slope} we can infer that at NLO
interference terms are indeed negative for the on-shell $\ttbar H$
process. 

The sensitivity formula, \eqref{eq:yukawa-fromfit}, can be used to
assess the impact of perturbative corrections on the extraction of
$y_t$. This is roughly half as large as the corrections reported
in~\cref{fig:tth-xsec-scan}. As already observed at the cross section
level, the shifts in the extracted $y_t$ value that result from the
inclusion of NLO corrections and off-shell contributions have
comparable size and opposite sign. The magnitude of the individual
effects amounts to a few per cent at $\ValGeV{800}$ and grows up to
about 10\% at full CLIC energy.

\subsection{Polarization effects}
\label{ssec:polarization}

We complete our study of inclusive cross sections for leptonic
top-pair and Higgs associated top-pair production with an
investigation of possible beam polarization effects on these
processes. Beam polarization is a powerful tool at linear colliders to
disentangle contributing couplings and to reduce
backgrounds~\cite{1307.8102} or improve the measurement of the top
Yukawa coupling~\cite{1409.7157}. In tables~\ref{tab:tt-polarized}
and~\ref{tab:tth-polarized} inclusive LO and NLO cross sections with
different polarization settings as suggested by the favored ILC
running scenarios~\cite{Barklow:2015tja} and two different collider
energies are listed for the on-shell processes \eett and \eetth,
respectively. While cross sections vary strongly with the beam
polarization, the K-factors are unaffected. These results confirm the
naive expectation that NLO QCD corrections fully factorize with
respect to the beam polarization due to the uncolored initial state. 
On the other hand, one can view the constant K-factors in
tables~\ref{tab:tt-polarized} and~\ref{tab:tth-polarized} as
validation of the polarization dependent \wz-\ol-interface via a BLHA
extension. The factorization also holds when top-quark decays are
considered and we refrain from showing polarized cross sections for
off-shell production processes.

\begin{table}[htbp]
 \def\arraystretch{1.1}
  \caption{LO and NLO inclusive cross sections for \eett with possible
    ILC beam polarization settings at $\sqrts = \ValGeV{800}$ and
    $\ValGeV{1500}$.} 
  \label{tab:tt-polarized}
   \begin{center}
  \begin{tabular}{r r c c c c c r r c}
    \toprule{}%
    & & & \multicolumn{3}{c}{$\sqrts = \ValGeV{800}$} & & \multicolumn{3}{c}{$\sqrts = \ValGeV{1500}$}
    \\
    $P(e^-)$ & $P(e^+)$ & & $\sigma^{\text{LO}}[\fb]$ & $\sigma^{\text{NLO}}[\fb]$ & K-factor & & $\sigma^{\text{LO}}[\fb]$ & $\sigma^{\text{NLO}}[\fb]$ & K-factor \\
    \midrule{}%
    $0\%  $ & $0\%  $ & & 253.7 & 272.8 & 1.075 & & 75.8 & 79.4 & 1.049 \\
    $-80\%$ & $0\%  $ & & 176.5 & 190.0 & 1.077 & & 98.3 & 103.1 & 1.049 \\
    $80\% $ & $0\%  $ & & 176.5 & 190.0 & 1.077 & & 53.2 & 55.9 & 1.049 \\
    $-80\%$ & $30\% $ & & 420.8 & 452.2 & 1.074 & & 124.9 & 131.0 & 1.048\\
    $-80\%$ & $60\% $ & & 510.7 & 548.7 & 1.074 & & 151.6 & 158.9 & 1.048\\
    $80\% $ & $-30\%$ & & 208.4 & 224.5 & 1.077 & & 63.0 & 66.1 & 1.049\\
    $80\% $ & $-60\%$ & & 240.3 & 258.9 & 1.077 & & 72.7 & 76.3 & 1.049\\
    \bottomrule{}
  \end{tabular}
   \end{center}
\end{table}
\begin{table}[htbp]
 \def\arraystretch{1.1}
  \caption{LO and NLO inclusive cross sections for \eetth with
    possible ILC beam polarization settings at $\sqrts = \ValGeV{800}$
    and $\sqrts = \ValGeV{1500}$.} 
  \label{tab:tth-polarized}
  \begin{center}
  \begin{tabular}{r r c c c c c c c c}
    \toprule{}%
    & & & \multicolumn{3}{c}{$\sqrts = 800~\GeV$} & & \multicolumn{3}{c}{$\sqrts = 1500~\GeV$}
    \\
    $P(e^-)$ & $P(e^+)$ & & $\sigma^{\text{LO}}[\fb]$ & $\sigma^{\text{NLO}}[\fb]$ & K-factor & & $\sigma^{\text{LO}}[\fb]$ & $\sigma^{\text{NLO}}[\fb]$ & K-factor \\
    \midrule{}%
    $0\%  $ & $0\%  $ & & 2.358 & 2.337 & 0.991 & & 1.210 & 1.064 & 0.879 \\
    $-80\%$ & $0\%  $ & & 1.583 & 1.571 & 0.992 & & 1.576 & 1.381 & 0.876 \\
    $80\% $ & $0\%  $ & & 1.584 & 1.571 & 0.992 & & 0.843 & 0.746 & 0.885 \\ 
    $-80\%$ & $30\% $ & & 3.988 & 3.950 & 0.990 & & 2.003 & 1.757 & 0.877 \\
    $-80\%$ & $60\% $ & & 4.840 & 4.795 & 0.991 & & 2.429 & 2.128 & 0.876 \\
    $80\% $ & $-30\%$ & & 1.860 & 1.846 & 0.992 & & 0.996 & 0.879 & 0.883 \\
    $80\% $ & $-60\%$ & & 2.134 & 2.120 & 0.993 & & 1.148 & 1.018 & 0.886 \\
    \bottomrule{}
  \end{tabular}
\end{center}
\end{table}

\section{Numerical predictions for differential distributions}
\label{sec:differential_predictions}

Theoretically -- but also experimentally -- leptonic \ttbar and \tth
production and decay are very similar. Therefore, a sound
understanding of \ttbar production and decay in the continuum, where
experimentally a large amount of data can easily be accumulated, is a
necessary prerequisite for precision measurements of the top Yukawa
coupling in the \eetth process. In this section we discuss
differential predictions for \eett and \eetth at
$\sqrt{s}=\ValGeV{800}$ including NLO QCD corrections and off-shell
effects in the decays. We also present predictions for the
forward-backward asymmetry in \eett. 

\subsection{Top-pair production and decay}
\label{sec:nlo_tt}

\begin{figure*}[tbp]
\centering
   \includegraphics[width=\relplotwidth\textwidth]{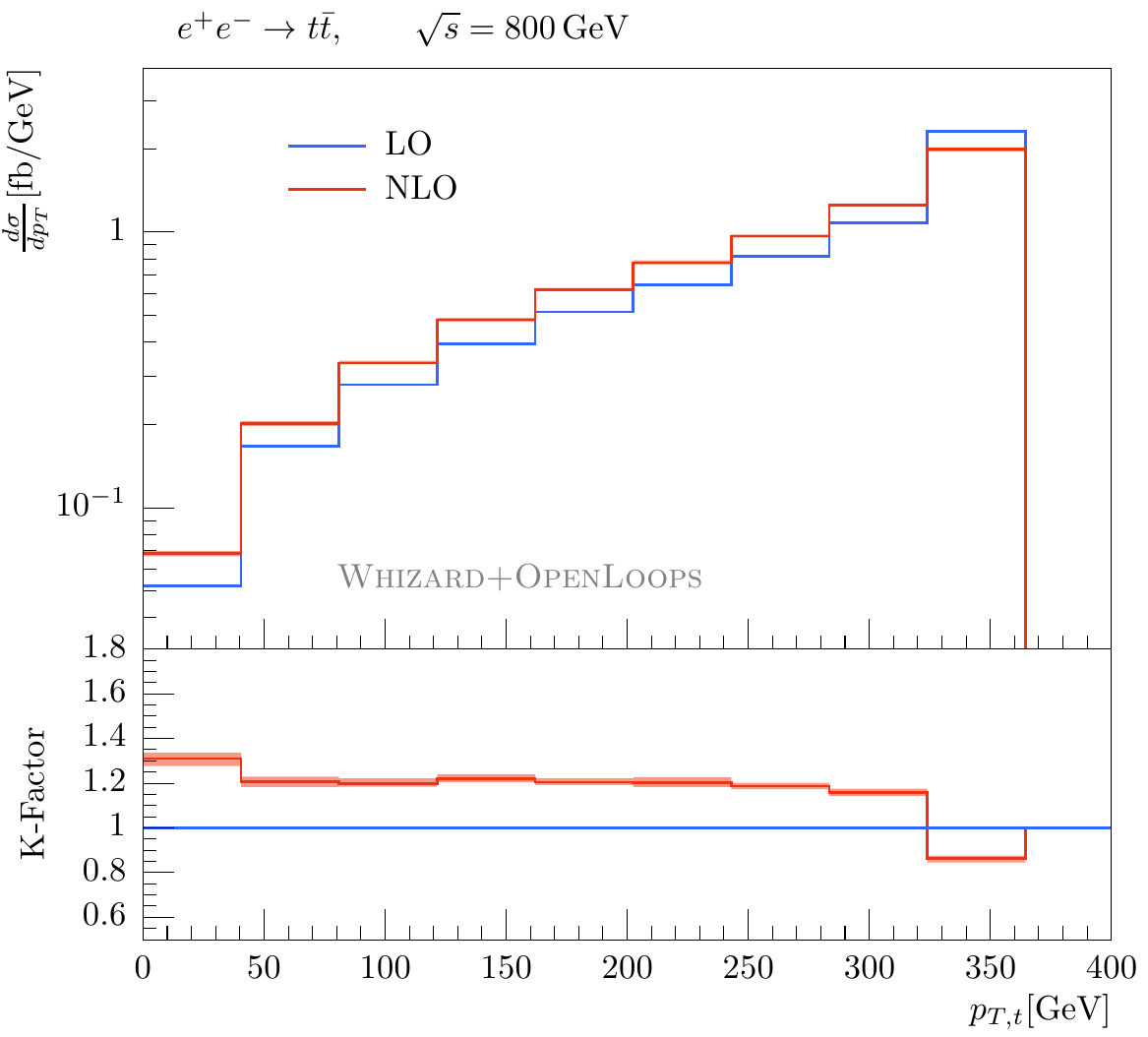}
         \quad 
      \includegraphics[width=\relplotwidth\textwidth]{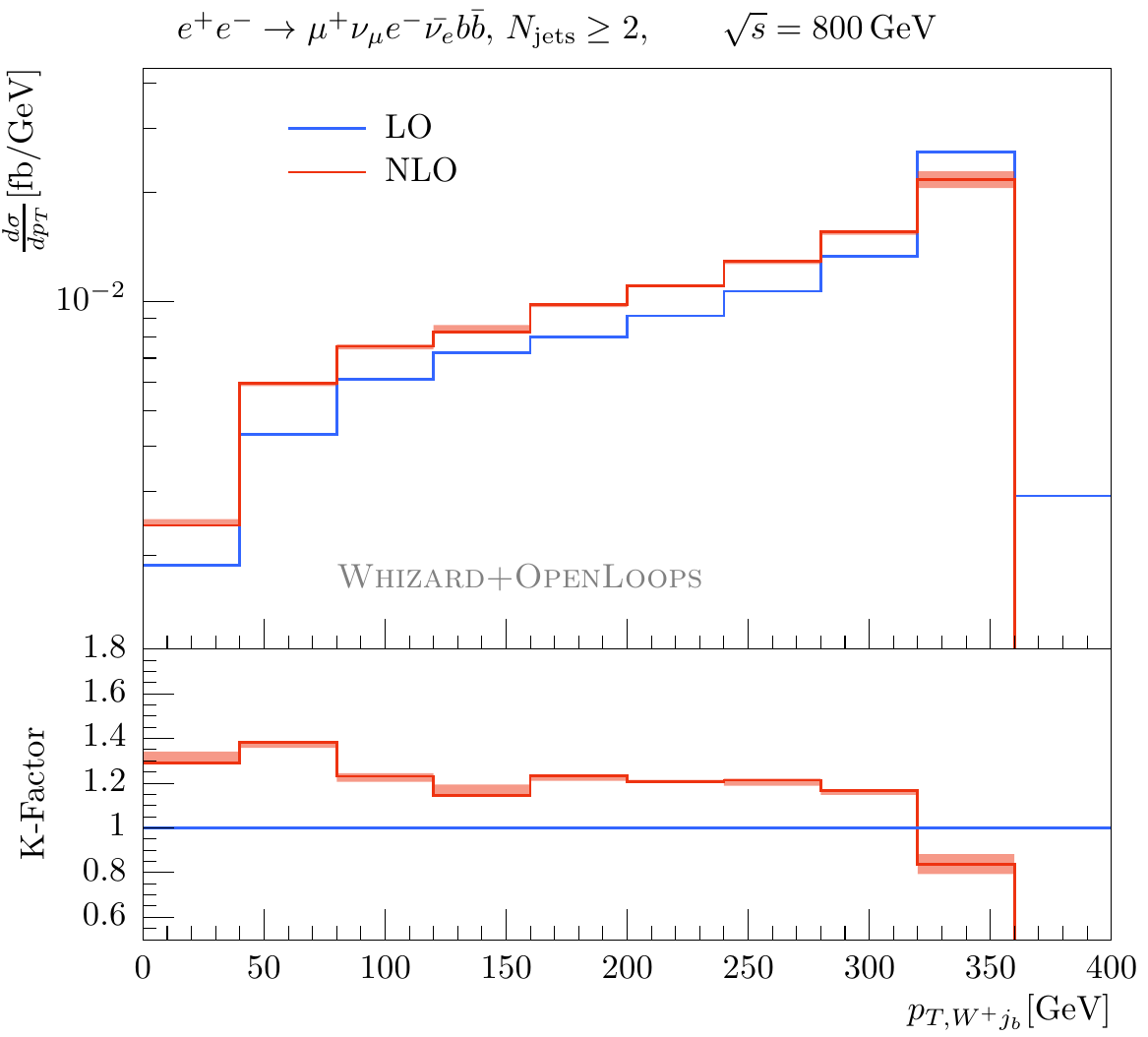}
   \caption{Differential distributions in the transverse momentum of
     the top quark in \eett (left) and the reconstructed top quark in
     \eellllbb (right).  Shown are LO (blue) and NLO (red) predictions
     together with the corresponding K-factors and NLO scale
     uncertainties. 
}
\label{fig:tt-pT}
\label{fig:tt-firstplot}
\end{figure*}

We start our analysis of differential distributions for top-pair production and
decay considering in \cref{fig:tt-pT} the top-quark transverse momentum
distribution for the on-shell process \eett and the corresponding off-shell
process \eellllbb including leptonic decays. For the latter the top quark is
reconstructed from its leptonic decay products at Monte Carlo truth level, i.e.
$\pTtoprec = p_{{\rm T},\ell^+ \nu \jb }$. 
Despite the different normalization of the two distributions, 
due to the fact that the on-shell process does not include leptonic branching
ratios,
the LO and NLO shapes are very similar below the Jacobian peak located
at around $\ValGeV{350}$. This peak with its large event density is smeared out by 
the NLO corrections, in particular due to kinematic shifts induced by the real gluon radiation, 
yielding corrections at the level of $-20\%$ at the peak and around $+20\%$ below the
peak.
%
%
For the on-shell process the phase-space above the Jacobian peak is
kinematically not allowed at LO and gets only sparsely populated at
NLO. In contrast, for the off-shell process this kinematic regime is
allowed already at LO. The observed sizeable corrections in the
transverse momentum of the intermediate top quarks also translate into
relevant corrections in the directly observable transverse momentum of
the final state leptons, as shown in appendix \ref{app:tt_further} 
(\cref{fig:tt-pTl}). Namely, we find corrections up to $-40\%$ and up
to $-30\%$ for the hardest and second hardest lepton, respectively. In
a realistic setup, where experimental selection cuts have to be
applied on the leptons, such effects become also relevant for the
fiducial cross section in precision top physics. 

\begin{figure*}[tbp]
\centering
   \includegraphics[width=\relplotwidth\textwidth]{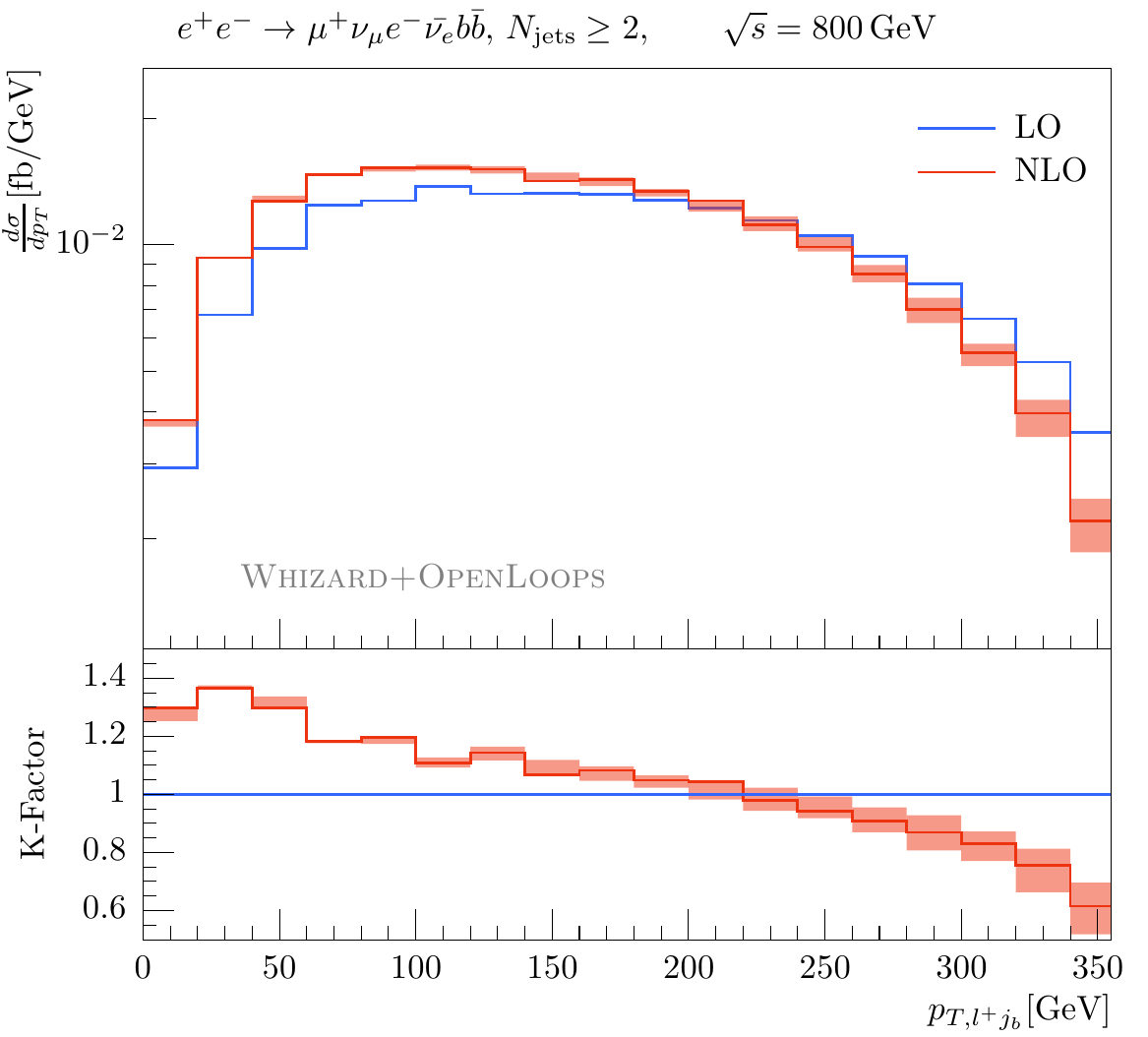}
   \quad
      \includegraphics[width=\relplotwidth\textwidth]{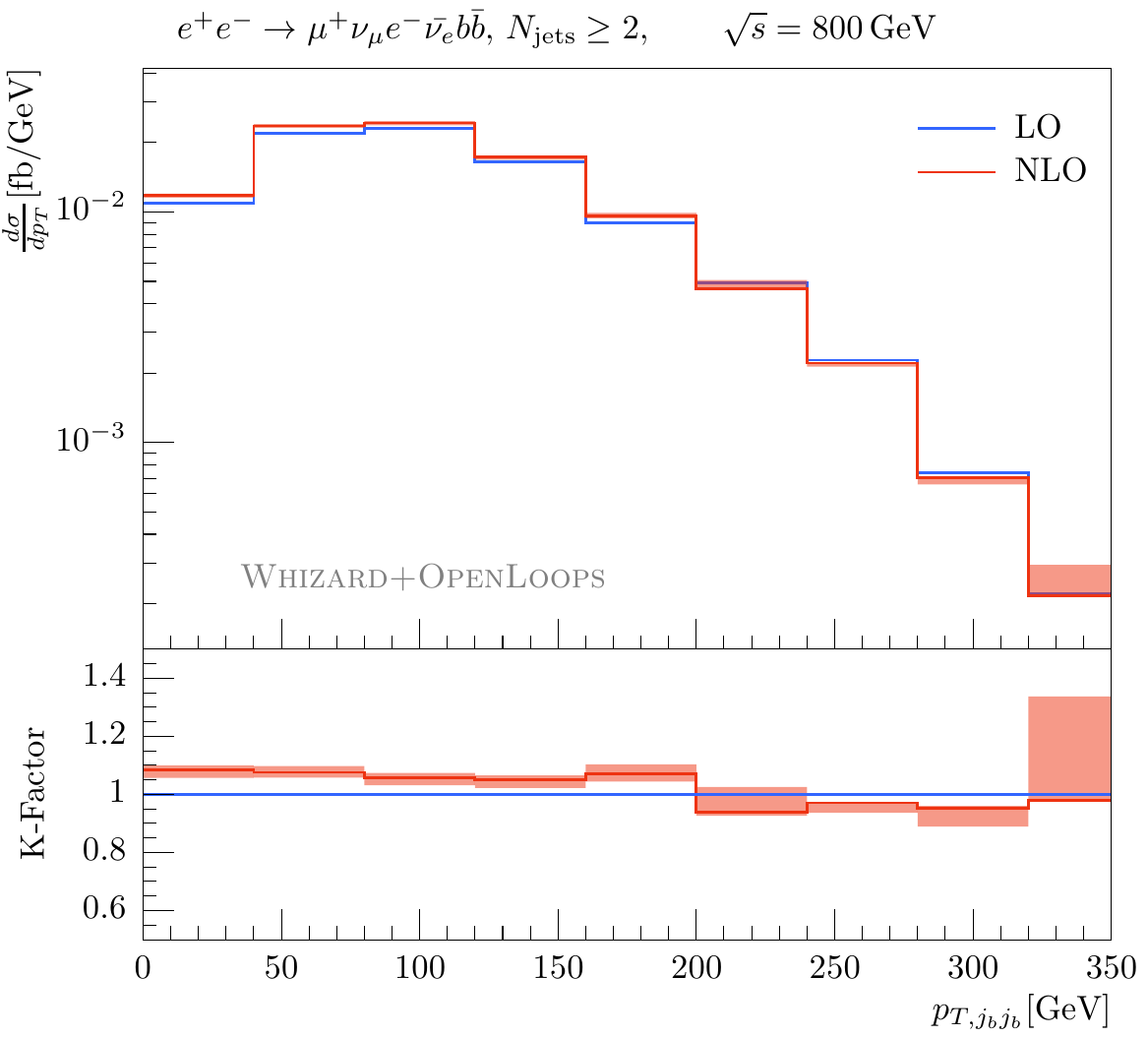}
   \caption{
Transverse momentum distribution of the bottom-jet--lepton system (left), \pTblp, and of the \jb--\jbbar system (right), \pTbb, in \eellllbb. 
Curves and bands as in Fig.~\ref{fig:tt-firstplot}.
}
\label{fig:tt-pTbl}
\end{figure*}

Experimentally, \pTtoprec is not directly measurable, as in the
considered leptonic decay mode the top quark cannot be 
exactly reconstructed due to the two invisibly escaping neutrinos. As
a proxy we can, however, construct and measure the transverse momentum
of the $b$-jet--lepton system, \pTblp.  Corresponding predictions for
\eellllbb are shown in \cref{fig:tt-pTbl} (left). Here we observe a
tilt of the NLO shape with respect to the LO one, yielding corrections
up to $20\%$ for small \pTblp and up to $-40\%$ for large \pTblp. 
In contrast, the transverse momentum distribution of the \jb--\jbbar
system -- as shown on the right of \cref{fig:tt-pTbl} -- only receives
mild QCD corrections at the level of $10\%$.

\begin{figure*}[tbp]
  \centering
  \includegraphics[width=\relplotwidth\textwidth]
                {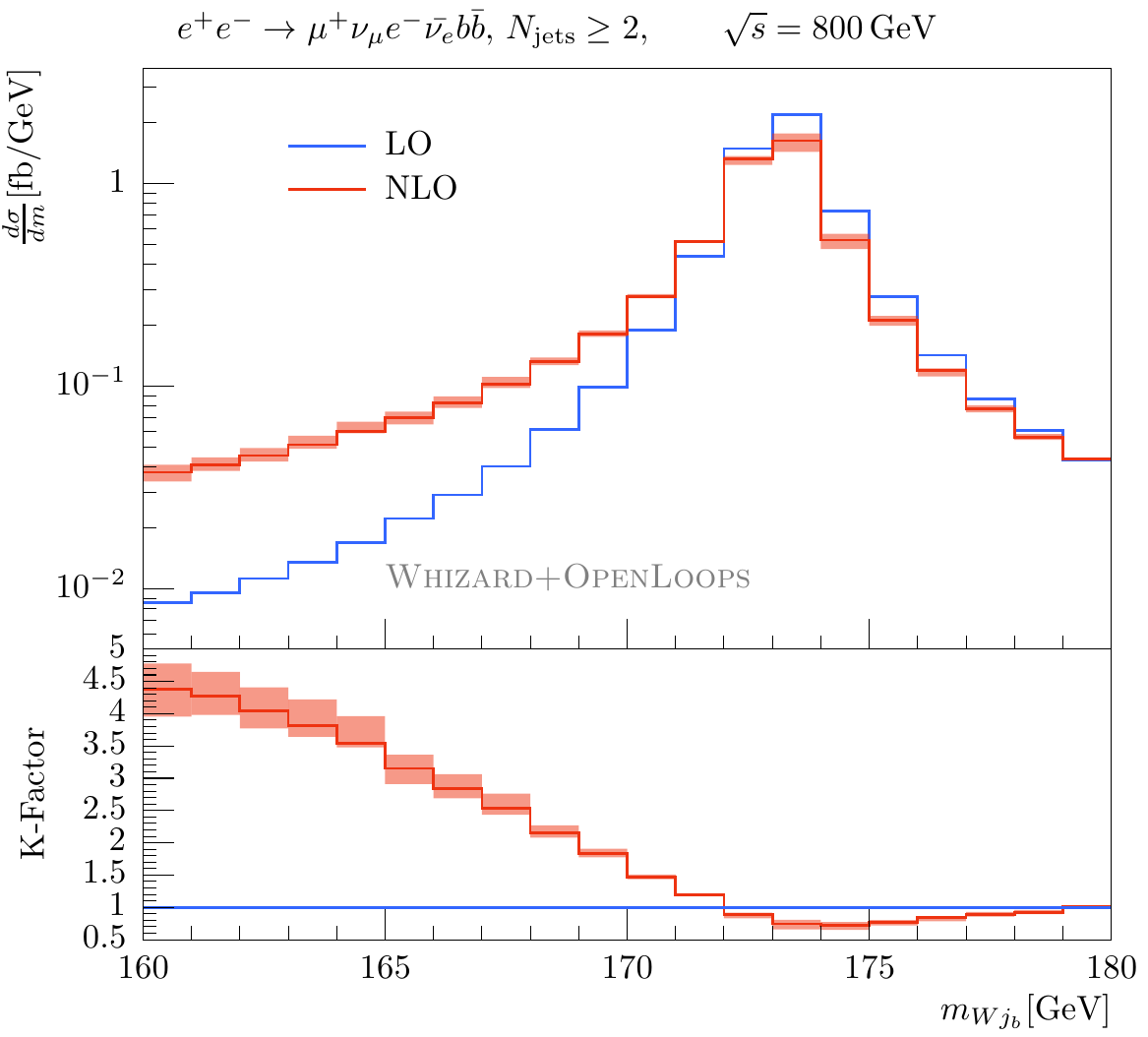}
                \quad
  \includegraphics[width=\relplotwidth\textwidth]
                  {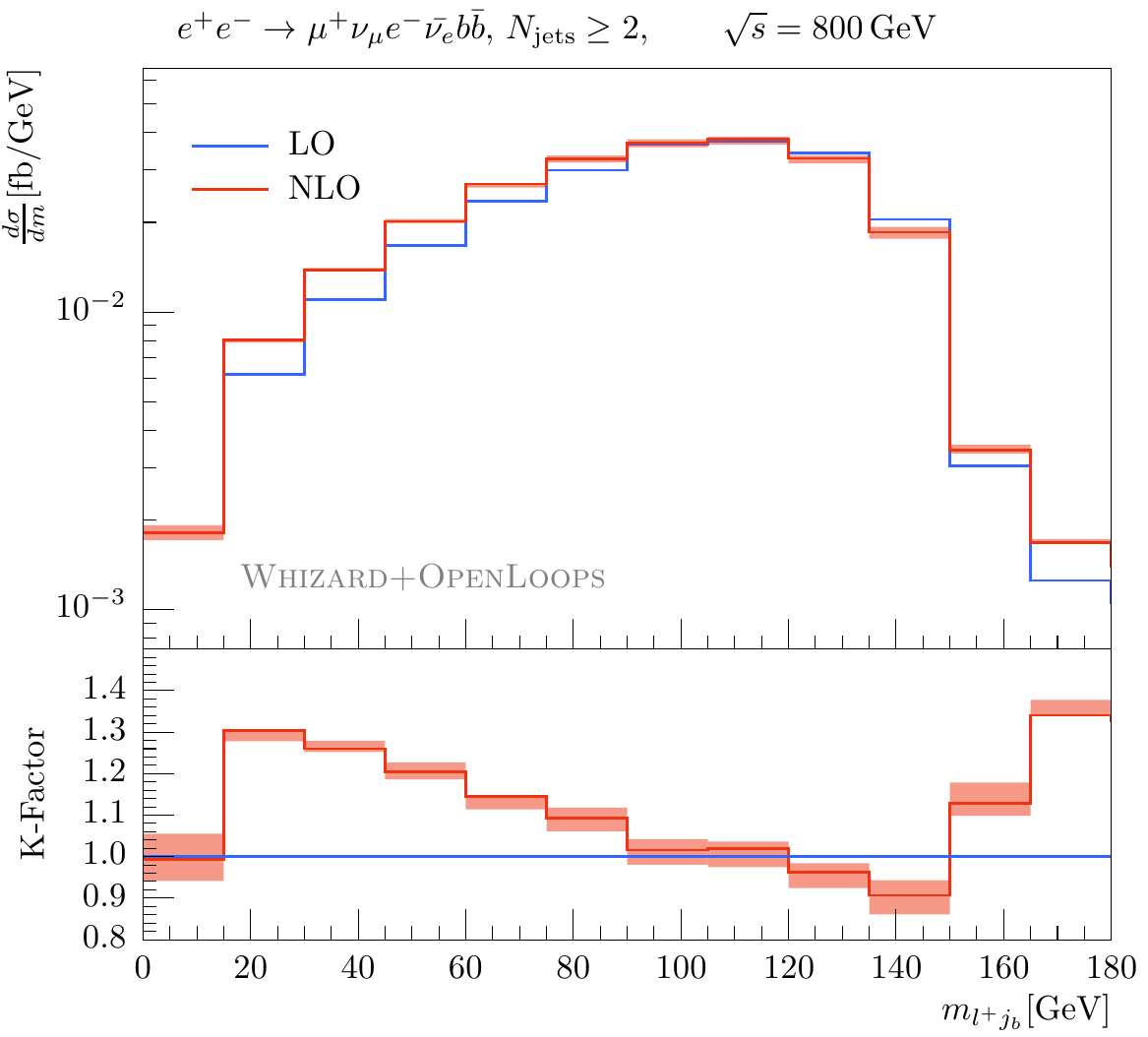}
   \caption{Reconstructed top invariant mass (left) and invariant mass
     of the $b$-jet--$\ell^+$ system in  \eellllbb.  
     Curves and bands as in Fig.~\ref{fig:tt-firstplot}.}
\label{fig:BW-inv}
\end{figure*}

One of the observables of prime interest is the kinematic mass of the
top resonance. In \cref{fig:BW-inv} we 
show on the left the reconstructed invariant top-quark mass, $\mtoprec =
m_{\ell^+ \nu \jb }$, 
where the $\ell^+ \nu \jb$ system is identified based on Monte Carlo truth.
At LO and close to the peak, this distribution corresponds to the Breit-Wigner
that arises due to the propagator.
Off-shell effects and non-resonant contributions become visible a couple of GeV
away from the pole and tend to increase the background.
At NLO we observe a drastic shape distortion compared to LO -- in particular below the
resonance peak. These NLO shape distortions are very sensitive to the
cone size of the employed jet algorithm.
They can be attributed to QCD radiation that escapes the $b$-jet forming either
a separate light jet or being recombined with the other b-jet. 
The reconstructed invariant top-quark mass is thus on average significantly
shifted compared to the top-quark resonance. Similar shape distortions have also
been observed in \Rcite{1511.02350} as well as at the LHC
\cite{Frederix:2016rdc,Kraus:2016ayb}.

Again, the perfectly reconstructed top-quark mass is not directly measurable due to the escaping neutrinos. 
However, we can resort to the invariant mass of the $b$-jet and the associated charged lepton. 
In fact, this distribution can be used to measure the top-quark mass via~\cite{Kharchilava:1999yj, Beneke:2000hk, Biswas:2010sa}
\begin{equation}
  \label{eq:top_estimator}
  m_t^2 = m_W^2 + \frac{2\langle \mbl^2\rangle}{1 - \langle\thetabl\rangle}\,,
\end{equation}
where $\langle \mbl^2\rangle$ and $\langle\thetabl\rangle$ are the
mean values of the corresponding invariant mass and angular
distributions. Predictions for the \mblp invariant mass distribution
are shown on the right of \cref{fig:BW-inv}. The position of the
kinematic edge at around $\mblp \approx \ValGeV{150}$ is unaffected by
the NLO QCD corrections, however, below the edge we observe
significant shape effects with corrections varying between $-10\%$ and
$+20 \%$.
Finally, as shown in appendix~\ref{app:tt_further}
(\cref{fig:BL-Theta}), we want to note that QCD radiative corrections
to the angular separation \thetablp entering the top-quark mass
estimator of eq.~\eqref{eq:top_estimator} are negligible.

\subsection{Forward-backward asymmetries}
\label{ssec:afb}
At a future lepton collider the top quark forward-backward asymmetry $A_{FB}$
is defined as
\begin{equation}
  \label{eqn:afb}
  A_{FB} = \frac{\sigma(\cos{\theta_t}>0) - \sigma(\cos{\theta_t}<0)}
    {\sigma(\cos{\theta_t}>0) + \sigma(\cos{\theta_t}<0)} \,,
\end{equation}
where $\theta_t$ is the angle between the positron beam axis 
and the outgoing top-quark. This asymmetry
can be measured with a precision below $2\%$~\cite{1307.8102}.
The SM prediction for $A_{FB}$
is non-zero due to interference contributions between s-channel $Z$- and
$\gamma^*$-exchange in the dominant production
process~\cite{Bernreuther:2006vp}. Various new physics models can
substantially alter the SM prediction (for an overview
cf.~\cite{Richard:2014upa}) and thus, a precise determination of
$A_{FB}$ serves as a stringent probe for new physics.~\footnote{A
  similar asymmetry can also be defined and measured at hadron
  colliders, where the dominant top-production channels are of  QCD
  type, such that within the SM the LO forward-backward asymmetry is
  zero. 
At the Tevatron a non-vanishing $A_{FB}$ was
measured~\cite{Aaltonen:2008hc,Aaltonen:2011kc,Abazov:2011rq}, posing
a long-standing puzzle, which was finally solved within the SM by
taking QCD corrections up to NNLO~\cite{Czakon:2014xsa} and NLO EW
corrections~\cite{Hollik:2011ps} into account.} 

\begin{figure*}[tbp]
\centering
   \includegraphics[width=\relplotwidth\textwidth]{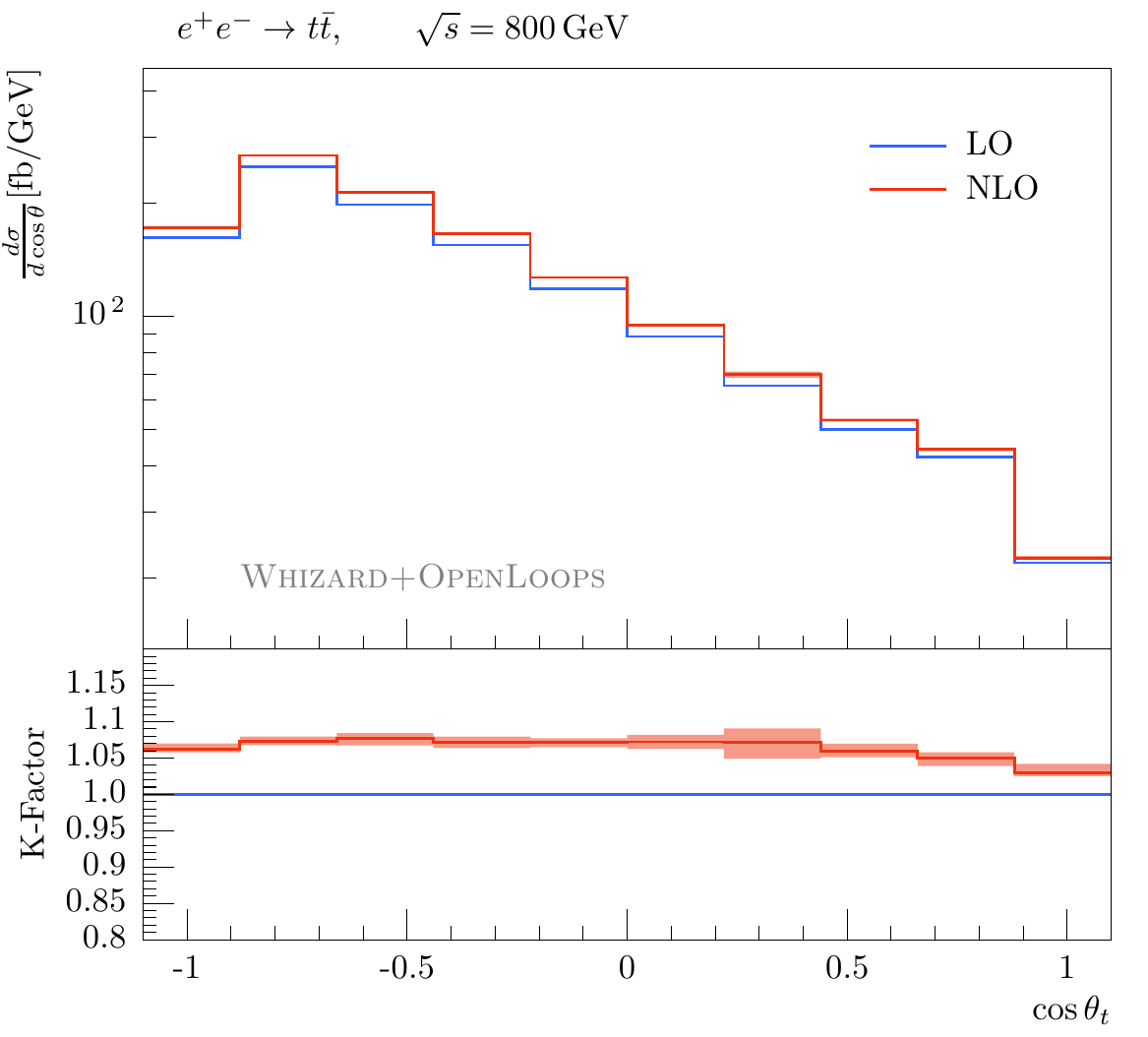}
\quad
     \includegraphics[width=\relplotwidth\textwidth]{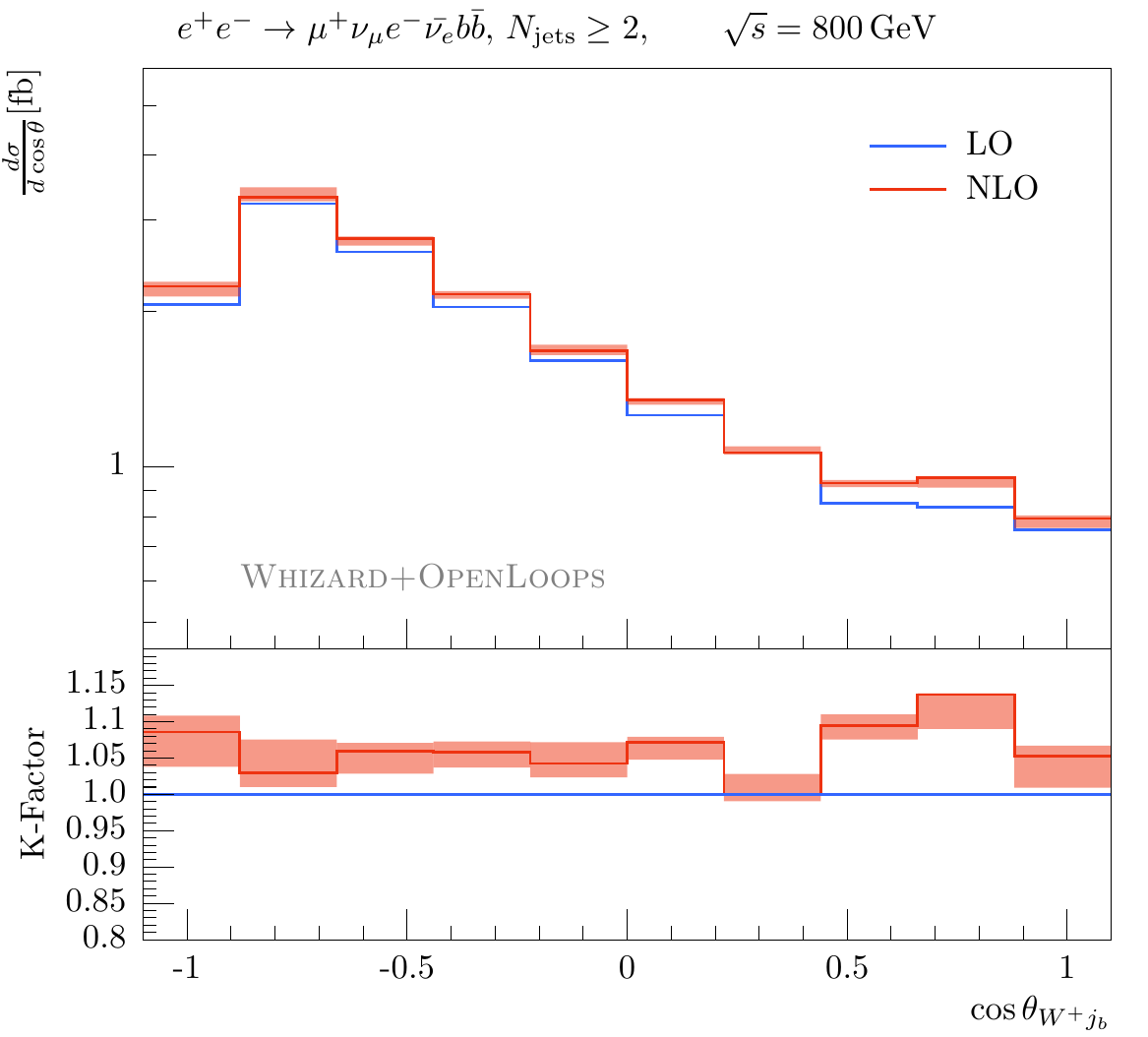}
   \caption{Differential distributions in the azimuthal angle of the top quark in \eett (left) and \eellllbb (right). 
Curves and bands as in Fig.~\ref{fig:tt-firstplot}.
}
\label{fig:tt-theta}
\end{figure*}

\begin{table}[htbp]
 \def\arraystretch{1.2}
    \caption{Forward-backward asymmetries of the top quark, $A_{FB}$, and the anti-top quark, $\bar A_{FB}$.
}
   \label{tab:AFB-top}
   \begin{center}
      \begin{tabular}{cl c c  c  c}
        \toprule{}%
	& $\epem \to$ & & $A_{FB}^{\text{LO}}$ & $A_{FB}^{\text{NLO}}$ & $A_{FB}^{\text{NLO}} / A_{FB}^{\text{LO}}$ \\
        \midrule{}%
	\multirow{ 4}{*}{$A_{FB}$}	&$t\bar{t}$ & & -0.535 & -0.539 & 1.013 \\
	&$\bbww$ & & -0.428 & -0.426 & 0.995 \\
	&$\mu^+ e^- \nu_\mu \bar{\nu}_e b \bar{b}$ & & -0.415 & -0.409 & 0.986 \\
        &$\mu^+ e^- \nu_\mu \bar{\nu}_e b \bar{b}$, without neutrinos & &
	-0.402 & -0.387 & 0.964 \\
        \midrule{}
	\multirow{ 4}{*}{$\bar A_{FB}$}	& $t\bar{t}$ & & 0.535 & 0.539 & 1.013 \\
	&$\bbww$ & & 0.428  & 0.426  & 0.995 \\
	&$\mu^+ e^- \nu_\mu \bar{\nu}_e b \bar{b}$ & & 0.415  & 0.409 & 0.986 \\
        &$\mu^+ e^- \nu_\mu \bar{\nu}_e b \bar{b}$, without neutrinos & &
	0.377 & 0.350 & 0.928  \\
      \bottomrule{}
      \end{tabular}
   \end{center}
\end{table}

In \cref{fig:tt-theta} we show the underlying distribution in
the angle of the (reconstructed) top quark with respect to the beam axis 
for on-shell top-pair production and the corresponding 
off-shell process \eellllbb. The prediction of a non-zero
forward-backward asymmetry at lepton colliders is apparent in
fig.~\ref{fig:tt-theta} and  the shape of this distribution is hardly
affected by radiative corrections, which yield an almost constant
K-factor of about 1.05.  

For $\cos\theta_{W^+\jb} \lesssim 0.75$, the angular distribution of
the reconstructed top quark in \eellllbb correlates with the on-shell
prediction. However, for $\cos\theta_{W^+\jb} \gtrsim 0.75$, there is
an enhancement of events, which can be attributed to single-top
background diagrams. This has a significant effect on the
reconstructed top forward-backward asymmetry, which is reduced by
about $20\%$, see the \eellllbb and \eewwbb predictions in table
\ref{tab:AFB-top}.  

In table \ref{tab:AFB-top} we list LO and NLO predictions for the
forward-backward asymmetry $A_{FB}$ (and the corresponding asymmetry
for the anti-top quark), considering different treatments of the
top-quark off-shellness. In \eellllbb, either the top-quark is
reconstructed at MC truth level or the information of the neutrino
momenta is dropped. NLO QCD corrections to $A_{FB}$ can be sizeable
(up to a few percent), but are small compared to the changes
associated with increasing the final-state multiplicity and taking
into account all off-shell and non-resonant effects. Note that if the neutrino momenta
are omitted, the relation $A_{FB} = -\bar{A}_{FB}$ is not fulfilled
any more, both at LO and NLO. This can also observed directly in the
angular distribution of $lj_b$-pairs, see \cref{fig:BL-Theta} in
appendix \ref{app:tt_further}, where there is a slightly more
pronounced dip at the lower edge of $\cos\theta_{l^-j_{\bar{b}}}$ than
at the one of $\cos\theta_{l^+j_b}$. The differences come from
combinatorial issues in the event reconstruction, where the neutrino
momentum in the MC truth information allows to determine the top
helicity and hence the flight direction of the lepton
(cf.~e.g. \cite{Garcia:2013svh}). Such information is unavailable when the
neutrino kinematics are omitted.

\subsection{Higgs associated top-pair production and decay}
\label{sec:nlo_tth}

\begin{figure*}[tbp]
   \includegraphics[width=\relplotwidth\textwidth]{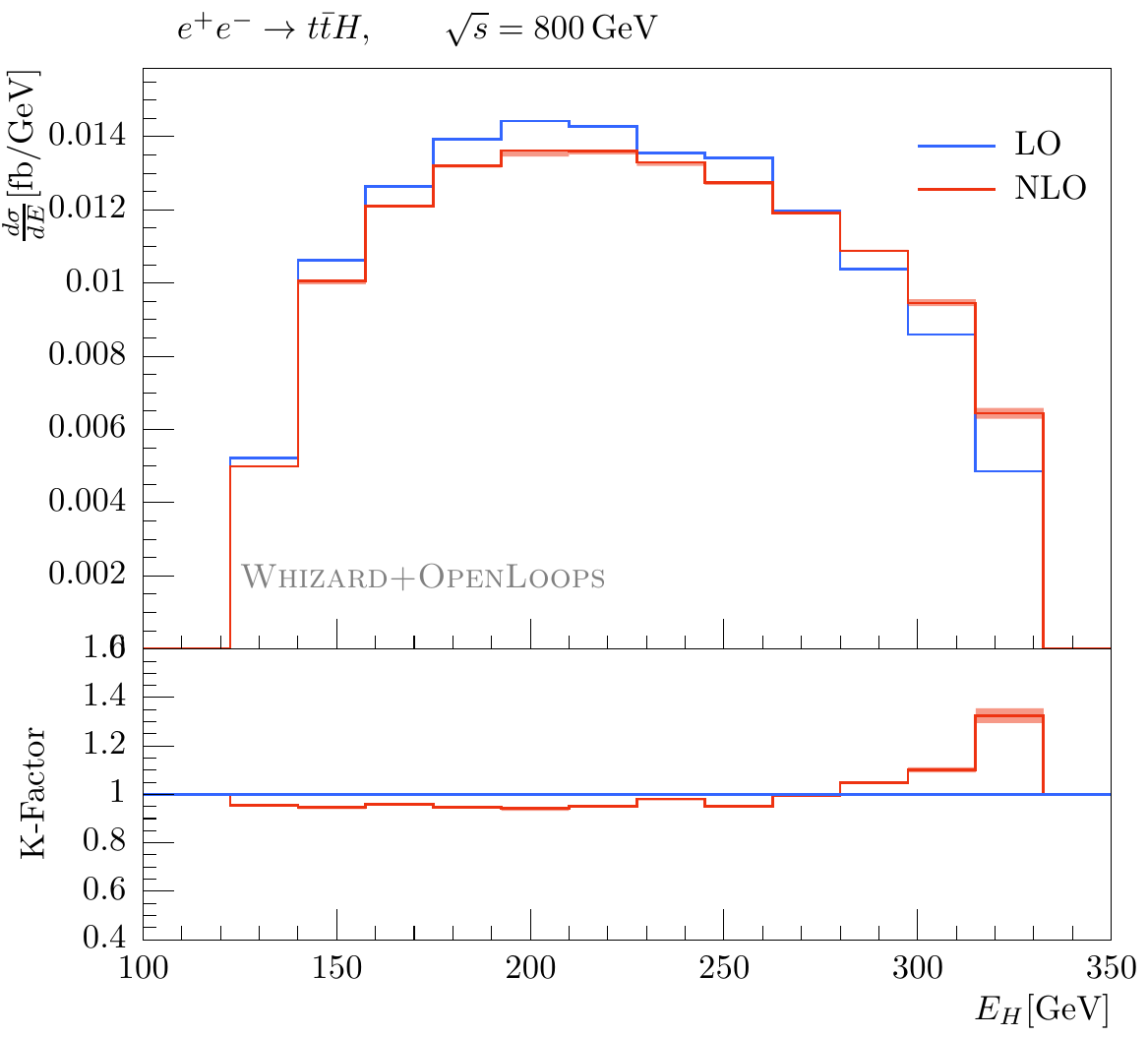}
         \quad
   \includegraphics[width=\relplotwidth\textwidth]{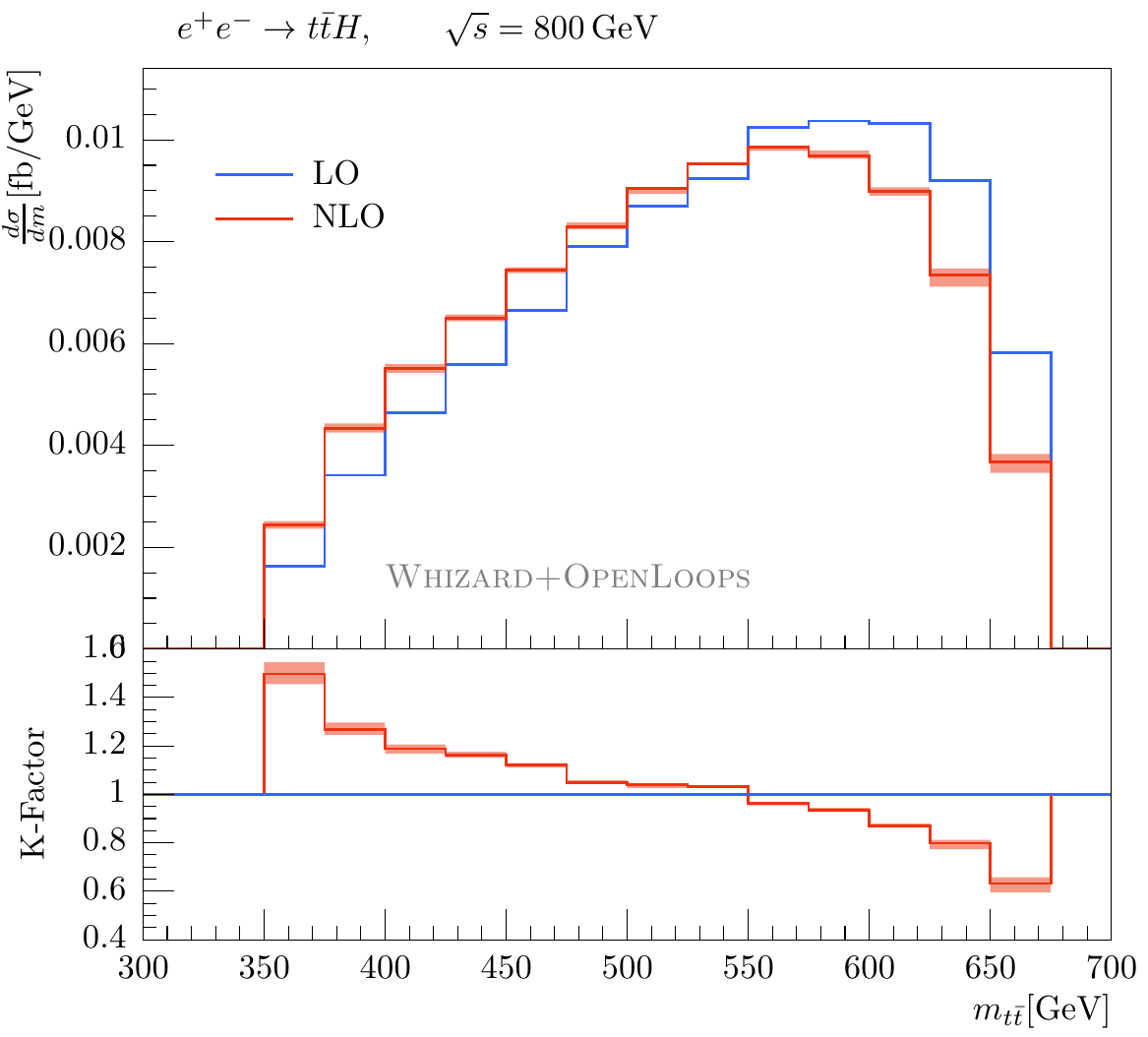}
   \caption{The energy of the Higgs boson, $E_H$, and the invariant mass of the top-quark pair, $m_{t\bar{t}}$, in \eetth. Curves and bands as in Fig.~\ref{fig:tt-firstplot}.}
\label{fig:tth-E-h-m-tt}
\end{figure*}

We start our analysis of differential Higgs associated top-pair
production by considering in \cref{fig:tth-E-h-m-tt} the energy of
the Higgs boson, $E_H$, and the invariant mass of the $t\bar{t}$
system, $m_{t\bar{t}}$,  in the on-shell process \eetth. The energy of
the Higgs boson is the key observable to identify $t\bar{t}$ threshold
effects, and it is of great phenomenological 
relevance for realistic experimental analyses including Higgs boson
decays. From the point of view of $t\bar{t}$ dynamics, the Higgs acts
as a colorless recoiler, reducing the effective center-of-mass energy
for the $t\bar{t}$ system. For $m_{t\bar{t}} \to 2m_t \approx
\ValGeV{346.4}$ the top-quark pairs are more and more
non-relativistic, yielding large logarithmic enhancements in the loop
matrix elements. In fact, the energy of the Higgs boson  and the
top-pair invariant mass are at LO directly related by 
\begin{equation}
  \label{eq:EH-ET-relation}
  E_{H} = \frac{1}{2\sqrts} \left(s + m_H^2 - m_{t\bar{t}}^2\right)\,.
\end{equation}
Thus, small $t\bar{t}$ invariant masses correspond to large Higgs
energies. And indeed, for large Higgs energies and small \mtt, in
\cref{fig:tth-E-h-m-tt} we observe sizeable positive NLO QCD
corrections up to $+35\%$ and $+50\%$ for the $E_{H}$ and
$m_{t\bar{t}}$ distributions, respectively. Such large NLO QCD
corrections should ideally be resummed for a precise theoretical
prediction. 

For the on-shell process \eetth the lower kinematic bound of the $E_H$
distribution is given by $E^{\rm min}_H = m_H = \ValGeV{125}$ and its
upper bound by $E^{\rm max}_H = \ValGeV{335}$, which follows from 
$m_{t\bar{t}}^{\rm min}=2 m_t$.
Noteworthy, for small Higgs boson energies we observe an apparent
mismatch of the NLO QCD corrections with respect to large top-pair
masses. While for small Higgs boson energies the K-factor flattens out
to an almost constant value of about $0.95$, the K-factor for the
top-pair invariant mass distribution monotonically decreases to a
minimum value of about $0.60$. As a matter of fact, the Higgs boson
energy distribution is a fully inclusive observable that is completely
independent of the clustering applied to final state QCD radiation. On
the other hand, the \mtt distribution does not include hard gluon
radiation off the $t\bar t$ system, while soft and collinear gluons
are recombined with the top quarks. The resulting systematic shift in
the \mtt distribution towards lower values results in the observed
differences with respect to the $E_H$ distribution.

\begin{figure*}[tbp]
\centering
     \includegraphics[width=\relplotwidth\textwidth]{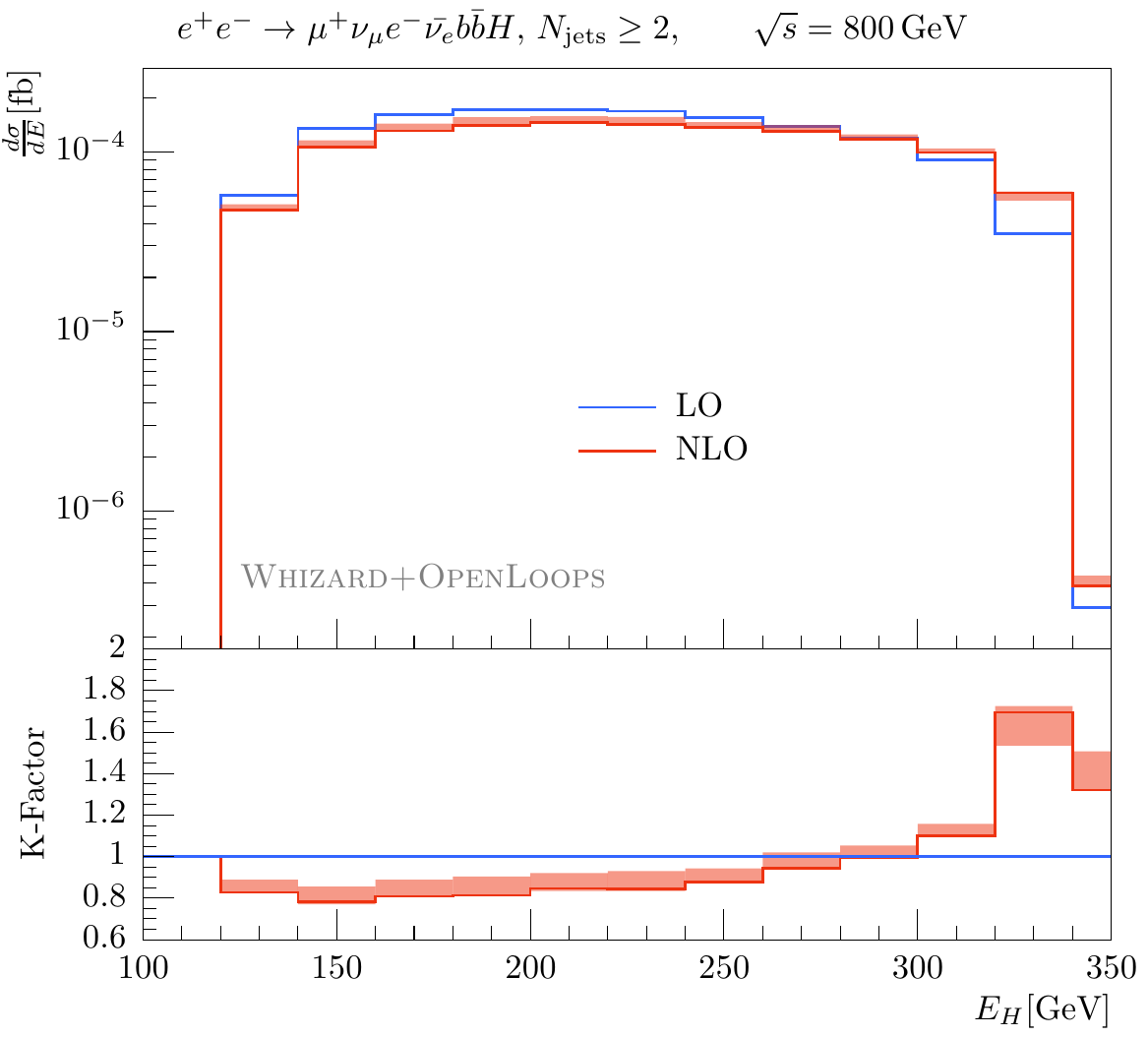}
         \quad 
     \includegraphics[width=\relplotwidth\textwidth]{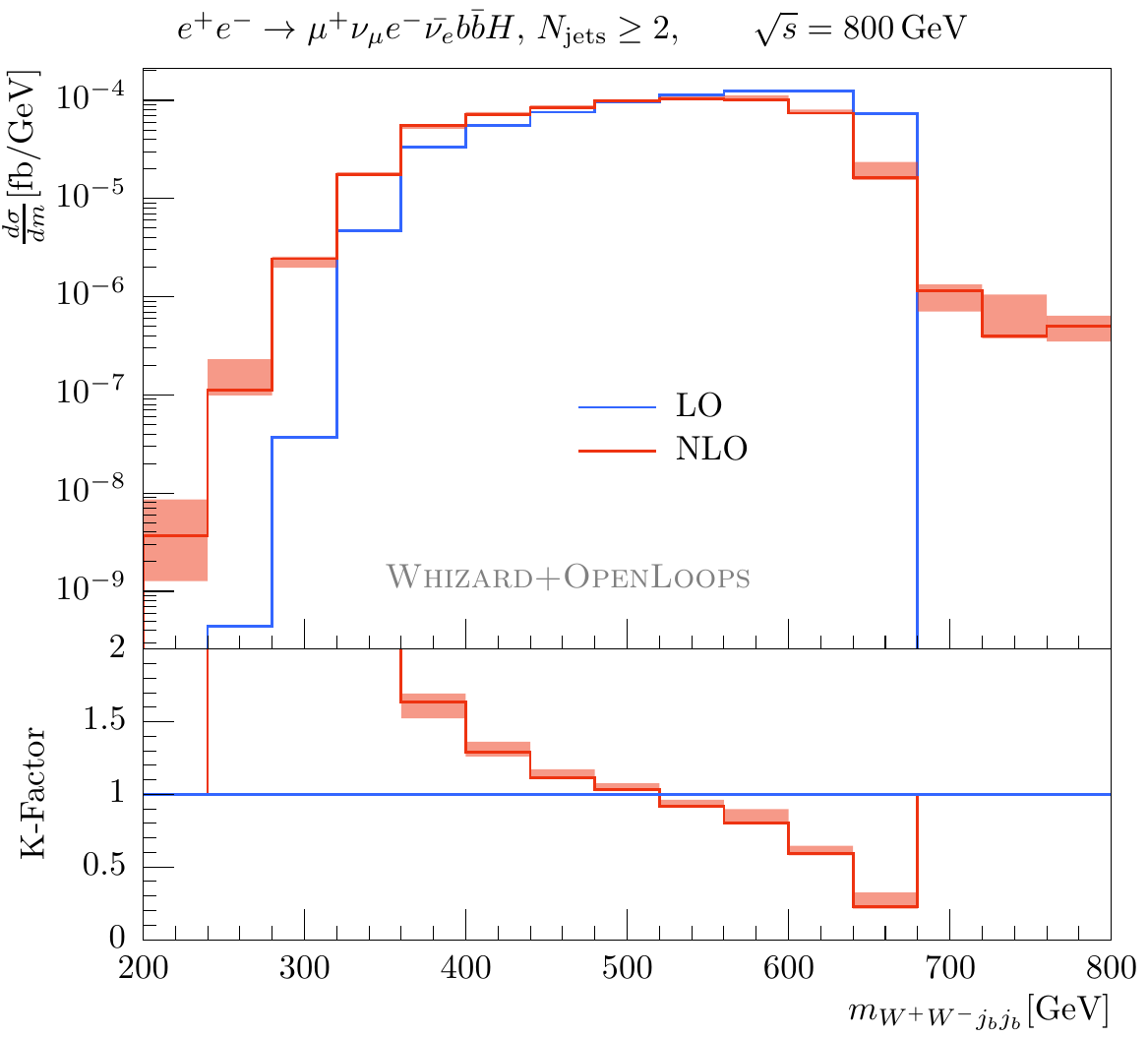}
   \caption{
   The energy of the Higgs boson, $E_H$, and the invariant mass of the reconstructed top-quark pair, \mttrec, in \eellllbbH. Curves and bands as in Fig.~\ref{fig:tt-firstplot}.
}
\label{fig:Hbblnulnu-E-h-E-BWpBWm}
\end{figure*}

The corresponding distributions for the off-shell process \eellllbbH
are shown in \cref{fig:Hbblnulnu-E-h-E-BWpBWm}. Again, we observe a
strong enhancement for large Higgs boson energies and small
reconstructed top-pair masses, together with a strong suppression for
large reconstructed top-pair masses. In contrast to the on-shell
process, already at LO kinematic boundaries are washed out due to
off-shell and non-resonant contributions. In particular, the $E_H$
distributions range to energies above $\ValGeV{335}$, with strongly
increasing NLO corrections. 
The \mttrec distribution at LO falls off quickly below $\mttrec=2
m_t$, while at NLO it reaches to very small values. As already
discussed in the context of \cref{fig:BW-inv}, this phase-space 
region is
populated at NLO due to kinematic shifts of the reconstructed masses
originating from the recombination of radiation from different stages
of production and decay.

\begin{figure*}[tbp]
\centering
   \includegraphics[width=\relplotwidth\textwidth]{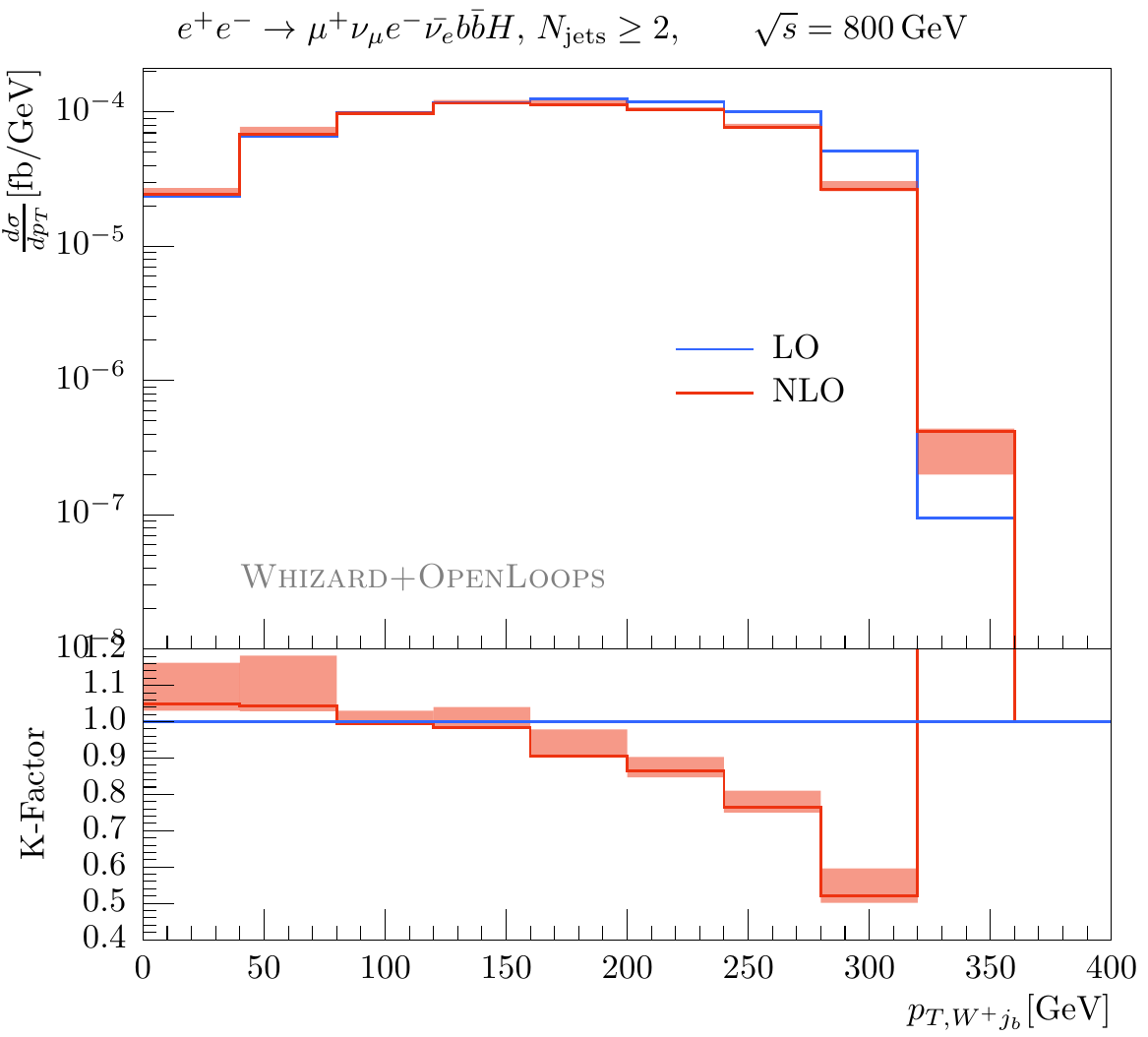}
   \quad 
     \includegraphics[width=\relplotwidth\textwidth]{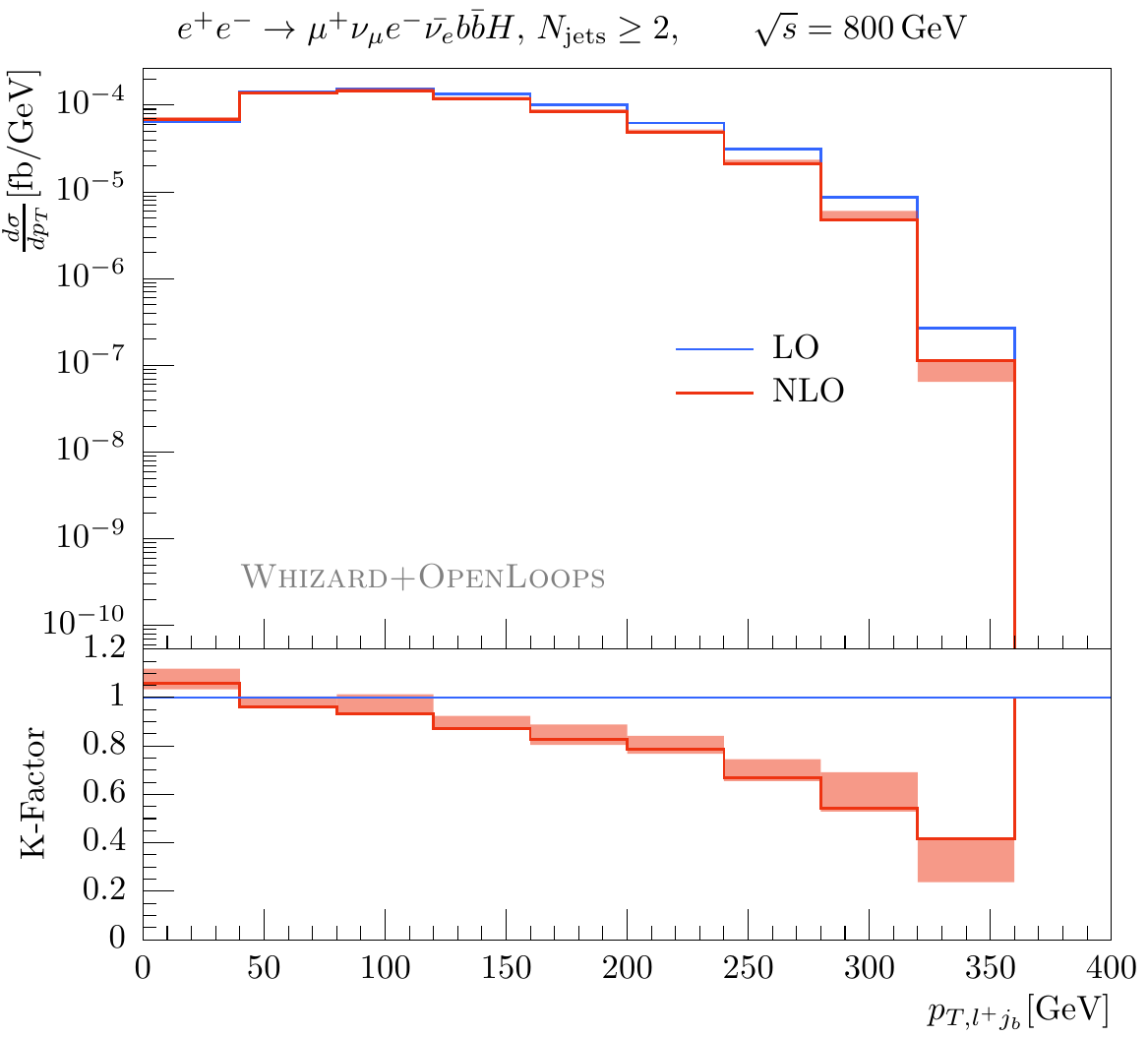} 
   \caption{Transverse momentum distributions of the reconstructed top quark (left) and of the 
   bottom-jet--lepton system (right), \pTblp, in \eewwbbH. Curves and bands as in Fig.~\ref{fig:tt-firstplot}.}
\label{fig:tth-top-pT}
\end{figure*}

In \cref{fig:tth-top-pT} we show the transverse momentum distribution
of the reconstructed top quark and the directly observable
bottom-jet--lepton system in the off-shell process \eellllbbH. 
Comparing these distributions with the corresponding ones for top-pair
production, shown in \cref{fig:tt-pT} and \cref{fig:tt-pTbl}, we
observe distinct (LO) shape differences. Instead of a pronounced peak
in the \pTtoprec distribution we observe a plateau between about
\ValGeV{100} and \ValGeV{250}. At larger transverse momenta, the
distribution drops sharply to its kinematical bound at around
\ValGeV{325}. NLO QCD corrections shift both the \pTtoprec and the
\pTblp distribution towards smaller values inducing shape effects up
to $-50\%$ at large \pTtoprec and up to $-60\%$ at large \pTblp.

\begin{figure*}[tbp]
\centering
   \includegraphics[width=\relplotwidth\textwidth]{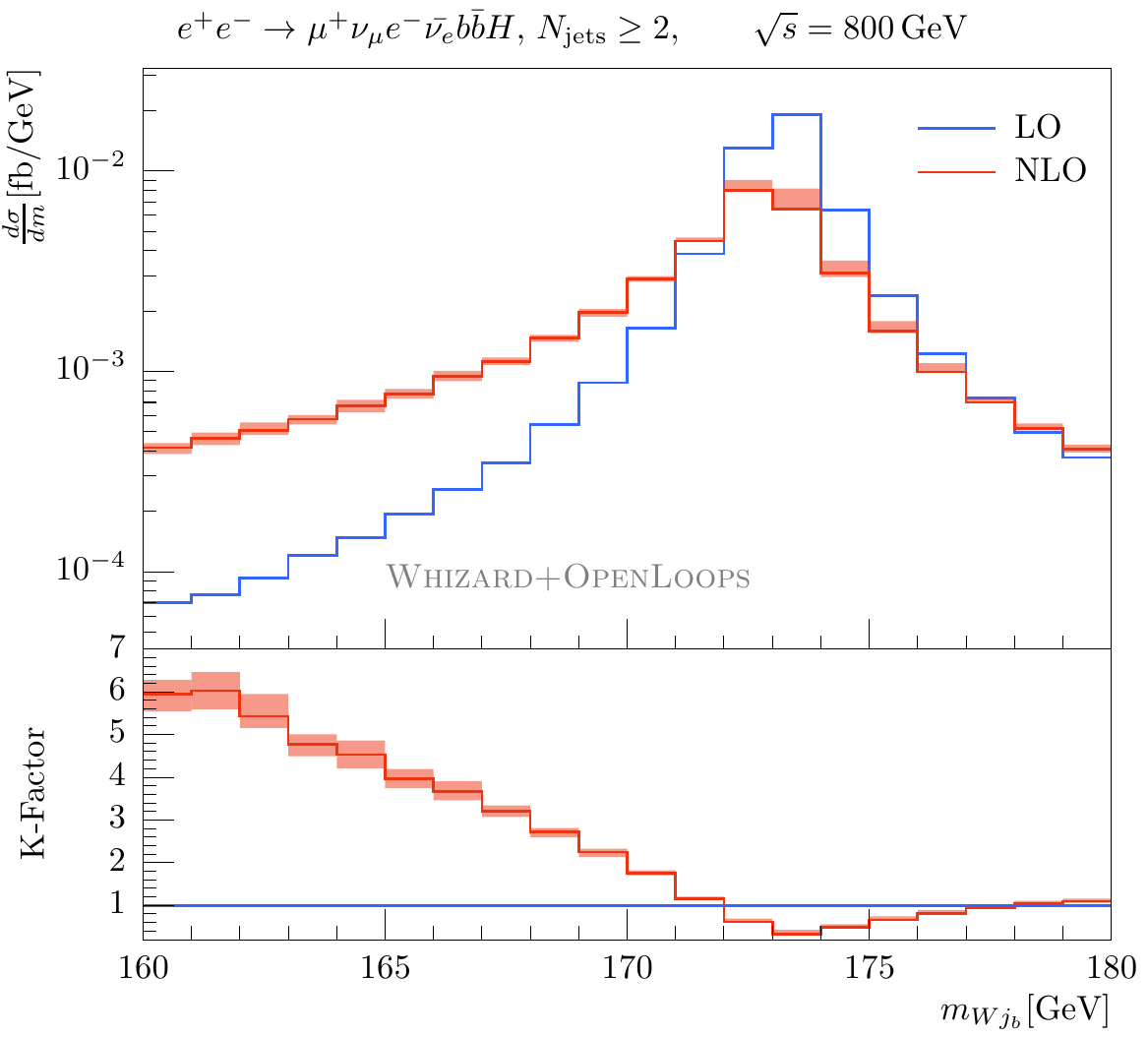}
         \quad 
   \includegraphics[width=\relplotwidth\textwidth]{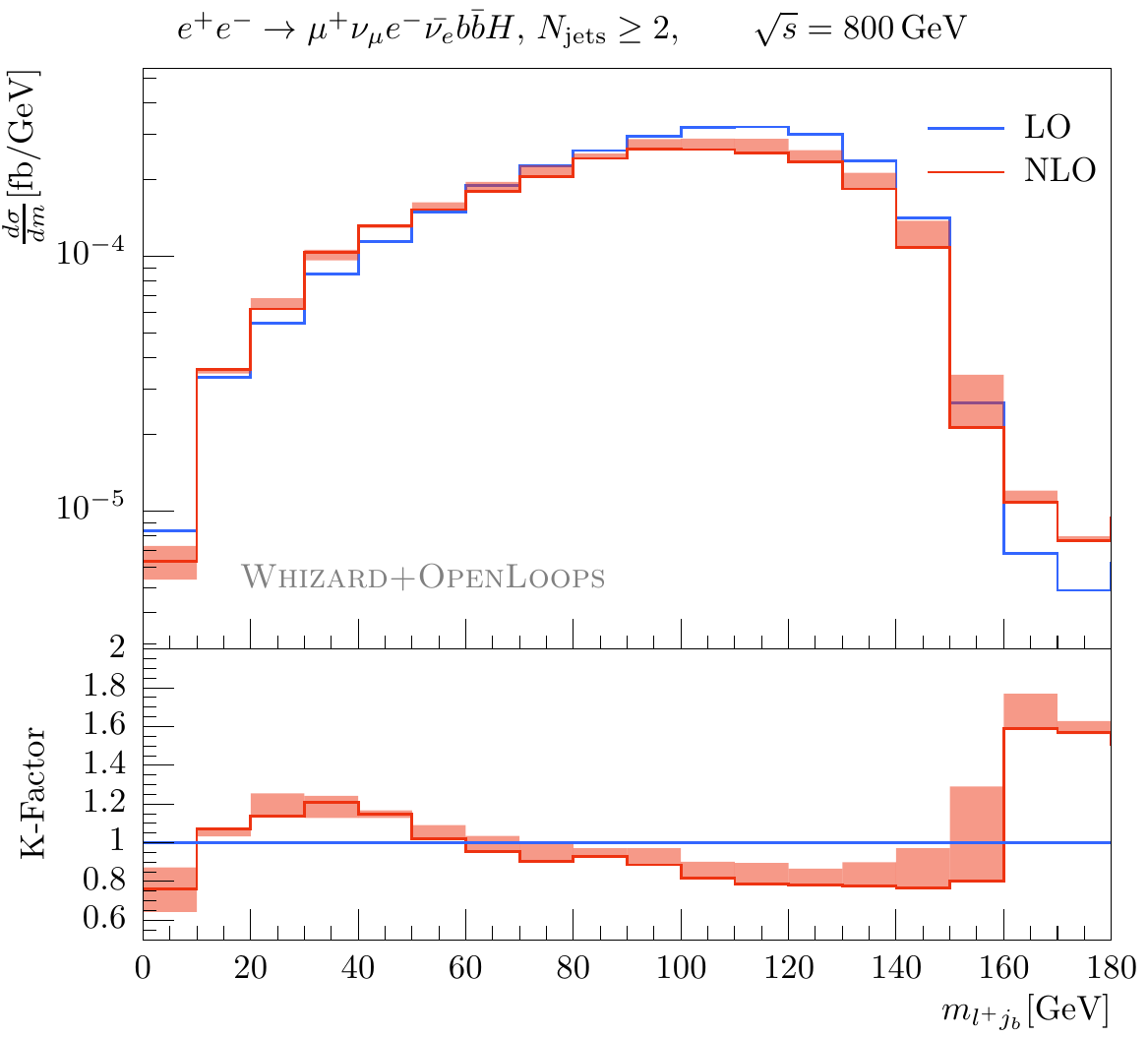}
   \caption{Reconstructed top invariant mass (left) and invariant mass
     distribution of the $b$-jet--$\ell^+$ system (right) in
     \eellllbbH. Curves and bands as in Fig. \ref{fig:tt-firstplot}. 
}
\label{fig:tth-BW-inv}
\end{figure*}

Finally, in \cref{fig:tth-BW-inv} we turn to the reconstructed
kinematic top mass, \mtoprec, and its directly observable relative,
\mblp. We observe similar NLO shape distortions as already discussed
in the case of top-pair production, shown in \cref{fig:BW-inv}. For
$\mtoprec < m_t$, i.e. below the top resonance, we observe a strong
NLO enhancement that translates to $20\%$ shape corrections in the
case of the \mblp distribution. As already noted before, the size of
these corrections strongly depends on the details of the employed jet
clustering. 

Further differential distributions are shown in the appendix,
cf. \cref{fig:tth-Higgs-pT}.

\section{Conclusions}
\label{sec:conclusions}

In this article, we have presented the first application of the \wz{}
NLO framework based on a process-independent interface between \wz{}
and the amplitude generator \ol{}. 
We have presented a precision study for a future high-energy lepton
collider considering for the first time at NLO QCD the processes
$\epem \to \llllbb$ and $\epem \to \llllbbH$, i.e. off-shell top-pair
and Higgs associated top-pair production. 
Finite-width effects for intermediate top quarks and $W$ bosons,
single-top and non-resonant contributions as well as their
interferences together with spin correlations have been taken into
account consistently at NLO. 

We have presented a study of inclusive cross sections varying as a
function of the center-of-mass energy considering different
approximations for the top off-shellness and an in-depth study at the
differential level for $\sqrts = \ValGeV{800}$. Off-shell effects play
an important role even for the inclusive cross sections as 
the narrow-width approximation does not suffice to describe
interference effects and background diagrams at energies far above
threshold. 

NLO QCD corrections also influence the dependence of the cross section
on the top Yukawa coupling for the $t\bar t H$ processes, which has direct
consequences for the achievable accuracy in measuring this coupling.
In particular, we have shown that the NLO QCD corrections induce
negative interference terms yielding a deviation from the quadratic
Yukawa coupling dependence of the cross section ($\kappa^{\rm NLO} <
0.5$), both in the on-shell treatment of \ttbarH production and the
corresponding off-shell process.

Many facets of the strong physics potential of linear lepton colliders
are based on polarized lepton beams. In order to describe beam
polarization, the Binoth Les Houches Accord for the interface between
\wz{} and \ol{} was generalized. As expected, we found that beam
polarization has no effect on the relative size of NLO QCD
corrections. It is however important to incorporate  a treatment of
polarization effects in the NLO framework of \wz{}, as well as QED
initial state radiation (and to a lesser 
extent also beamstrahlung, which always factorizes), in order to allow
for realistic Monte Carlo simulations in the environment of a lepton
collider. In particular, as soon as one includes higher-order
electroweak corrections, a consistent treatment of beam polarization
is mandatory.  

In addition to these inclusive studies, NLO QCD effects on
differential observables have been investigated. Our results  
show that NLO effects can yield (clustering-dependent) corrections of
up to $\pm 50\%$ for a variety of observables. Even larger corrections
occur due to non-relativistic top-threshold effects. To obtain reliable
predictions in these threshold regions, a resummed calculation is
required. The easy-to-use automation of threshold matching for
on-shell top quarks as well as the matching of this calculation to the
relativistic continuum will be supported by \wz~in the near
future. Off-shell effects are most relevant in top-mass related
observables, which is of crucial importance for the determination of
$m_t$. On the other hand, Higgs observables are mostly unaffected by
off-shell contributions, but can be influenced significantly by NLO
QCD corrections. Studying the top-quark forward-backward asymmetry, we
found the effect due to off-shell contributions dominating over NLO
QCD corrections.

The study at hand was performed at the fixed-order NLO QCD level. The
matching to parton showers based on an independent implementation of the
\textsc{POWHEG} method within \wz{} will be considered in the future. This will
allow for realistic experimental analyses including resummation of soft and
collinear radiation at the NLO+LL level. Besides the parton shower matching also
NLO electroweak corrections and their interplay with the QCD corrections should
be considered in the future as they are well known to play an important role for
the considered processes.

In addition to its phenomenological relevance, the presented calculation
demonstrates the flexibility of \wz{} for NLO QCD computations at lepton
colliders and the smooth interplay with \ol{}. All distributions can be
reproduced easily with these publicly available tools and finely adjusted to the
experimental requirements.

\section*{Acknowledgments}

We are indepted to Stefan Kallweit for providing us with independent
cross checks for the results 
presented in this paper, based on the Monte Carlo program {\sc Munich}.
We thank Tom\'{a}\v{s} Je\v{z}o for fruitful discussions about
the resonance-aware method. We are also grateful to Andrew
Papanastasiou for providing us with details of the calculation
presented in~\cite{1511.02350} performed within
\textsc{Madgraph5\_aMC@NLO}. Moreover, we wish to thank Ansgar Denner,
Stefan Dittmaier and Lars Hofer for providing us with pre-release
versions of the one-loop tensor-integral library {\sc Collier}.
This research was supported in part by the Swiss National Science Foundation
(SNF) under contracts BSCGI0-157722 and PP00P2-153027, 
by the Research Executive Agency of the European Union
under the Grant Agreement
PITN--GA--2012--316704 ({\it HiggsTools}), 
and by the Kavli Institute for Theoretical
Physics through the National Science Foundation's Grant No. NSF PHY11-25915.

\appendix
\section{Details of the Resonance-aware IR subtraction}
\label{sec:bbmumu_example}

\subsection{Soft Mismatch}
The additional soft mismatch component, which restores parts of the real-subtracted
correction not covered by the resonance-aware FKS mappings, integrates
the expression~\cite{1509.09071}
\begin{multline}
  \label{eq:soft_mismatch}
  R^{\mis}_{\alpha_r} = \int d\Phi_B \int_0^\infty d\xi \int_{-1}^1 dy
  \int_0^{2\pi} d\phi \frac{s\xi}{(4\pi)^3} 
  \Biggl\{\mathcal{R}^{\rm soft}_{\alpha_r} \left(\exp\left[-\frac{2k \cdot k_{\rm
          res}}{k_{\rm res}^2}\right] - \exp\left[-\xi\right]\right) \\ 
  - \frac{32\pi\alpha_s C_F}{s\xi^2} \mathcal{B} 
  \left(\exp\left[-\frac{\bar{k}_{\rm em} \cdot k_{\rm res}}{k_{\rm res}^2}
      \frac{k^0}{\bar{k}_{\rm em}^0}\right] - \exp\left[-\xi\right]\right)
  (1-\cos\theta)^{-1}\Biggr\}\,,
\end{multline}
which is evaluated for
each individual singular region, denoted by the index $\alpha_r$
here. Correspondingly, $\mathcal{R}_{\alpha_r}^{\rm soft}$  
is the soft limit of the real matrix element in this region, as it is
also used in the computation of real subtraction terms, and
$\mathcal{B}$ the underlying Born matrix element. $k$ and $k_{\rm
res}$ are the momenta of the radiated gluon and the intermediate
resonance, respectively, and $\bar{k}_{\rm em}$ is the  momentum of
the emitter in the Born phase space. The integration is performed over
the whole real phase space, which factorizes into the Born phase space
$d\Phi_B$ and the real-radiation variables $\xi$, $y$, and
$\phi$. Note that, in contrast to the traditional FKS subtraction,
where $\xi = 2k^0 / \sqrts \leq 1$, a generalized $\xi \in [0,
\infty)$ is used, which originates from using integral identities. 
\begin{figure}
  \centering
  \includegraphics[width=\relplotwidth\textwidth]{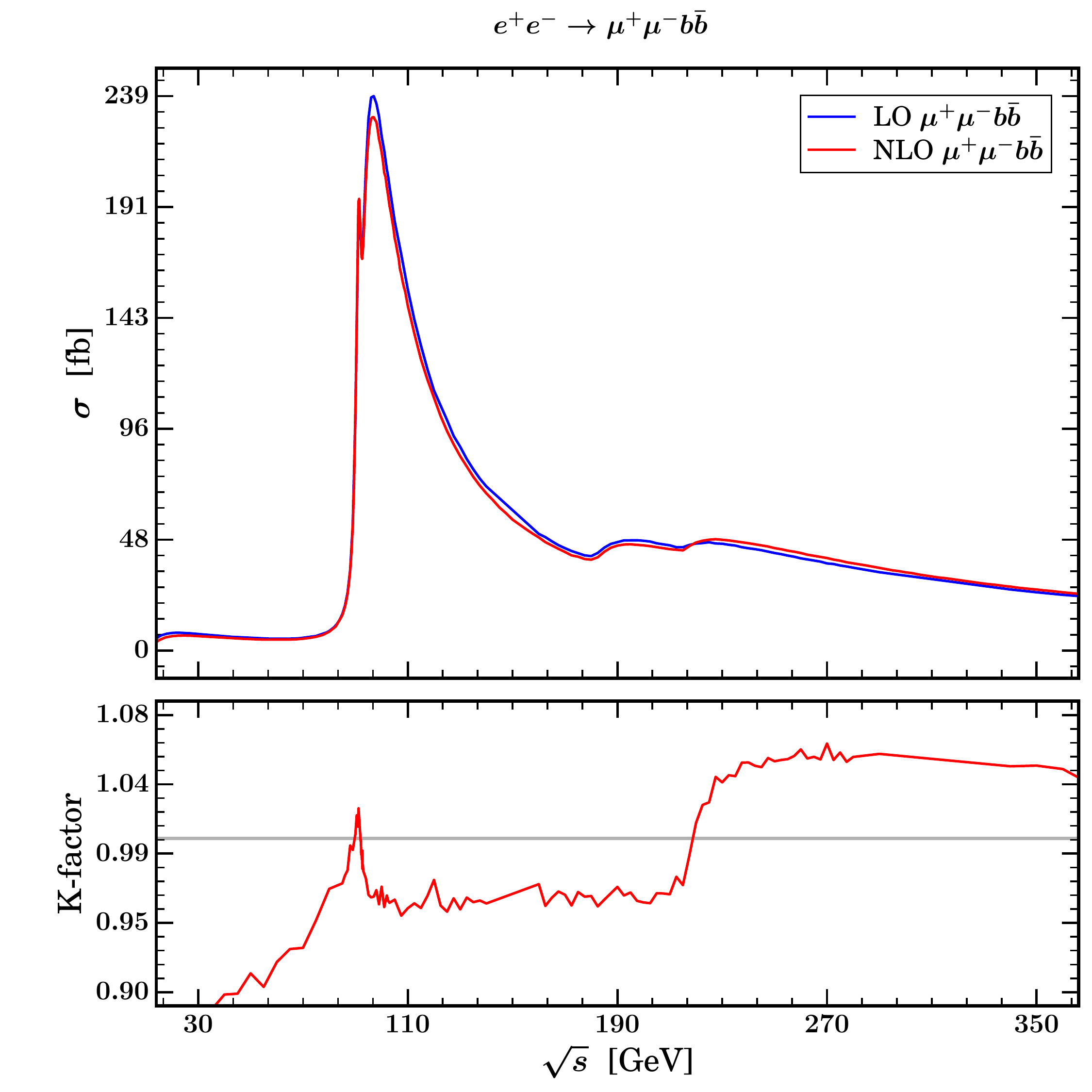}
  \caption{Total cross section of the process $e^+ e^- \to \mu^- \mu^+ 
    b \bar{b}$ at LO and NLO using resonance-aware FKS
    subtraction. In contrast to the validation described in the text,
    here the physical muon mass and Higgs width have been used.} 
  \label{fig:bbmumu_xsec}
\end{figure}
Therefore, the soft mismatch has to be evaluated with its own phase
space and must be treated as a separate integration component in
\wz{}, additionally to Born, Real and Virtual.

\subsection{Validation and Efficiency}

We have checked our implementation of resonance-aware FKS subtraction
extensively using  the production of two massive quarks in
association with two muons as a benchmark process, i.e.  $\epem \to b
\bar{b} \mu^+ \mu^-$. This process has only one resonance topology
with two different resonance histories, $Z \rightarrow b\bar{b}$ and
$H \rightarrow b\bar{b}$, comprising $Z$ pair production as well as
Higgsstrahlung.  We have set $m_b = \ValGeV{4.2}$, so that
collinear divergences do not occur. For the validation of our
implementation, in order to avoid any sort of cuts, we have
set the muon mass equal to $m_\mu = \ValGeV{20}$.
To ensure a converging integration also in the case of the non
resonance-aware approach, we have approximated the limit $\Gamma_H \to
\infty$ by numerically fixing the Higgs width to $\Gamma_H = \ValGeV{1000}$.  
In this way, the standard subtraction can be compared to the improved
one, see table~\ref{tab:sigma_mumubb}, where $\sigma_{\text{real}}$
denotes the full real-subtracted matrix element and 
$\sigma_{\text{mism}}$ the result of the integration of the soft
mismatch component. Adding the real and soft-mismatch components
for the resonance-aware FKS subtraction, perfect agreement with the
real radiation component of the resonance-unaware subtraction is found. Here, we want to emphasize the significantly 
higher number of integration calls required in the standard approach
to reach the same accuracy as in the resonance-aware subtraction
scheme.  
\begin{table}[htbp]
  \caption{Real-subtracted integration component and, in the case of
  resonance-aware subtraction, soft mismatch, for $\Gamma_H = \ValGeV{1000}$. A
  fictitious muon mass $m_\mu = \ValGeV{20}$ has been used to avoid
  cuts.  All other parameters are as in section~\ref{ssec:input_parameters}.} 
\label{tab:sigma_mumubb}
  \begin{center}
  \begin{tabular}{c c c c}
    \toprule{}
    & $\sigma_\text{real} [\fb] $ & $\sigma_\mis [\fb]$ & $n_\text{calls}$\\
    \midrule{}%
    standard & $-1.90485 \pm 0.99 \%$ & n/a & $5 \times 100000$ \\
    resonances & $-9.15077 \pm 0.52 \%$ & $-0.97930 \pm 0.94 \%$ & $5 \times
    20000 (\text{real}) + 5\times 20000 (\mis)$\\
    \bottomrule{}
  \end{tabular}
  \end{center}
\end{table}

Fig.~\ref{fig:bbmumu_xsec} shows a scan of the total cross
section. For this scan, we used the physical muon mass and Higgs
width. There are two distinct peaks at $m_Z$ and $m_Z + 2m_b$, as
well as two less pronounced enhancements at $m_Z + m_H$ and $2m_Z$. NLO
QCD corrections are in the range of $+5\%$ for $\sqrts > 2m_Z$ and
approximately $-4\%$ for $m_Z + 2m_b < \sqrts < 2m_Z$. Below $\sqrts =
m_Z$, the K-factor is significantly smaller than 1.

\section{Further NLO predictions for \eellllbb}
\label{app:tt_further}

\begin{figure*}[htp]
  \centering
  \includegraphics[width=0.41\textwidth]
                  {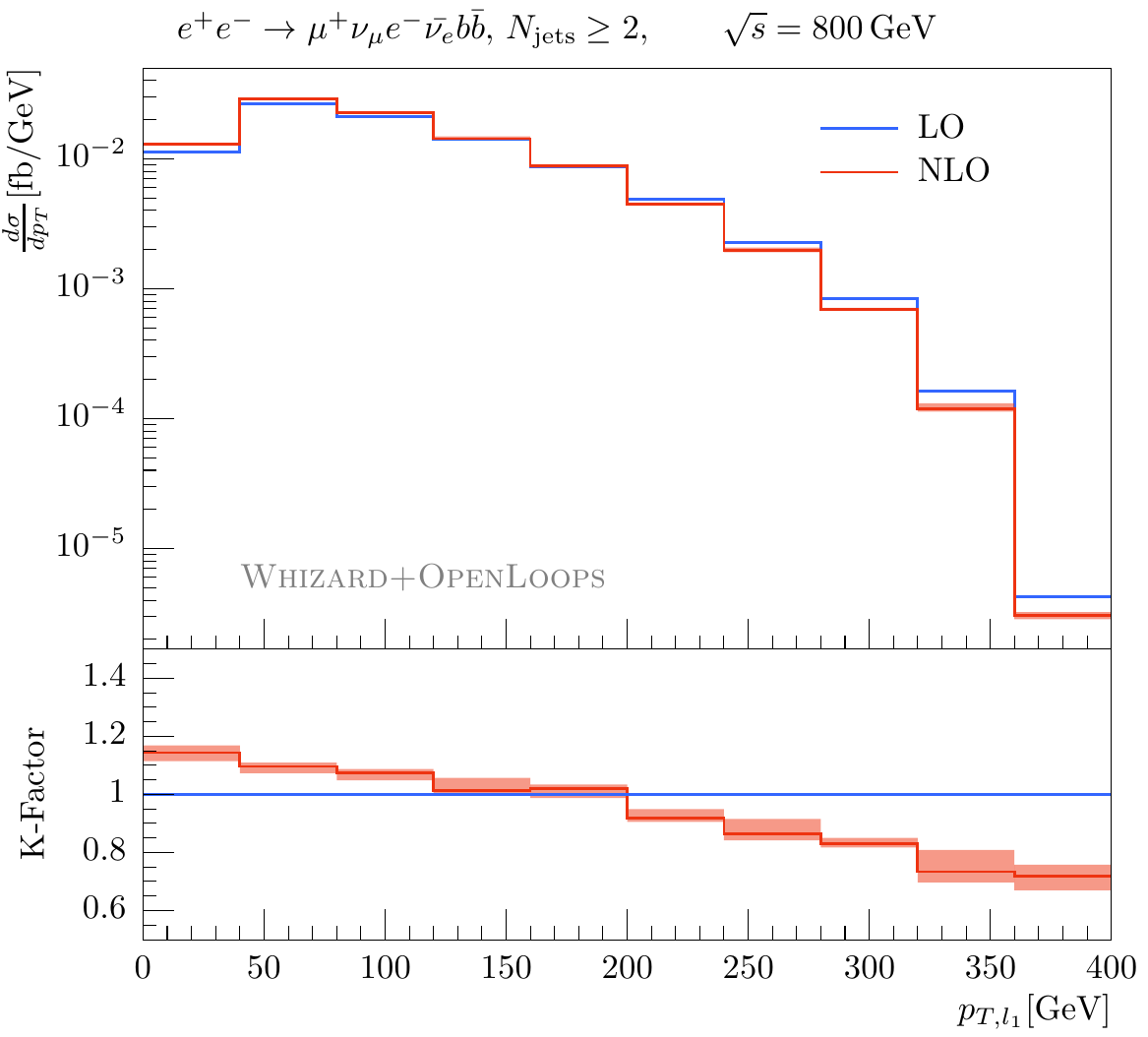}
  \quad
  \includegraphics[width=0.41\textwidth]
                  {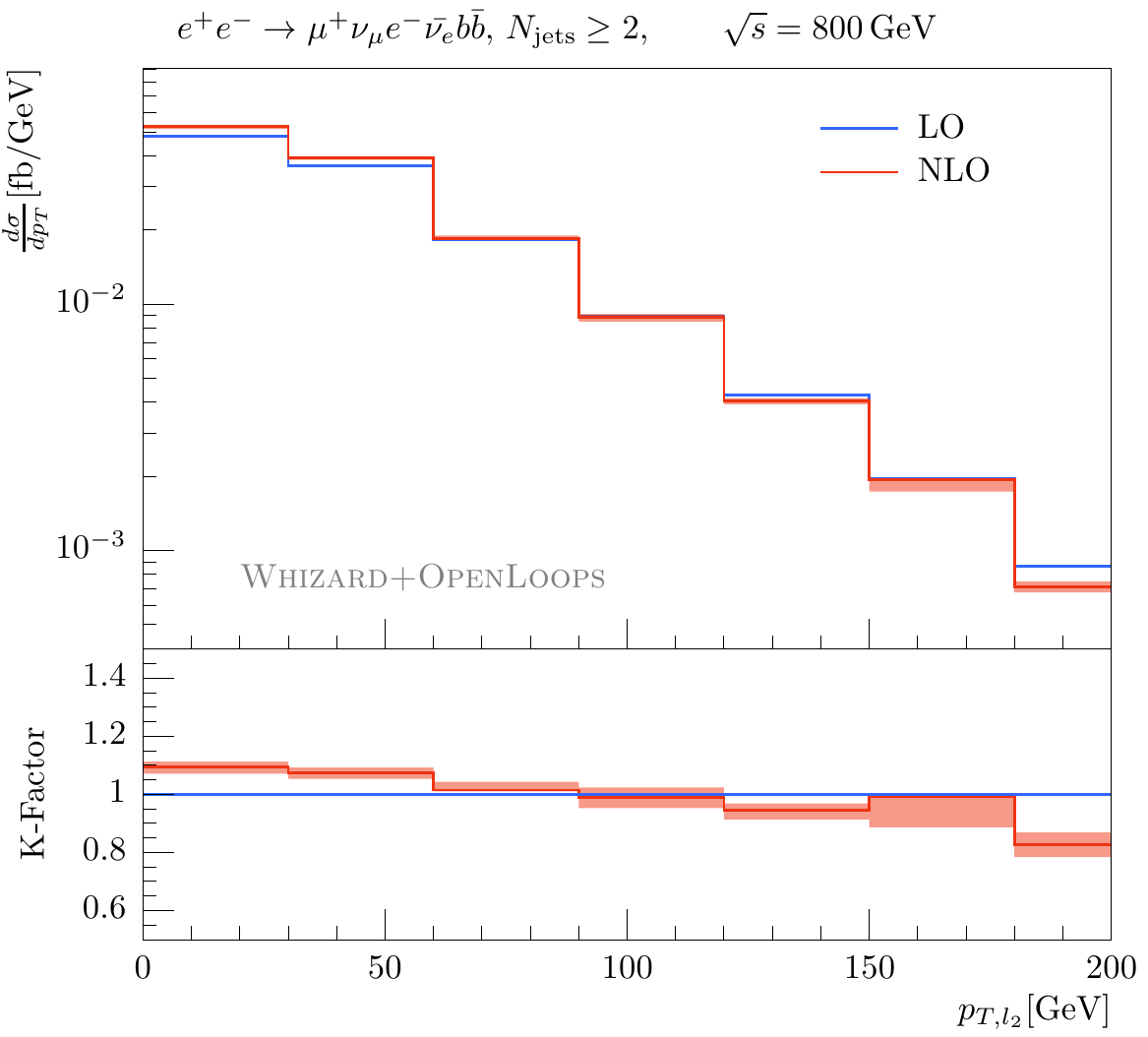}
  \\
  \caption{
    Transverse momentum distributions of the hardest and second hardest
    lepton in \eellllbb. Curves and bands as in
    Fig.~\ref{fig:tt-firstplot}.} 
  \label{fig:tt-pTl}
\end{figure*}

\begin{figure*}[htp]
  \centering
  \includegraphics[width=\relplotwidth\textwidth]
                  {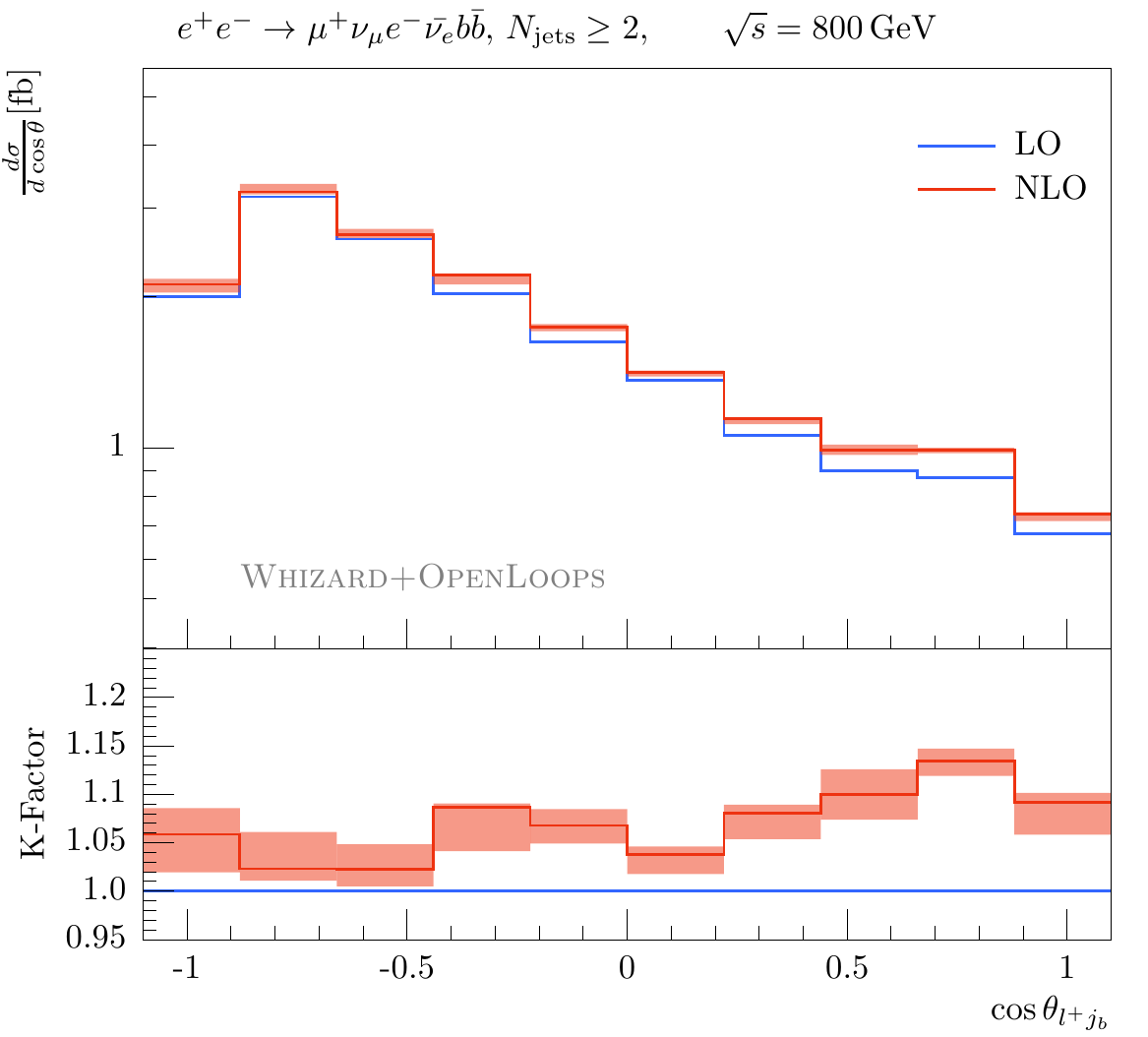}
  \quad
  \includegraphics[width=\relplotwidth\textwidth]
                  {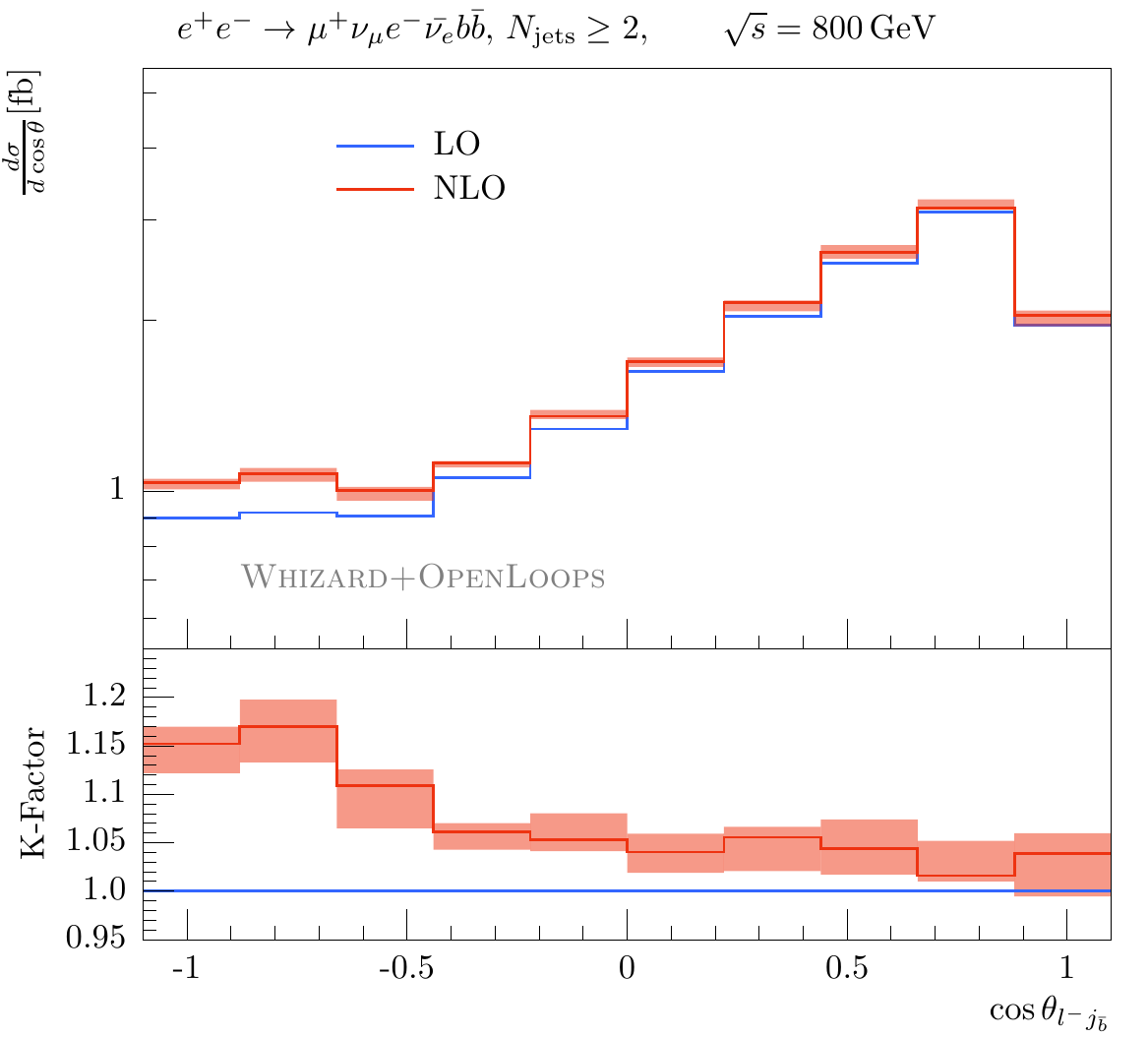}
  \caption{Distributions differential in the angular separations
    \thetablp (left) and \thetablm (right) for \eellllbb.  
    Curves and bands as in Fig.~\ref{fig:tt-firstplot}.}
  \label{fig:BL-Theta}
\end{figure*}

\clearpage

\section{Further NLO predictions for \eellllbb}
\label{app:tth_further}

\begin{figure*}[ht]
  \centering
  \includegraphics[width=\relplotwidth\textwidth]
                  {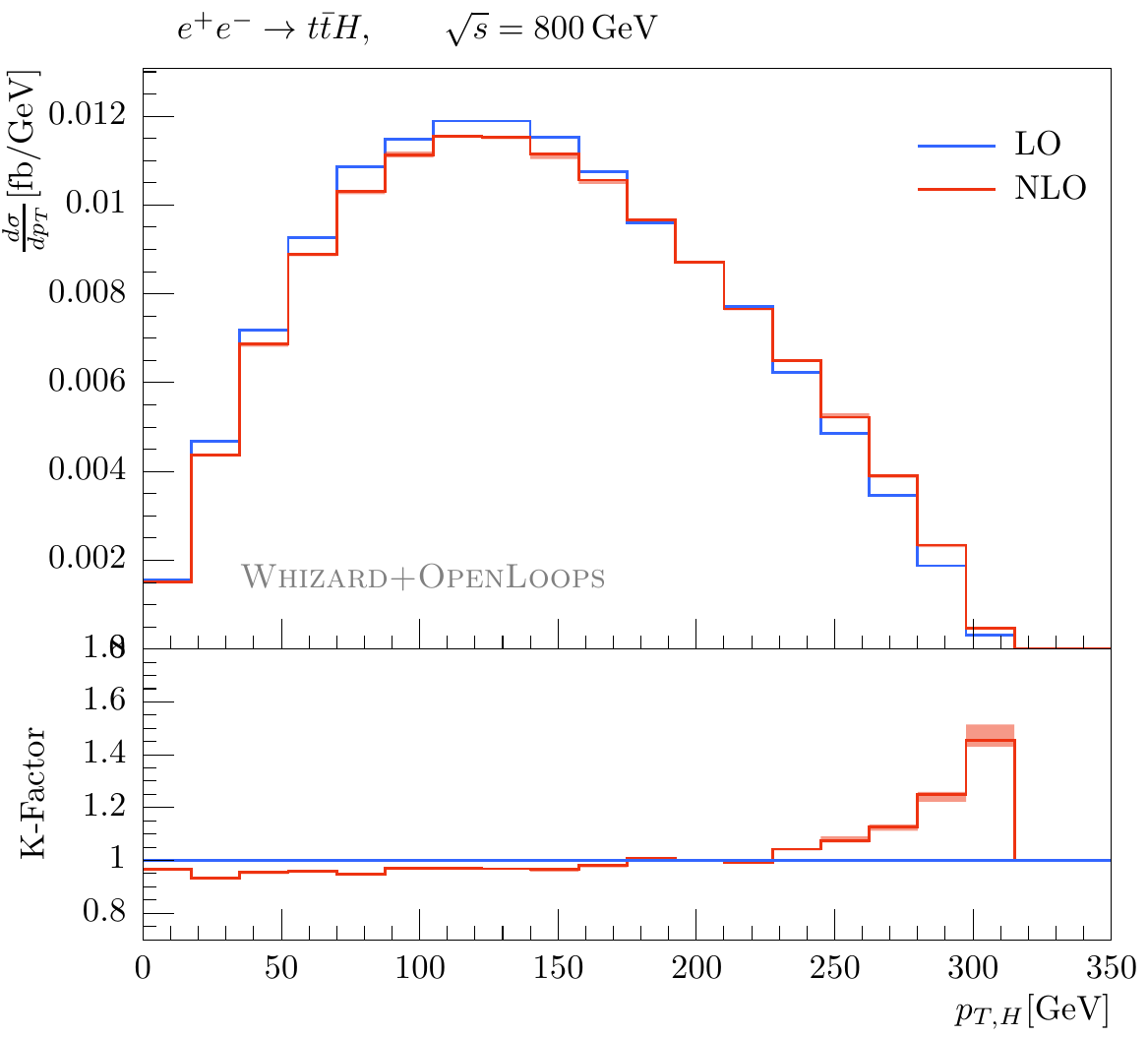} 
  \quad
  \includegraphics[width=\relplotwidth\textwidth]
                  {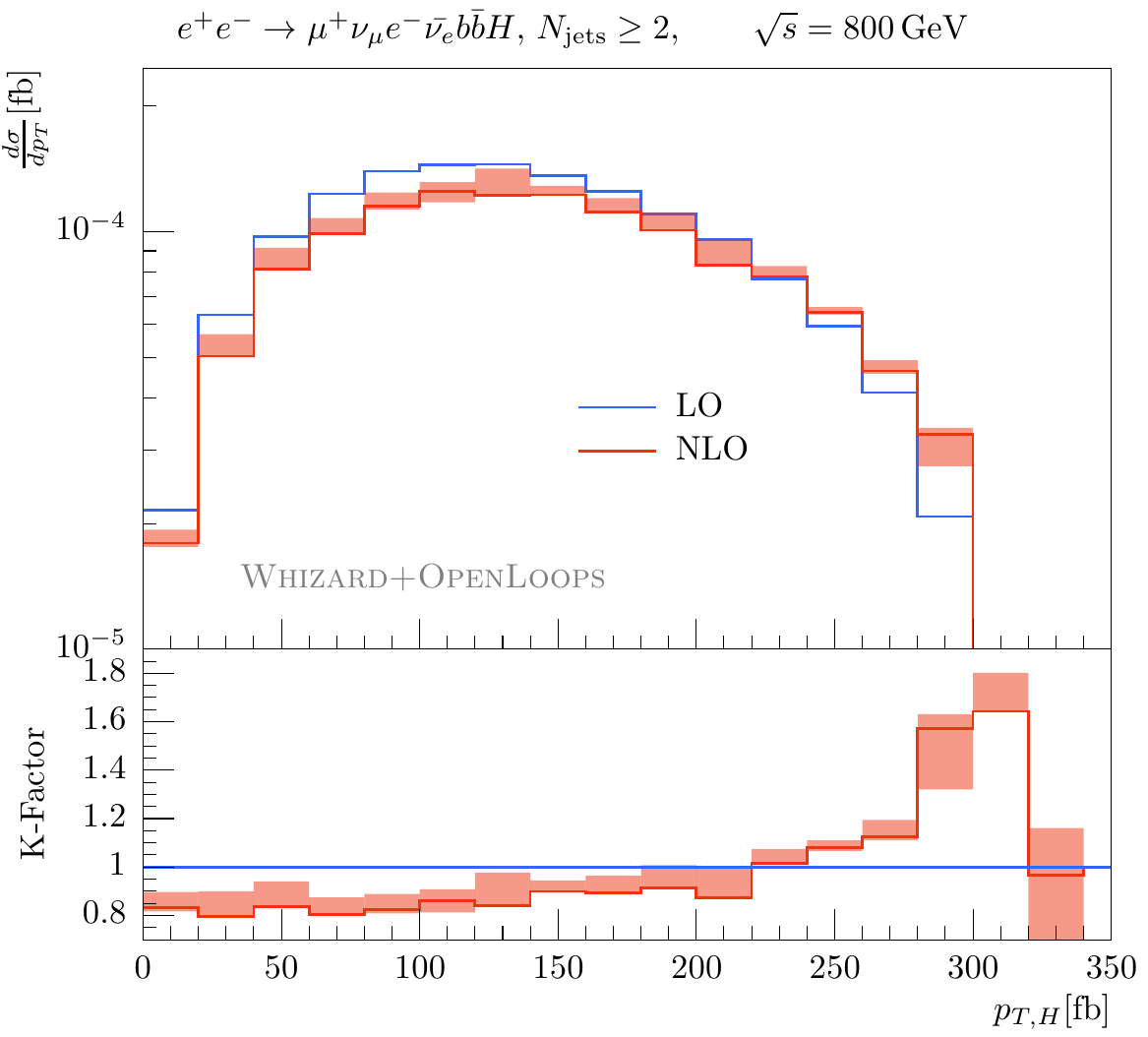}
  \caption{Transverse momentum distributions of the Higgs boson in
    \eetth (left) and in \eewwbbH (right). Curves and bands as in
    Fig.~\ref{fig:tt-firstplot}. } 
  \label{fig:tth-Higgs-pT}
\end{figure*}
\baselineskip15pt

\clearpage

\printbibliography
\end{document}